\documentclass{emulateapj}

\usepackage{enumitem}
\usepackage{color}
\usepackage{multirow}

\newcommand{\msun}{\ensuremath{\rm M_\odot}}

\newcommand{\kms}{km s$^{-1}$}

\newcommand{\lya}{Ly$\alpha$}
\newcommand{\ha}{H$\alpha$}
\newcommand{\hb}{H$\beta$}
\newcommand{\hg}{H$\gamma$}

\newcommand{\wla}{\ensuremath{W_{\rm Ly\alpha}}}

\newcommand{\ebv}{E\ensuremath{(B-V)}}

\newcommand{\super}[1]{\ensuremath{^\textrm{\scriptsize{#1}}}}
\newcommand{\sub}[1]{\ensuremath{_\textrm{\scriptsize{#1}}}}
\shorttitle{Rest-Frame Optical Spectroscopy of \lya-Emitters}
\shortauthors{R. F. Trainor et al.}

\begin{document}


\title{The Rest-Frame Optical Spectroscopic Properties of Ly$\alpha$-Emitters at
  $Z\sim2.5$: The Physical Origins of Strong \lya\ Emission\altaffilmark{1}}

\author{Ryan F. Trainor\altaffilmark{2}}
\affil{Department of Astronomy, University of California, Berkeley, 501 Campbell Hall, Berkeley, CA 94720; trainor@berkeley.edu}
\author{Allison L. Strom}
\author{Charles C. Steidel}
\affil{Cahill Center for Astrophysics, MC 249-17, 1200 E California Blvd, Pasadena, CA 91125}
\and \author{Gwen C. Rudie}
\affil{Carnegie Observatories, 813 Santa Barbara Street, Pasadena, CA 91101}

\altaffiltext{1}{Based on data obtained at the W.M. Keck Observatory, which is operated as a scientific partnership among the California Institute of Technology, the University of California, and NASA, and was made possible by the generous financial support of the W.M. Keck Foundation.}
\altaffiltext{2}{Miller Fellow.}

\begin{abstract}
We present the rest-frame optical spectroscopic properties of 60 faint ($R\sub{AB} \sim
27$; $L \sim 0.1 L_*$) Ly$\alpha$-selected galaxies (LAEs) at
$z\approx 2.56$. These LAEs also have rest-UV spectra of their \lya\
emission line morphologies, which trace the effects of interstellar and
circumgalactic gas on the escape of \lya\ photons. We find that the
LAEs have diverse rest-optical spectra, but their average spectroscopic properties
are broadly consistent with the extreme low-metallicity end of the
populations of continuum-selected galaxies selected at
$z\approx2-3$. In particular, the LAEs have extremely high
[\ion{O}{3}] $\lambda$5008/\hb\ ratios (log([\ion{O}{3}]/\hb) $\sim$
0.8) and low [\ion{N}{2}] $\lambda$6585/\ha\ ratios
(log([\ion{N}{2}]/\ha) $<1.15$). Coupled with a detection of
the [\ion{O}{3}] $\lambda$4364 auroral line, these measurements indicate that
the star-forming regions in faint LAEs are characterized by 
 high electron temperatures ($T_e\approx1.8\times10^4$K), low
 oxygen abundances (12 + log(O/H) $\approx$ 8.04,
 $Z\sub{neb}\approx0.22Z_\odot$), and high excitations with respect to their more luminous 
continuum-selected analogs. Several of our faintest LAEs have line ratios
consistent with even lower metallicities, including six with 12 + log(O/H) $\approx$ 6.9$-$7.4
  ($Z\sub{neb}\approx0.02-0.05Z_\odot$). We interpret these observations in light of new models
of stellar evolution (including binary interactions) that have been
shown to produce long-lived populations of hot, massive
stars at low metallicities. We find that strong, hard ionizing
continua are required to reproduce our observed line ratios,
suggesting that faint galaxies are efficient producers of ionizing
photons and important analogs of reionization-era galaxies. Furthermore,
we investigate the physical trends accompanying \lya\ emission across
the largest current sample of combined \lya\ and rest-optical galaxy
spectroscopy, including both the 60 KBSS-\lya\ LAEs and 368 more
luminous galaxies at similar redshifts. We find that the net \lya\ emissivity  
(parameterized by the \lya\ equivalent width) is strongly correlated
with nebular excitation and ionization properties and 
weakly correlated with dust attenuation, suggesting that metallicity plays a
strong role in determining the observed properties of these galaxies
by modulating their stellar spectra, nebular excitation,
and dust content.
\end{abstract}


\section{Introduction} \label{sec:intro}

The optical emission lines of galaxies can provide
a wealth of information regarding the properties of young stars and the
regions in which they form. The historical accessibility of optical wavelengths
has enabled large spectroscopic surveys to measure these lines in
local star-forming regions and thereby identify and calibrate their relationships
with the physical properties of galaxies.  These properties include star-formation rates,
elemental abundances, gas temperatures and densities, ionization
states, and the nature of sources of ionizing radiation within
galaxies and gaseous clouds. In particular, the \citet{bal81} line
ratios (including the N2-BPT ratios [\ion{N}{2}]/\ha\ and
[\ion{O}{3}]/\hb) are frequently used to discriminate between
ionization by star formation and/or active galactic nuclei (AGN), and
these same line ratios provide a measure of gas-phase metallicity
among star-forming galaxies \citep{dop00}. 

At $z\gtrsim1$, however, these well-studied transitions shift into
infrared bands, where efficient, multiplexed spectrocopic survey
instruments have only recently become available (e.g., {\it Keck}/MOSFIRE
[\citealt{mcl12}]; {\it VLT}/KMOS [\citealt{sharples13}]). In the last few
years, these spectrometers have enabled the collection of the first
statistical samples of galaxies at $z\approx2-3$ with rest-optical
spectra, in particular through the Keck Baryonic Structure Survey
(KBSS; \citealt{ste14,str16}) and the MOSFIRE Deep Evolution
Field survey (MOSDEF; \citealt{kriek15}).

One of the earliest results of these new surveys has been that the nebular
emission line ratios of high-redshift galaxies occupy a distinct
locus in the N2-BPT parameter space with respect to typical
low-redshift galaxies (e.g., from the Sloan Digital Sky Survey; SDSS). Earlier limited rest-optical spectroscopy had 
hinted at this trend \citep{sha05a,erb06a,liu08,brinchmann08}, but the
KBSS and MOSDEF surveys have demonstrated that the population of bright
continuum-selected galaxies (LBGs; $L\sim L_*$) is shifted toward high
[\ion{O}{3}]/\hb\ ratios at fixed [\ion{N}{2}]/\ha\ (or equivalently,
high [\ion{N}{2}]/\ha\ at fixed [\ion{O}{3}]/\hb) with respect to
the SDSS star-forming galaxy locus
\citep{ste14,ste16,shapley15,san15,str16}.

While this trend has been firmly established for LBGs, its physical origins
remain under debate. Several studies have suggested that this ``BPT
offset'' may be explained by high N/O ratios at $z\approx2-3$
\citep{masters14,masters16,san15}, which would produce a
shift toward high [\ion{N}{2}]/\ha. However, \citet{str16} 
present detailed rest-optical spectroscopy of $\sim$250 LBGs over a
wide range of stellar masses and star-formation rates (SFRs), finding
behavior consistent with the low-redshift N/O vs. O/H relation. Rather,
\citet{str16} (along with \citealt{ste14,ste16}) suggest that the
N2-BPT offset corresponds primarily to a change in the excitation
state of nebular gas, which can be explained by the presence of stars
with hotter, harder spectra than typical of stars in the local Universe.

At the same time, new studies of massive stars have
indicated that the rotation and binarity of stars play important roles
in shaping their evolution and emission (e.g.,
\citealt{eld09,brott11,levesque12}). These effects can cause stars to
exhibit longer lifetimes and  higher-temperature spectral shapes than
their single or slowly-rotating counterparts. Furthermore, these
effects may naturally be more pronounced at high redshifts, where low
photospheric metallicities produce weaker winds and less associated loss
of angular momentum. Similarly, lower-metallicity stars produce harder
spectra regardless of their rotation rate, and metallicity may influence stellar
binarity through the cooling and fragmentation of star-forming clouds.

In light of these trends,
\citet{ste16} used extremely deep composite LBG spectra in the rest-UV
and rest-optical to measure a suite of emission lines and their
ratios, finding that stellar spectra from the Binary Population and
Spectral Synthesis (BPASSv2; \citealt{eld16,sta16}) models are able to
reproduce the full set of measured values. Conversely, the softer
spectra of single-star models, including non-binary BPASSv2 models and
those from the Starburst99 model suite \citep{lei14}, do not fulfill
these constraints irrespective of the assumed N/O ratio.

In addition to these observational constraints, model spectra similar
to those produced by binary evolution models are preferred by other
observations as well as theoretical grounds. Most obvious is that
approximately 70\% of local massive stars are expected to experience mass
transfer in binary systems (e.g., \citealt{sana12}), so single star models are unlikely to provide
an accurate description of their evolution. Secondly, simulations of
ionizing photon production have extreme difficulty generating the
large escape fractions needed to reionize the Universe, in part
because the stars remain buried in their optically thick birth clouds
for longer than the $\lesssim$3 Myr lifetimes expected for strong
sources of ionizing photons in typical stellar models (e.g., \citealt{ma15}). Models
that maintain populations of hot, luminous stars for 3$-$10 Myr
(including the aformentioned binary models) continue to produce
ionizing photons after their birth clouds have been cleared away by
stellar feedback, which allows the photons to escape to large
radii. \citet{ma16_fesc} demonstrate that simulations from
the Feedback In Realistic Environments (FIRE; \citealt{hop14}) project that
include the BPASSv2 stellar models are able to reproduce the high
escape fractions of ionizing photons required by other constraints on
reionization (e.g., \citealt{rob15}). Modeling the epoch of
reionization (EoR) therefore requires further understanding of the
stellar populations of these galaxies.

The specific galaxies that dominated the EoR, however,
were less luminous and less mature than those selected as LBGs at
$z\approx2-3$. One method of identifying galaxies with properties more
analgous to the EoR population is by selecting on the basis of \lya\ line
emission. As this technique does not require a detection in the
stellar continuum, it naturally selects fainter, younger, and
lower-mass galaxies than typical samples of LBGs (e.g.,
\citealt{gaw06}). We have also shown through the KBSS-\lya\ survey
\citep{tra15} that these \lya-emitting galaxies (LAEs) have
lower covering fractions of interstellar and/or circumgalactic gas,
which allows a higher fraction of their UV photons to
escape. Furthermore, there is evidence from radiative-transfer
simulations that the physical conditions which facilitate \lya\
emission are also conducive to ionizing photon escape
(\citealt{dijkstra16}, but cf. \citealt{yajima14}).

LAEs
comprise an increasing proportion of galaxies as redshift increases up to the EoR at
$z\approx6$ (e.g., \citealt{stark10,stark11}), above which the
fraction drops due to \lya\ absorption and scattering by the neutral
intergalatic medium \citep{pentericci11,schenker12,ono12}. However, the
{\it intrinsic} \lya\ emissivity of galaxies (divorced from their intergalactic and
circumgalactic attenuation) likely continues to increase toward the
earliest cosmic times, given the trends of \lya\ equivalent width with galaxy
luminosity and UV spectral slope (e.g., \citealt{schenker14}). For all
of these reasons, LAEs at $z\approx2-3$ are an ideal laboratory for
probing the physical properties of star-formation in EoR analogs.

In addition, \citet{str16} find that the high-redshift 
LBGs most offset from the SDSS N2-BPT locus are those with below-average
masses and high specific star formation rates, indicating that low-mass LAEs provide
key probes of the changing nebular conditions and stellar populations
at high redshifts as well. Some previous studies of \lya-emitting galaxies
have affirmed this interpretation: \citet{mcl11} measured strong
[\ion{O}{3}] emission in two
LAEs at $z\approx3.1$, \citet{finkelstein11} found strong
[\ion{O}{3}]/\hb\ and weak [\ion{N}{2}]/\ha\ in two LAEs at
$z\approx2.4$, and \citet{nak13} measured high [\ion{O}{3}]/[\ion{O}{2}]
ratios in two LAEs and in a composite spectrum of four additional
LAEs at $z\approx2.2$. \citet{song14} present a sample of 16 LAEs,
including 10 with rest-frame optical spectroscopy, which show 
similarly extreme line ratios consistent with low gas-phase and stellar
metallicities. However, the relatively small sample sizes of these  
studies have made it difficult to establish the average properties of
high-redshift LAEs. Furthermore, the LAEs in these studies
have luminosities similar to those of typical LBG samples
($\sim$10$\times$ that of the KBSS-\lya\ LAEs), which 
eliminates many of the advantages of continuum-faint LAEs for selecting
EoR analogs.

Galaxy selections based on emission lines other than \lya\ have also
yielded similar populations. \citet{hag16} find that
optical-emission-line-selected galaxies exhibit a wide range of
properties consistent with LAEs (although cf. \citealt{oteo15}, who
find that \ha-selected galaxies are more massive and redder than LAEs). \citet{mas14} present 22 Extreme
Emission-Line Galaxies (EELGs) selected by \ha\ or [\ion{O}{3}]
emission in grism spectroscopy, which show 
high excitations (log([\ion{O}{3}]/\hb) $\approx 0.7$) and low
metallicities ($Z\sub{neb}=0.05-0.3Z_\odot$). \citet{masters14}
present a similar grism-selected sample of 26 galaxies with a comparable
N2-BPT offset to the LBG samples discussed above. Both of these EELG samples lie at slightly
lower redshifts ($z\approx1.3-2.3$) than the KBSS, KBSS-\lya, and MOSDEF
galaxies. The EELGs also have atypically high specific
star-formation rates (sSFR), with stellar and dynamical masses similar or slightly higher
than those of KBSS-\lya\ LAEs ($M\sub{dyn}\sim10^9$\msun) and SFRs
3$-$10$\times$ higher, similar to typical LBGs (SFR $\sim$ 30 \msun\
yr$^{-1}$). Most of these EELGs also do not have \lya\ measurements
(being observationally difficult at $z\lesssim2$), which prohibits
the comparison of nebular properties with \lya\ production and escape.

This paper, therefore, has two primary objectives: 1) to extend our
understanding of the nebular properties of galaxies toward low stellar
masses and faint luminosities by measuring the rest-frame optical
spectra of {\it typical} LAEs; and 2) to determine how the \lya\
emissivities of galaxies are related to the physical properties of their
stellar populations and star-forming regions. Toward this aim, we present deep nebular
emission line spectra of 60 LAEs from the KBSS-\lya\ sample
presented in \citet{tra15}, as well as a comparison sample of 368 LBG
spectra from the KBSS \citep{ste14,str16}, and rest-UV spectra
covering the \lya\ transition of each galaxy.

In Sec.~\ref{sec:obs}, we describe the rest-optical LAE spectra presented
in this paper, as well as the LBG spectra and rest-UV LAE spectra
originally presented in previous work. Sec.~\ref{sec:em} describes our
emission line measurements from the individual object spectra and the
creation and fitting of composite LAE spectra. In
Sec.~\ref{sec:ratios}, we present diagnostic line ratios 
including the N2-BPT diagram in order to compare the nebular LAE
properties to those of the LBG sample. In Sec.~\ref{sec:lyaneb}, we
discuss the variation of \lya\ equivalent width across the combined
sample of LAEs and LBGs to infer the relationship between \lya\
emission (and absorption) and physical galaxy properties including
dust content and nebular excitation. Sec.~\ref{sec:models} includes
our photoionization modeling of the rest-optical line ratios and the
resulting constraints on the properties of the gas and 
stellar populations (including binary star models) in the LAE
star-forming regions. Finally, we provide additional physical
interpretation of our results in comparison with previous work in
Sec.~\ref{sec:discussion} and a summary and conclusions in
Sec.~\ref{sec:conclusions}. Because the majority of our observed
properties are line ratios, our results have little dependence on the
assumed cosmology, but we assume a $\Lambda$CDM universe with
($\Omega_m$, $\Omega_\Lambda$, $H_0$) = (0.3, 0.7, 70 km s$^{-1}$ Mpc$^{-1}$)
when necessary.

\section{Observations} \label{sec:obs}

\subsection{LAE sample} \label{subsec:photobs}

LAEs were selected from the KBSS-\lya\ (\citealt{tra15}; hereafter
T15) sample of LAEs. The KBSS-\lya\ is a survey for \lya-emitting
objects in the Keck Baryonic Structure Survey (KBSS; \citealt{rud12a,ste14,str16})
fields, which are centered on hyperluminous QSOs at $z\sim2-3$. The
spectroscopic properties of the KBSS-\lya\ LAE sample are 
described in detail in T15 and \citet{tra13}, and the full photometric
parent sample of objects 
will be described in an upcoming paper (R. Trainor et al., in
prep.). Briefly, the LAEs are selected via imaging in narrowband
filters selecting \lya\ emission near the redshift of the QSO in each
KBSS field. These narrowband images are combined with deep continuum
images to isolate objects with strong \lya\ emission, typically defined as a
rest-frame photometric equivalent width
$W\sub{\lya}>20$\AA.\footnote{As in T15, we note that this definition
  is useful both for consistency with previous samples and to rule out
low-redshift [\ion{O}{2}] emitters, which rarely exhibit
$W\sub{[\ion{O}{2}],obs}\gtrsim 70$\AA. However, we include 5 sources
with $0 <$ \wla\ $< 20$\AA\ whose narrowband detections and
multi-line spectroscopic redshifts confirm their identities as
\lya-emitting galaxies at $z\approx2.5$.} The selected objects
have a limiting narrowband magnitude $m\sub{NB,AB}=26.5$,
corresponding to an integrated \lya\ line flux $F\sub{\lya} >
10^{-17}$ erg s$^{-1}$ cm$^{-2}$ for a continuum-free source, or
$F\sub{\lya} \gtrsim 6\times10^{-18}$ erg s$^{-1}$ cm$^{-2}$ for a typical LAE with
$W\sub{\lya}\approx40$\AA. The LAEs have typical continuum magnitudes
$\mathcal{R}\sub{AB}(6930$\AA$)\approx 27$, corresponding to $L\sim0.1
L_{*}$ at $z\sim 2.6$ from the luminosity functions of
\citet{red08}. As discussed in T15, the primary contaminants to the
sample (as estimated by rest-UV spectroscopy of the putative \lya\
line and surrounding wavelengths) are low-redshift [\ion{O}{2}] emitters or
intermediate-redshift AGN with high equivalent
width \ion{C}{4} $\lambda \lambda$1549,1551 or \ion{He}{2}
$\lambda$1640 line emission, and the estimated contamination rate of
the photometric sample is 3\%. Given that all the LAEs and LBGs
presented in this paper have multiple spectroscopic line detections,
we expect that there are no such contaminants in this sample. 

The properties of the individual LAEs presented in this paper are
summarized in Table~\ref{table:laes}. The majority of the LAEs lie in the Q2343
field ($z\sub{QSO}=2.573$) of T15, with the remainder selected from a new field
(Q1603) centered on the hyperluminous QSO HS1603+3820
($z\sub{QSO}=2.551$). Because this QSO lies at a very similar redshift to
the Q2343 QSO, the Q1603 LAEs were selected with the same NB4325
filter used for the Q2343 field, as described in T15. Properties
of HS1603+3820 and its surrounding field are presented in
\citet{tra12}. This field has slightly shallower narrowband imaging
(by $\sim$0.5 mag) than the other KBSS-\lya\ fields, resulting in a limiting \lya\
luminosity $\sim$60\% higher, but the properties of the Q1603
LAEs appear to be otherwise consistent with the remainder of the KBSS-\lya\ sample.

As discussed in T15, the KBSS-\lya\ LAEs have typical dynamical masses
$\langle M\sub{dyn} \rangle \approx 8\times 10^{8}$ \msun, and our
measurements of stacked LAE spectral energy distributions (SEDs;
R. Trainor et al., in prep.) imply stellar masses $M_*\sim10^8-10^9$ \msun.

\begin{deluxetable*}{lcccccccc}
\tablecaption{MOSFIRE LAE Properties}
\tablewidth{0pt}
\tablehead{
Object Name & $z\sub{sys}$\tablenotemark{a} & RA & Dec & \phs $\mathcal{R}$ & $W\sub{\lya}$\tablenotemark{b}
& Bands & [\ion{O}{3}] $\lambda$5008\tablenotemark{c} & \hb\tablenotemark{c} }

\startdata
Q1603-NB1036 & 2.5412 & 16:04:43.35 & +38:11:16.47 &  $>$27.3  & \phn58.8 &  H  & 20.6$\pm$2.1\phn &  2.2$\pm$1.8 \\ 
Q1603-NB1365 & 2.5491 & 16:04:54.62 & +38:14:01.55 &  \phs25.0 & \phn84.4 &  H  & 54.6$\pm$2.1\phn &  9.0$\pm$1.7 \\ 
Q1603-NB1599 & 2.5471 & 16:04:54.89 & +38:13:11.81 &  $>$27.3  & 107.7 &  H  & 46.4$\pm$2.0\phn & $<$10.9 \\ 
Q1603-NB1700 & 2.5447 & 16:04:57.60 & +38:12:25.04 &  \phs24.7 & \phn53.0 &  H  & 69.6$\pm$3.6\phn & 14.2$\pm$3.7\phn \\ 
Q1603-NB1756 & 2.5541 & 16:05:03.28 & +38:12:54.03 &  \phs27.1 & 100.9 &  H  & 17.6$\pm$1.6\phn &  7.7$\pm$4.3 \\ 
Q1603-NB1934 & 2.5504 & 16:04:48.08 & +38:12:09.42 &  \phs25.0 & \phn20.3 &  H  & 79.6$\pm$3.1\phn & 18.4$\pm$1.8\phn \\ 
Q1603-NB1962 & 2.5514 & 16:04:48.79 & +38:12:29.30 &  \phs25.8 & \phn79.4 &  H  & 60.8$\pm$2.5\phn & 10.3$\pm$1.7\phn \\ 
Q1603-NB2182 & 2.5602 & 16:04:54.51 & +38:12:01.08 &  \phs24.4 & \phn\phn5.9 &  H  & 11.7$\pm$3.0\phn & $<$1.8 \\[5pt] 
Q2343-NB0193 & 2.5791 & 23:46:23.98 & +12:45:52.31 & \phs25.6 & \phn30.9 &  H  & 13.6$\pm$0.5\phn &  2.7$\pm$0.5 \\ 
Q2343-NB0280 & 2.5777 & 23:46:20.67 & +12:46:19.82 & \phs26.3 & \phn52.6 &  H  &  6.0$\pm$0.7 &  2.0$\pm$0.5 \\ 
Q2343-NB0308 & 2.5663 & 23:46:17.83 & +12:46:41.53 & $>$27.3 & \phn58.8 & H+K & 20.2$\pm$5.8\tablenotemark{d} &  1.6$\pm$1.1 \\ 
Q2343-NB0320 & 2.5817 & 23:46:17.52 & +12:46:45.78 & \phs25.4 & \phn25.4 &  H  & 19.7$\pm$2.1\tablenotemark{d} &  5.8$\pm$0.9 \\ 
Q2343-NB0345 & 2.5879 & 23:46:13.94 & +12:47:11.73 & \phs24.5 & \phn36.9 & H+K & 79.5$\pm$1.7\tablenotemark{d} & 13.4$\pm$1.1\phn \\ 
Q2343-NB0405 & 2.5866 & 23:46:22.51 & +12:46:19.06 & \phs27.0 & 134.8 & H+K &  \phn9.2$\pm$1.5\tablenotemark{d} & $<$0.3 \\ 
Q2343-NB0508 & 2.5602 & 23:46:32.73 & +12:45:33.09 & $>$27.3 & 156.6 &  H  &  9.6$\pm$0.9 &  1.2$\pm$0.4 \\ 
Q2343-NB0565 & 2.5634 & 23:46:14.01 & +12:47:35.29 & $>$27.3 & 313.9 & H+K & 20.5$\pm$1.3\phn & $<$1.5 \\ 
Q2343-NB0585 & 2.5451 & 23:46:31.42 & +12:45:47.79 & $>$27.3 & \phn46.7 &  H  & 25.0$\pm$0.4\phn &  4.2$\pm$0.5 \\ 
Q2343-NB0791 & 2.5742 & 23:46:33.99 & +12:50:28.08 & \phs25.7 & \phn37.4 & H+K & 25.8$\pm$0.6\phn &  9.9$\pm$0.9 \\ 
Q2343-NB0970 & 2.5609 & 23:46:33.43 & +12:50:12.93 & $>$27.3 & \phn44.1 &  H  &  8.2$\pm$1.2 &  1.5$\pm$0.8 \\ 
Q2343-NB1075 & 2.5610 & 23:46:33.23 & +12:50:04.84 & $>$27.3 & \phn23.7 &  H  &  2.4$\pm$0.8 &  0.6$\pm$0.5 \\ 
Q2343-NB1093 & 2.5610 & 23:46:36.75 & +12:49:40.93 & $>$27.3 & \phn54.9 &  H  &  6.5$\pm$1.3 &  0.9$\pm$0.8 \\ 
Q2343-NB1154 & 2.5763 & 23:46:26.23 & +12:50:38.86 & \phs23.7 & \phn54.1 & H+K &  5.9$\pm$0.4 & $<$1.0 \\ 
Q2343-NB1174 & 2.5476 & 23:46:24.14 & +12:50:49.14 & \phs25.0 & 236.6 & H+K & 39.6$\pm$0.7\phn &  6.0$\pm$1.6 \\ 
Q2343-NB1264 & 2.5475 & 23:46:27.33 & +12:50:18.73 & \phs25.8 & 108.2 &  H  & 18.8$\pm$0.5\phn &  4.8$\pm$1.7 \\ 
Q2343-NB1361 & 2.5590 & 23:46:34.09 & +12:48:03.47 & $>$27.3 & \phn90.6 & H+K & 10.1$\pm$1.7\phn &  1.6$\pm$1.0 \\ 
Q2343-NB1386 & 2.5654 & 23:46:36.23 & +12:47:46.48 & $>$27.3 & \phn25.8 & H+K &  9.9$\pm$1.3 & $<$0.9 \\ 
Q2343-NB1396 & 2.5817 & 23:46:27.44 & +12:48:39.87 & \phs25.3 & \phn67.7 &  H  &  9.8$\pm$1.0 & $<$1.0 \\ 
Q2343-NB1416 & 2.5590 & 23:46:34.04 & +12:47:57.04 & $>$27.3 & \phn20.7 & H+K &  3.7$\pm$1.1 &  0.7$\pm$0.6 \\ 
Q2343-NB1501 & 2.5590 & 23:46:37.77 & +12:47:24.28 & \phs27.1 & \phn35.2 & H+K & 21.9$\pm$1.4\phn &  2.4$\pm$0.8 \\ 
Q2343-NB1518 & 2.5860 & 23:46:40.75 & +12:47:02.10 & \phs26.2 & \phn22.3 &  K  & $-$ & $-$ \\ 
Q2343-NB1585 & 2.5648 & 23:46:35.59 & +12:47:28.53 & $>$27.3 & 303.4 & H+K & 10.3$\pm$0.9\phn &  1.8$\pm$0.8 \\ 
Q2343-NB1591 & 2.5438 & 23:46:32.56 & +12:47:47.05 & $>$27.3 & \phn88.8 &  H  &  4.2$\pm$0.4 &  2.1$\pm$0.6 \\ 
Q2343-NB1680 & 2.5795 & 23:46:12.50 & +12:49:45.07 & \phs26.0 & \phn43.5 &  H  &  5.7$\pm$0.5 &  1.2$\pm$0.6 \\ 
Q2343-NB1684 & 2.5807 & 23:46:18.98 & +12:49:03.47 & \phs26.9 & \phn24.0 &  H  &  4.0$\pm$0.7 &  0.9$\pm$0.6 \\ 
Q2343-NB1692 & 2.5604 & 23:46:39.30 & +12:46:54.50 & \phs26.8 & 359.2 & H+K & 17.6$\pm$1.5\phn &  2.2$\pm$0.8 \\ 
Q2343-NB1783 & 2.5767 & 23:46:28.71 & +12:47:51.04 & \phs26.6 & \phn35.0 & H+K & 29.9$\pm$1.0\phn &  4.4$\pm$1.1 \\ 
Q2343-NB1789 & 2.5450 & 23:46:29.21 & +12:47:48.05 & $>$27.3 & \phn88.2 & H+K & 15.1$\pm$0.5\phn &  3.8$\pm$0.5 \\ 
Q2343-NB1806 & 2.5954 & 23:46:20.09 & +12:48:43.60 & \phs26.8 & \phn19.3 &  K  & $-$ & $-$ \\ 
Q2343-NB1828 & 2.5727 & 23:46:25.25 & +12:48:08.61 & \phs25.8 & \phn26.7 & H+K & 21.7$\pm$0.5\phn &  3.5$\pm$0.4 \\ 
Q2343-NB1829 & 2.5754 & 23:46:25.12 & +12:48:07.80 & \phs25.6 & \phn32.1 & H+K & 35.0$\pm$0.8\phn & $<$5.5 \\ 
Q2343-NB1922 & 2.5653 & 23:46:12.61 & +12:49:17.22 & $>$27.3 & 142.3 &  H  &  8.2$\pm$1.1 & $<$0.7 \\ 
Q2343-NB2098 & 2.5647 & 23:46:12.35 & +12:49:10.62 & $>$27.3 & 256.0 &  H  &  5.2$\pm$1.3 &  3.3$\pm$1.3 \\ 
Q2343-NB2211 & 2.5756 & 23:46:22.90 & +12:47:43.45 & $>$27.3 & \phn66.2 & H+K & 24.6$\pm$0.6\phn & 11.7$\pm$3.7\phn \\ 
Q2343-NB2571 & 2.5791 & 23:46:23.29 & +12:46:58.64 & $>$27.3 & \phn29.4 & H+K &  3.7$\pm$0.6 & $<$1.0 \\ 
Q2343-NB2785 & 2.5785 & 23:46:31.59 & +12:49:31.38 & $>$27.3 & \phn14.2 &  K  & $-$ & $-$ \\ 
Q2343-NB2807 & 2.5446 & 23:46:40.22 & +12:48:27.64 & \phs23.4 & \phn24.6 &  H  & 143.4$\pm$9.1\tablenotemark{d}\phn & 45.2$\pm$4.2\phn \\ 
Q2343-NB2816 & 2.5770 & 23:46:27.44 & +12:49:53.12 & $>$27.3 & \phn58.5 & H+K &  4.7$\pm$0.6 &  2.5$\pm$0.7 \\ 
Q2343-NB2821 & 2.5753 & 23:46:20.59 & +12:50:24.94 & \phs27.1 & \phn32.4 & H+K &  6.6$\pm$0.7 & $<$3.4 \\ 
Q2343-NB2834 & 2.5683 & 23:46:37.77 & +12:48:44.48 & $>$27.3 & \phn\phn7.2 &  K  & $-$ & $-$ \\ 
Q2343-NB2835 & 2.5738 & 23:46:33.80 & +12:49:10.63 & $>$27.3 & \phn61.3 &  H  &  3.9$\pm$0.6 &  3.7$\pm$0.8 \\ 
Q2343-NB2875 & 2.5450 & 23:46:33.06 & +12:49:11.19 & $>$27.3 & 287.2 &  H  &  5.6$\pm$0.5 &  3.3$\pm$0.5 \\ 
Q2343-NB2929 & 2.5456 & 23:46:33.03 & +12:49:00.27 & \phs26.5 & \phn23.7 &  H  & 13.4$\pm$0.6\phn &  3.6$\pm$0.9 \\ 
Q2343-NB2957 & 2.5768 & 23:46:37.05 & +12:47:47.85 & $>$27.3 & \phn23.3 &  H  &  3.4$\pm$0.4 &  1.4$\pm$0.4 \\ 
Q2343-NB3061 & 2.5765 & 23:46:21.30 & +12:50:03.51 & \phs27.2 & \phn57.6 & H+K & 12.5$\pm$0.5\phn &  2.4$\pm$0.7 \\ 
Q2343-NB3122 & 2.5648 & 23:46:19.50 & +12:50:12.91 & \phs26.9 & \phn80.4 &  H  & 43.2$\pm$1.2\phn &  4.3$\pm$1.2 \\ 
Q2343-NB3157 & 2.5445 & 23:46:39.46 & +12:48:01.31 & \phs25.0 & \phn21.9 &  H  & 95.3$\pm$0.7\phn & 17.2$\pm$0.8\phn \\ 
Q2343-NB3170 & 2.5475 & 23:46:39.81 & +12:47:50.11 & $>$27.3 & \phn90.0 &  H  & 12.0$\pm$0.6\phn &  2.9$\pm$2.3 \\ 
Q2343-NB3231 & 2.5713 & 23:46:24.76 & +12:49:24.97 & $>$27.3 & 110.4 & H+K & 27.6$\pm$0.8\phn &  5.2$\pm$0.4 \\ 
Q2343-NB3252 & 2.5709 & 23:46:31.98 & +12:48:35.32 & \phs26.1 & \phn32.9 &  H  & 10.3$\pm$1.3\phn &  1.5$\pm$0.6 \\ 
Q2343-NB3292 & 2.5631 & 23:46:25.35 & +12:49:09.51 & $>$27.3 & \phn13.4 &  K  & $-$ & $-$
\enddata
\tablenotetext{a}{Systemic redshift measured from the $H$ or $K$
  nebular line spectrum.}
\tablenotetext{b}{Rest-frame \lya\ equivalent width in \AA\ measured
  from the narrowband and broadband photometry.}
\tablenotetext{c}{Observed line flux in $10^{-18}$ erg s$^{-1}$ cm$^{-2}$.}
\tablenotetext{d}{[\ion{O}{3}] $\lambda$5008 flux inferred from the
  [\ion{O}{3}] $\lambda$4960 line assuming a 3:1 ratio.}
\label{table:laes}
\end{deluxetable*}

\subsection{Rest-optical spectroscopic observations} \label{subsec:mosobs}

Rest-frame optical spectra of the KBSS-\lya\ LAEs were obtained using
the Multi-Object Spectrometer For InfraRed Exploration (MOSFIRE;
\citealt{mcl10,mcl12}) on the Keck 1 telescope. Initial observations
(including all of the $K$-band spectroscopy presented here) were
obtained over the course of the KBSS-MOSFIRE survey \citep{ste14}. The
data were reduced using the spectroscopic reduction pipeline provided
by the instrument team, and the details of the observing strategies
and reduction are described in \citet{ste14}. All spectra were
observed using 0\farcs7 slits, yielding $R\approx 3660$ or
$\sigma\sub{inst}\approx35$ \kms. 

A total of 28 LAEs were detected in the $K$-band over the course of
the KBSS-MOSFIRE observations, all in the Q2343 field (this sample was
described in T15). As these LAEs
are selected at $z\approx2.56$, the observed $K$-band spectra cover
the range $5380$\AA\ $\lesssim \lambda\sub{rest} \lesssim 6730$\AA, which includes the
\ha\ $\lambda$6564 line and the [\ion{N}{2}] $\lambda\lambda$6549,6585
doublet. Because these spectra were primarily obtained as secondary
observations on KBSS-MOSFIRE masks, they have exposure times that vary
significantly: our deepest $K$-band spectra have 4.9 hour
integrations, while our shallowest have 1 hour integrations. The
average exposure time is 2.7 hours, for a total of 83 object-hours of
integration. 

Limited $H$-band spectroscopy was also obtained during KBSS-MOSFIRE
observations (12 LAEs, 36 object-hours of integration; T15). The majority
of the $H$-band spectra presented here were obtained on 15-16
September 2015 in clear weather. These observations included 40 LAEs
(7 of which had been previously observed) in the Q2343 field with a
typical exposure time of 4 hours per object 
and a total of 173.4 object-hours of exposure. Another 8 LAEs in the
Q1603 field have measurements of at least one nebular line from a
single mask observed for 1 hour on 16 September 2015, producing a
final sample of 55 LAEs with nebular line detections in the $H$-band
and a total of 218.4 object-hours of integration. 

At $z\approx2.56$, the
MOSFIRE $H$-band covers $4100$\AA\ $\lesssim \lambda\sub{rest} \lesssim 5060$\AA. The
spectra therefore typically include the \hg\ and \hb\ transitions of
hydrogen, the $[$\ion{O}{3}$]$ $\lambda$4364 auroral emission line,
and the [\ion{O}{3}] $\lambda\lambda$4960,5008 doublet (although only
\hb\ and [\ion{O}{3}] $\lambda\lambda$4960,5008 are detected in individual spectra; see
Sec.~\ref{subsec:individual}). At $z\sub{QSO}\approx 2.56$, the
$[$\ion{O}{3}$]$ $\lambda$5008 line falls near the atmospheric
transparency cutoff and the half-power point of the MOSFIRE $H$-band filter,
which lowers the accuracy of our flux calibration with respect to
bluer wavelengths. In addition, some LAEs in the Q2343 field have
slightly higher redshifts than the QSO, such that their [\ion{O}{3}]
$\lambda$5008 emission does not fall on the detector. For these
reasons, the [\ion{O}{3}] $\lambda$4960 line is used to infer the true
[\ion{O}{3}] $\lambda$5008 flux when the two fluxes are inconsistent
with the expected ratio $f\sub{5008}/f\sub{4960}=3$. While this issue
affects only 5 individual LAEs at a significant level (Table~\ref{table:laes}), it is
particularly clear in our composite $H$-band spectrum
(Fig.~\ref{fig:Hbandwide}; Table~\ref{table:lines}), where the
apparent line ratio $f\sub{5008}/f\sub{4960}=2.5$ underpredicts the
expected ratio by 17\%. For this reason, line ratios in this paper
measured from the composite [\ion{O}{3}] $\lambda$5008 line (e.g., O3 $\equiv$ log([\ion{O}{3}]
$\lambda$5008/\hb); Table~\ref{table:ratios}) are computed from the {\it corrected} [\ion{O}{3}]
$\lambda$5008 flux, which is defined to be
$f\sub{5008,corr}\equiv 3\times f\sub{4960}$.

\begin{deluxetable}{cl}
\tablecaption{Emission Line Ratios}
\tablewidth{0pt}
\tablehead{
Ratio & Definition 
}

\startdata
O3 & log([\ion{O}{3}] $\lambda$5008/\hb) \\
N2 & log([\ion{N}{2}] $\lambda$6585/\ha)  \\
O3N2 & O3$-$N2 \\
O32 & log([\ion{O}{3}] $\lambda\lambda$4960,5008)/[\ion{O}{2}] $\lambda\lambda$3727,3729) \\
R23 & log([\ion{O}{3}] $\lambda\lambda$4960,5008)+[\ion{O}{2}]
$\lambda\lambda$3727,3729/\hb) \\
R\sub{O3} & log([\ion{O}{3}] $\lambda$4364/[\ion{O}{3}] $\lambda\lambda$4960,5008)
\enddata
\tablecomments{The $\lambda\lambda$ notation refers to the sum of both
  lines.}
\label{table:ratios}
\end{deluxetable}

For both $H$-band and $K$-band spectra, slit corrections are estimated
for each mask according to the procedure developed for KBSS-MOSFIRE
\citep{str16}. The majority of LAE spectra included here were
observed on masks which include a bright star, for which the
integrated stellar flux in each band is compared to the photometric magnitude
to determine the flux correction for point sources on the mask. Given
that the typical KBSS-\lya\ LAEs are spatially unresolved, their slit
losses are well-approximated by this factor. For masks observed without
a calibration star, mask correction factors are estimated by
comparing the relative fluxes of objects observed on several different
masks. The optimal set of mask-specific correction factors are
calculated via an MCMC procedure that adjusts each correction factor
to achieve the maximum degree of internal consistency among all flux
measurements of objects appearing on multiple masks (using
the observations of calibration stars to set the overall normalization of
slit corrections). When there are not enough data to reliably
estimate the slit correction for a single object, we adopt a
correction factor of 2, the median of those observed. Slit
corrections for all the LAE spectra included in this sample range from
1.53 to 2.34.

\subsection{\lya\ spectroscopic observations} \label{subsec:lrisobs}

Rest-UV spectroscopy of the KBSS-\lya\ LAEs and a comparison sample of
KBSS LBGs is described in detail in T15. In Sec.~\ref{sec:lyaneb}, we use 
the same comparison sample of T15, but with the
addition of LBGs displaying \lya\ absorption (the T15 comparison
sample was restricted to those spectra displaying detectable \lya\
emission). Each of the LAE and LBG rest-UV spectra in the T15 sample
were obtained with LRIS-B \citep{oke95,ste04} using the 600 lines/mm
grism, yielding a resolution $R\approx1300$. In this paper, we
also include KBSS LBG spectra observed with the LRIS 400 lines/mm
grism ($R\approx800$). These spectra are described in detail in
\citet{ste10}. We limit our analysis to the set of 368 KBSS LBGs with
both rest-UV spectra covering the \lya\ line and rest-optical spectra
covering several emission lines in the MOSFIRE $H$ and $K$ bands
(described in detail in Sec.~\ref{subsec:bptlya}).

The \lya\ emission or absorption profiles in the LAE and LBG spectra are
identified using the line-detection algorithm described in T15, and
spectroscopic \lya\ equivalent widths (in emission or absorption) are
calculated comparing the directly-integrated line profile to the local
continuum level on the red side of the \lya\ line (1222\AA
$<\lambda\sub{rest}<$ 1240\AA). Unlike the 
detailed shape of the \lya\ line, the 
integrated \lya\ equivalent width is insensitive to the
spectral resolution, which allows us to use the larger sample of
low-resolution KBSS LBG rest-UV spectra.

As discussed in T15, the spectroscopic \lya\ equivalent widths ($W\sub{\lya,spec}$)
measured for individual or stacked rest-UV spectra are not directly
comparable to photometric \lya\ equivalent width measurements
($W\sub{\lya,phot}$) measured from narrowband and broadband
photometry. This difference arises in part from the scattering of \lya\
photons in the ISM and CGM of their host galaxies, leading to larger
inferred physical sizes in the \lya\ line with respect to the continuum
\citep{ste11,momose14} and therefore a relative underestimate of \lya\ flux in slit
spectroscopy. In T15, we find that this differential slit loss leads
to an overall underestimate of the \lya\ equivalent width in LAE spectra,
such that $W\sub{\lya,phot}\approx 2 W\sub{\lya,spec}$ (with
substantial scatter based on the size of the source). Similarly,
\citet{ste11} find a differential slit loss of 3$-$5$\times$ for
LBGs. In general, $W\sub{\lya,phot}$ has the advantage of including
all the \lya\ flux, as well as being more easily measured for faint
sources (for which the continuum flux is typically not detected in
individual spectra), but it also cannot be measured for the majority of LBGs that
do not fall within the narrowband-selected redshift range. As such,
we use $W\sub{\lya,spec}$ to compare the LAE and LBG samples, but
$W\sub{\lya,phot}$ is used to select individual high-$W\sub{\lya}$ LAEs, and we
denote the choice of measurement explicitly where values are
presented. In either case, the values of \wla\ always correspond to a
measurement of the \lya\ equivalent width in the rest-frame of the galaxy.

\section{Emission line measurements} \label{sec:em}

\subsection{Individual object spectra} \label{subsec:individual}

Rest-frame optical spectra for individual LAEs were fit using the IDL
program MOSPEC \citep{str16}, as described in T15. For
$H$-band spectra, the \hb,  [\ion{O}{3}] $\lambda$4960, and  [\ion{O}{3}]
$\lambda$5008 lines are fit simultaneously with gaussian line profiles
and a quadratic fit to the continuum. All three emission lines are
constrained to have the same redshift ($z\sub{neb,H}$) and velocity
width ($\sigma_{v,\rm{H}}$), and the 
[\ion{O}{3}] $\lambda$4960 and [\ion{O}{3}] $\lambda$5008 lines are
constrained to a 1:3 flux ratio.\footnote{As noted in
  Sec.~\ref{subsec:mosobs}, the [\ion{O}{3}] $\lambda$5008 line lies
off the detector or has poor flux calibration in 5
sources, identified in Table~\ref{table:laes}. In these cases only the
$\lambda$4960 and \hb\ lines are fit.} Sky
lines are masked in the fitting 
process using the error vector extracted by MOSPEC, and uncertainties
are measured from the $\chi^2$-minimization for each of the free
parameters in the fit (i.e., \hb\ flux, combined [\ion{O}{3}] $\lambda
\lambda$4960,5008 flux, velocity width, redshift, and continuum). The $K$-band
spectra are fit similarly, with a simultaneous fit to the \ha\ and
[\ion{N}{2}] $\lambda\lambda$6549,6585 emission lines, yielding the
\ha\ flux, combined [\ion{N}{2}] $\lambda\lambda$6549,6585 flux (also
constrained to a 1:3 ratio),
redshift ($z\sub{neb,K}$), and velocity width
($\sigma_{v,\rm{K}}$). For each LAE, the $H$-band and $K$-band line flux
measurements are corrected by the mask-specific correction factor(s)
of the spectra contributing to the line measurement. In total, 27
spectra have $>$3$\sigma$ detections of \ha\, while 27 (55, 
50) have $>$3$\sigma$ detections of \hb\ ([\ion{O}{3}] $\lambda$4960,
[\ion{O}{3}] $\lambda$5008). No LAEs in our sample have detected [\ion{N}{2}]
$\lambda\lambda$6549,6585 emission, and the 2$\sigma$ upper limits on
the ratio N2 ($\equiv$ log([\ion{N}{2}] $\lambda$6585/\ha);
Table~\ref{table:ratios}) range from $-$0.3 to $-$1.0. 

\begin{figure}
\center
\includegraphics[width=\linewidth]{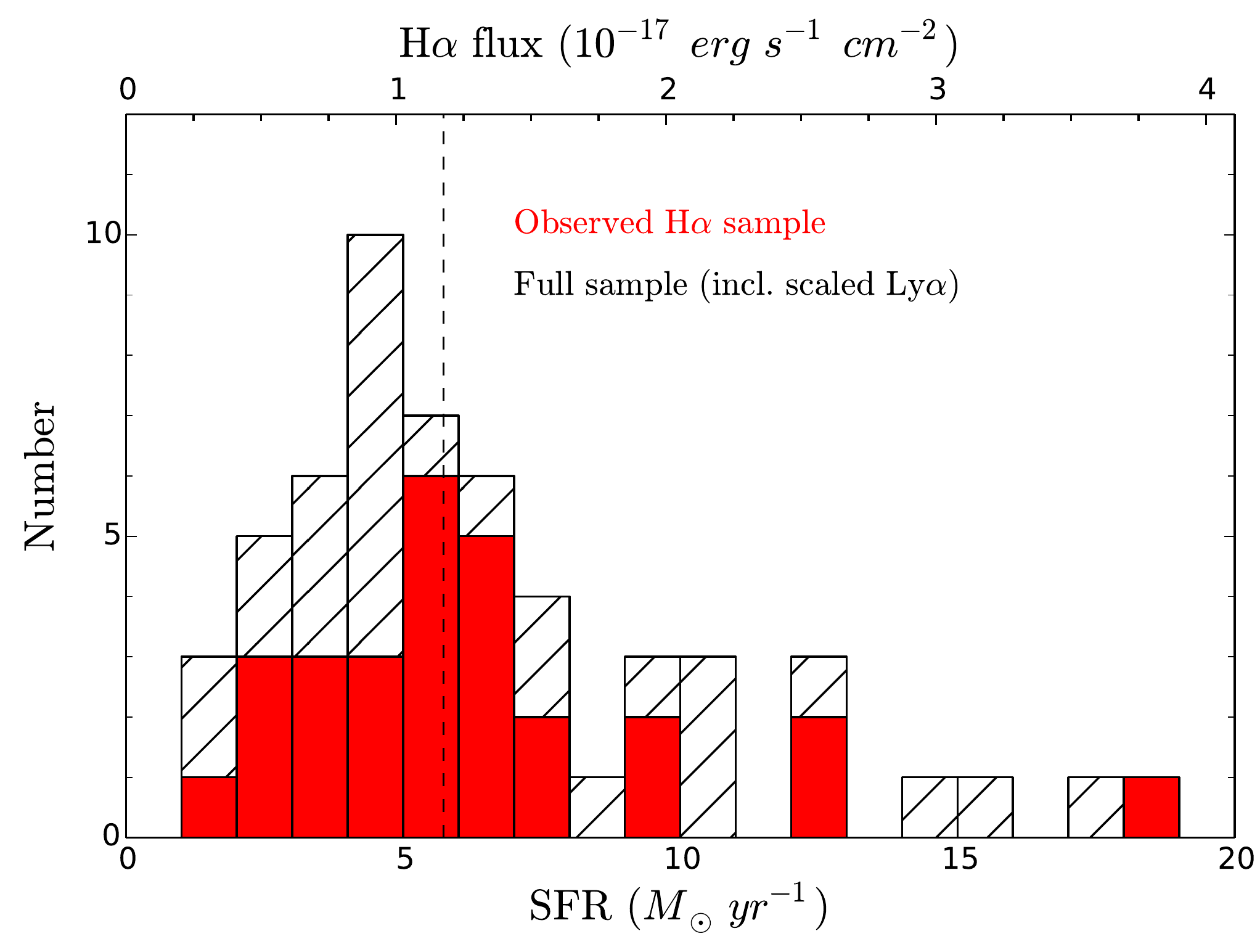}
\caption{The distribution of star-formation rates for
  the 60 LAEs presented in this paper. Red histogram gives the
  distribution of measured \ha\ fluxes and dust-corrected \ha\
  star-formation rates (SFR). The hatched histogram includes estimates for
  LAEs with no current $K$-band measurements; for these objects, the
  \ha\ flux is estimated from the measured \lya\ flux and the median
  \lya/\ha\ ratio among the objects with measured \ha\ fluxes. The
  \ha\ flux and SFR from the full LAE composite spectrum
  (Fig.~\ref{fig:Kband}) is shown by the dashed line.}
\label{fig:sfr}
\end{figure}

Star-formation rates (SFRs) are calculated for the 28 LAEs with \ha\
measurements using the \citet{ken98} relation, as displayed in
Fig.~\ref{fig:sfr}. We apply a uniform dust-correction based on the
average nebular reddening estimated in Sec.~\ref{subsec:balmer}
($\ebv=0.06$; Eq.~\ref{eq:extinction}) and a \citet{car89} extinction curve, which produces a 15\%
increase in the estimated star-formation rates. For the LAEs without
current \ha\ flux measurements, we estimate SFRs based on the
narrowband \lya\ flux, scaled by the median \lya/\ha\ flux ratio from
the objects with direct \ha\ measurements ($\langle
f\sub{\lya}/f\sub{\ha}\rangle=3.1$). The distribution of \ha- 
and \lya-derived SFRs are statistically indistiguishable, as shown in
Fig.~\ref{fig:sfr}. The median LAE \ha\ SFR = 5.3
\msun\ yr$^{-1}$, which is consistent with the value estimated from
the composite $K$-band spectrum (5.7$\pm$1.4 \msun\ yr$^{-1}$) and 
significantly lower than the \ha-derived SFRs typical of the KBSS
LBG sample we consider here (median SFR $=27$ \msun\ yr$^{-1}$;
\citealt{str16}).

\begin{figure}
\center
\includegraphics[width=\linewidth]{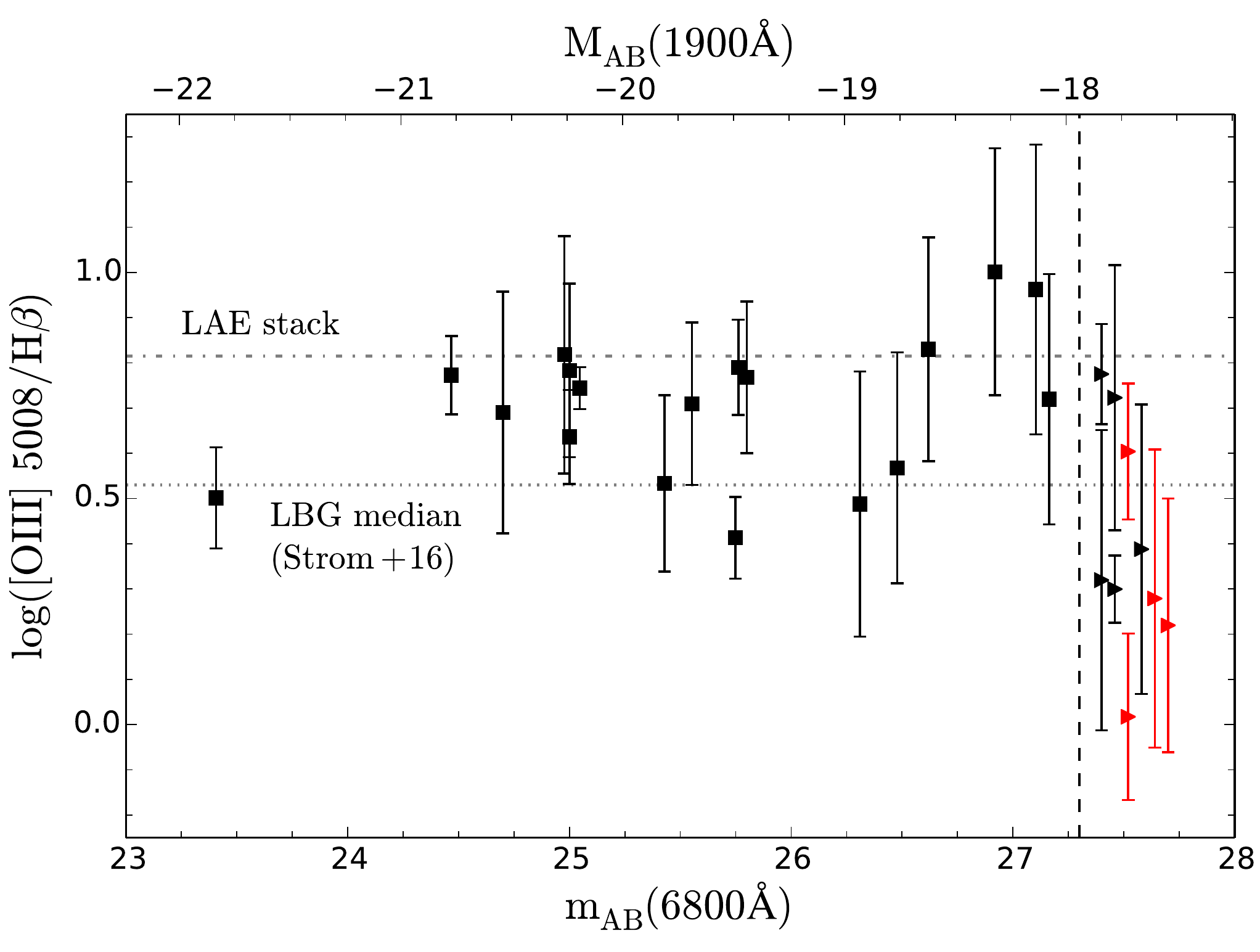}
\caption{The ratio O3 $\equiv$ log([\ion{O}{3}] $\lambda$5008/\hb) vs. $\mathcal{R}$-band
  magnitude for 27 LAEs with $>$3$\sigma$ detections of both \hb\ and
  [\ion{O}{3}] $\lambda$5008. Black squares denote LAEs with measured
  $\mathcal{R}$ magnitudes, while triangles denote LAEs fainter than the
  3$\sigma$ limit $\mathcal{R}>27.3$ (vertical dashed line), which are given
  arbitrary horizontal offsets for clarity. Red points are also
  undetected in deep HST/WFC F160W images sampling the rest-frame
  optical wavelengths (3$\sigma$ depth $m\sub{AB,F160W}>28.1$). The dot-dashed horizontal line
  is the best-fit value of O3 from our
  composite LAE spectrum (Fig.~\ref{fig:Hbandwide},
  Table~\ref{table:lines}), while the dotted horizontal line corresponds
  to the median value for all KBSS LBGs with $>$3$\sigma$ detections of both \hb\ and
  [\ion{O}{3}] $\lambda$5008. The LAEs generally have elevated values
  of O3 with respect to the LBGs, with the
  exception of the faintest LAEs. These faint objects are consistent with the
  turnover in O3 that occurs at
  extremely low oxygen abundances (12 + log(O/H) $\ll$ 8.3,
  $Z\sub{neb}\ll0.4Z_\odot$) as discussed in Sec.~\ref{subsec:lowz}.}
\label{fig:o3_rmag}
\end{figure}

Fig.~\ref{fig:o3_rmag} displays the O3 ratio for each of the 27 LAEs with
$>$3$\sigma$ detections of both \hb\ and at least one of [\ion{O}{3}]
$\lambda$5008 or [\ion{O}{3}] $\lambda$4960. Although the
uncertainties are large on many of the individual measurements, the ratios
are generally elevated with respect to the typical LBG value of O3
$\approx0.5$. An exception to this trend occurs for the LAEs with the
faintest continuum luminosities ($M\sub{AB}(1900$\AA$)\gtrsim-18$;
$L\lesssim0.06L_*$ from \citealt{red08}), which have lower O3 ratios
than the average LAEs or even the typical LBG values. These faintest
LAEs are undetected in our  $\mathcal{R}$ band images, and four of
them have no detection in deep, 8000 second images from {\it HST}/WFC3
in the F160W filter, setting a (point source) limiting magnitude
$m\sub{AB}(1.6\mu$m$)>28.1$ (3$\sigma$). The low O3 ratios of these sources
suggest that the faintest LAEs have either lower excitation states
than typical galaxies at these redshifts or very low metallicities
($Z\sub{neb}<0.1Z_\odot$). These objects are discussed in
more detail in Sec.~\ref{subsec:lowz}.

\subsection{Measurements from composite spectra} \label{subsec:stack}

\begin{figure*}[htbp]
\center
\includegraphics[width=\linewidth]{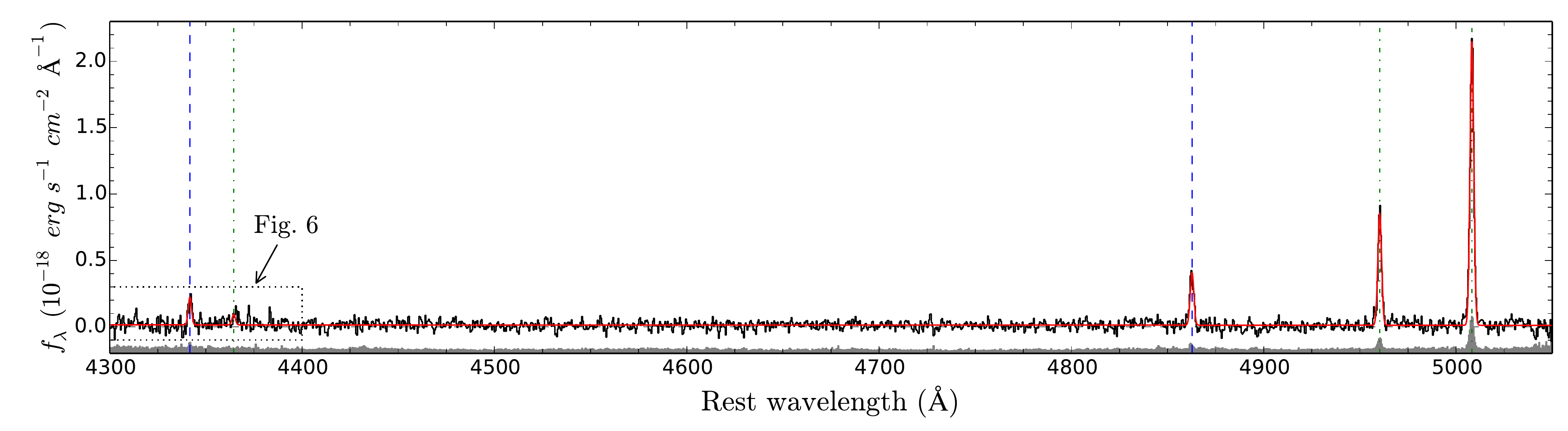}
\caption{Full composite $H$ band spectrum of 60 LAEs
  (218 object-hours). Red line is the fit to the composite spectrum as
  described in Sec.~\ref{subsubsec:fitstack}, and grey shaded region
  shows the 1$\sigma$ bootstrap errors on the composite spectrum,
  offset for clarity. Blue
  dashed lines correspond to the rest-wavelengths of \hg\ and \hb,
  while green dot-dashed lines denote the wavelengths of $[$\ion{O}{3}$]$
  $\lambda$4364 and [\ion{O}{3}] $\lambda\lambda$4960,5008
  (Table~\ref{table:lines}). Dotted-line box in the lower left
  contains the region of the spectrum that Fig.~6 displays in greater detail.}
\label{fig:Hbandwide}
\end{figure*}

\begin{figure}
\center
\includegraphics[width=\linewidth]{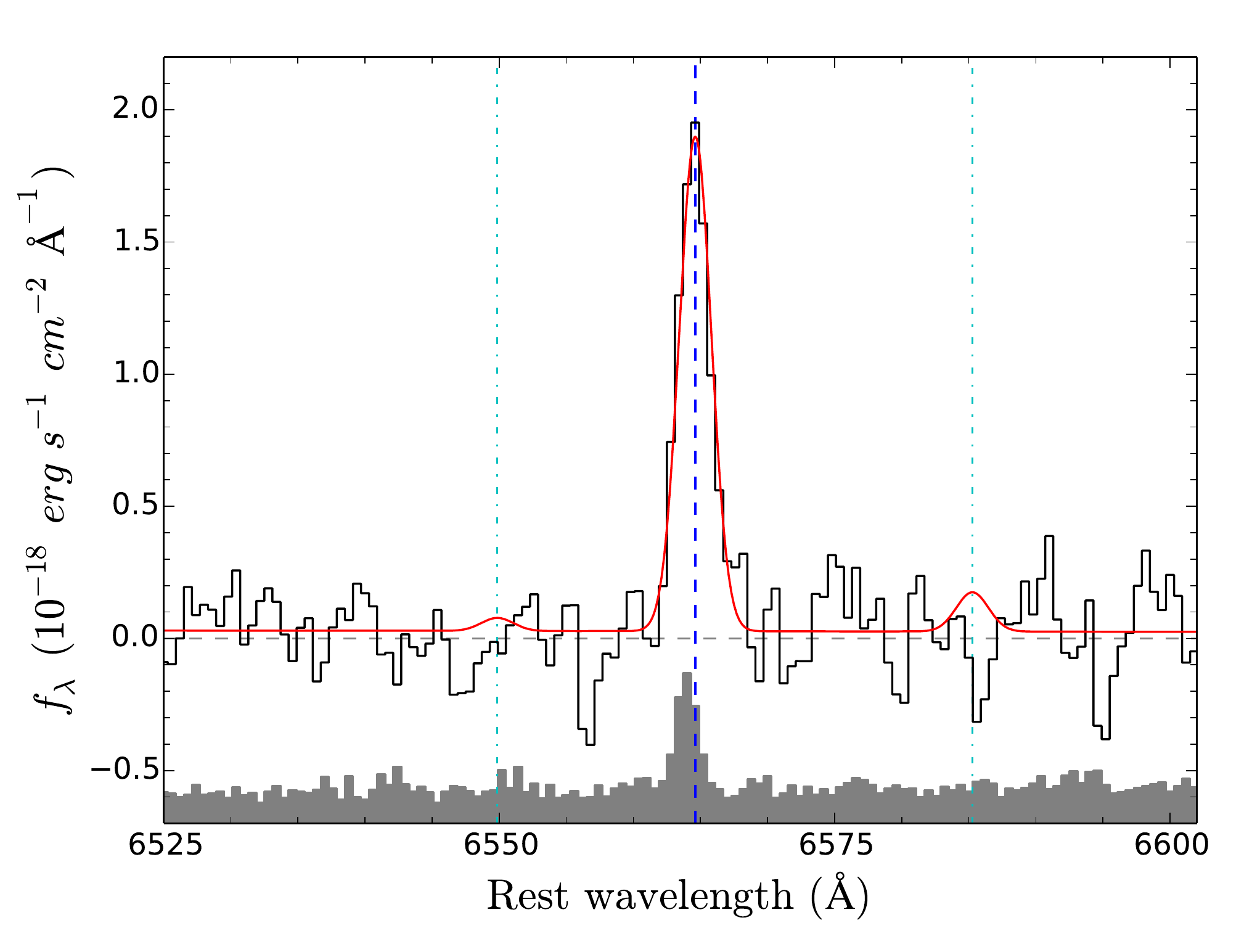}
\caption{Composite $K$ band spectrum of 28 LAEs (83 object-hours), as
  in Fig.~\ref{fig:Hbandwide}. Red line is the fit to the composite
  spectrum including the 2$\sigma$ upper limit 
  on the total [\ion{N}{2}] $\lambda\lambda$6549,6585 emission (cyan
  dot-dashed lines; Table~\ref{table:lines}). Blue dashed line indicates
  the rest wavelength of \ha.}
\label{fig:Kband}
\end{figure}

\subsubsection{Creation of composite spectra} \label{subsubsec:makestack}

In addition to our measurements of emission lines from individual
LAEs, we construct composite LAE spectra to obtain high-S/N
measurements typical of the population of LAEs and various sub-populations.

The $H$-band composite spectra are constructed as follows. Before the
individual spectra are combined, they are resampled to the 
same rest-frame wavelength scale of 0.45 \AA\ pix$^{-1}$ (equivalent
to MOSFIRE's native $H$-band pixel scale of 1.63 \AA\ pix$^{-1}$ in
the observed frame shifted to $z\approx 2.56$) spanning the wavelengths
4200\AA\ $< \lambda\sub{rest} <$ 5050\AA. An exposure mask is created to
isolate the spectral region that falls on the detector within the
MOSFIRE $H$ passband. In order to account for OH sky-line contamination, we
resample the error vector output by the MOSFIRE reduction pipeline
to the same wavelength scale as the science spectrum. The $H$-band
error vector is extremely flat across the band between the sky lines,
so we identify sky lines as those regions of the error spectrum that
exceed twice the median error value. The exposure mask is then set to
zero for these regions, such that contaminated pixels receive zero
weight in the final stack. Typically, 19$\pm$1\% of pixels in each
spectrum are removed by this algorithm. In addition, some of our
objects have particularly high background noise at
$\lambda\sub{rest}\lesssim4400$\AA; this is especially evident for the
Q1603 spectra, which have shorter exposure times and lie at lower
redshifts (such that these rest wavelengths fall nearer to the blue edge of the
MOSFIRE $H$ band). We therefore set the exposure mask to zero in
any regions of a given spectrum that lie at
$\lambda\sub{rest}<4400$\AA\ and have local noise properties
$>$2$\times$ the median noise level of the other $H$-band spectra at
the same rest-frame wavelengths.

Because the rest-optical
continuum is not detected in any individual LAE spectrum, we do not
scale the spectra by continuum magnitude before stacking. The final
$H$ composite spectrum for a given set of LAEs is then the
mean of all their $H$-band spectra, scaled by their mask-correction
factors and weighted by their corresponding exposure
masks (which reflect both the relative differences in exposure time
per object and the removal of contaminated or unobserved regions of each
spectrum). Our $H$-band spectra have very little difference in
exposure times, so each LAE receives approximately equal weight in the
final $H$-band composite. The composite spectrum of all 55 LAEs with
$H$-band spectroscopy is given in Fig.~\ref{fig:Hbandwide}.

Our $K$-band composite spectrum is constructed in a similar manner:
the individual object spectra are resampled to a constant pixel scale
of 0.6 \AA\ pix$^{-1}$ for 5400\AA\ $< \lambda\sub{rest} <$ 6800\AA\
(the native MOSFIRE $K$ pixel scale is 2.17 \AA\ pix$^{-1}$ in the
observed frame). The $K$-band error spectrum has less sky-line
contamination than the $H$ band, but it also has a strong wavelength
dependence at $\lambda\sub{obs}\gtrsim 2.1\mu$m, where the uncertainty
is dominated by thermal noise. In order to remove
sky-line-contaminated regions of the spectra before combining, we
construct a smoothed noise spectrum for each object using a 25-pixel
boxcar kernel and median averaging within the kernel. This process
produces a good approximation to the thermal noise profile with
minimal contributions from sky lines. Pixels where the error spectrum
is greater than 2.5$\times$ this local smoothed noise spectrum are then
identified as sky-line-contaminated regions and do not contribute to
the composite spectrum. Only 5$\pm$1\% of pixels are removed from each
$K$-band spectrum via this process. The composite spectrum of all 28 LAEs with
$K$-band spectroscopy is given in Fig.~\ref{fig:Kband}.

We estimate uncertainties in the $H$ and $K$ composite spectra by means of a
bootstrap procedure. For each sample of $N\sub{obj}$ LAE spectra used
to construct a composite spectrum, a bootstrap spectrum is generated
by constructing a bootstrap sample of size $N=N\sub{obj}$ (where
spectra are drawn randomly with replacement) averaging them into a bootstrap
composite spectrum using the same weighting and masking procedures
described above. This process is repeated 300 times to construct an
array of bootstrap composite spectra, and the uncertainty at each
pixel is estimated from the standard deviation of bootstrap composite
values at that pixel. In this way, the bootstrap uncertainties
represent the combination of measurement uncertainties and the true
variation of LAE spectra within each sample. These uncertainties are shown in
Figs.~\ref{fig:Hbandwide}~\&~\ref{fig:Kband} as a grey shaded region
below each spectrum.

In Sec.~\ref{sec:lyaneb}, we discuss composite spectra for high-\wla\
and low-\wla\ subgroups of our LAE sample. These composites are
constructed in a manner identical to the above, including the creation
of their boostrap uncertainty vectors. The groups are split at the
median photometric \lya\ equivalent width of our sample; LAEs with $\wla>57$\AA\
contribute to the high-\wla\ composite spectrum (30 $H$-band spectra,
12 $K$-band), while the LAEs with $\wla<57$\AA\
contribute to the low-\wla\ composite spectrum (25 $H$-band, 16 $K$-band).

\subsubsection{Fitting composite spectra} \label{subsubsec:fitstack}

\begin{deluxetable}{lcc}
\tablecaption{Emission Line Measurements}
\tablewidth{0pt}
\tablehead{
Transition & $\lambda\sub{rest}$ (\AA)\tablenotemark{a} &
  Flux ($10^{-18}$ cgs)\tablenotemark{b} 
}

\startdata
H$\gamma$ & 4341.67 &  1.63$\pm$0.54  \\
$[$\ion{O}{3}$]$ $\lambda$4364 & 4364.44 &  0.62$\pm$0.22  \\
H$\beta$ & 4862.72 &  3.44$\pm$0.58  \\
$[$\ion{O}{3}$]$ $\lambda$4960 & 4960.30 &  7.47$\pm$1.09  \\
$[$\ion{O}{3}$]$ $\lambda$5008 & 5008.24 & 18.92$\pm$2.72\tablenotemark{c}  \\
$[$\ion{O}{3}$]$ $\lambda$5008 (corr) & 5008.24 & 22.41$\pm$3.28\tablenotemark{d}  \\[5pt]
$[$\ion{N}{2}$]$ $\lambda$6549 & 6549.86 &  $<$ 0.31\tablenotemark{e,}\tablenotemark{f}  \\
H$\alpha$ & 6564.61 &  11.76$\pm$2.92\tablenotemark{e}  \\
$[$\ion{N}{2}$]$ $\lambda$6585 & 6585.27 &  $<$0.93\tablenotemark{e,}\tablenotemark{f}  
\enddata

\tablenotetext{a}{Rest-frame vacuum wavelength of transition.}
\tablenotetext{b}{Best-fit line flux ($10^{-18}$ erg s$^{-1}$
  cm$^{-2}$) in composite spectrum with  68\% confidence intervals from
  bootstrap analysis (Sec.~\ref{subsec:stack}).}
\tablenotetext{c}{Raw $[$\ion{O}{3}$]$ $\lambda$5008 flux measurement.}
\tablenotetext{d}{Corrected $[$\ion{O}{3}$]$ $\lambda$5008 flux
  estimate based on the measured $[$\ion{O}{3}$]$ $\lambda$4960 line
  flux (Sec.~\ref{subsec:mosobs}).}
\tablenotetext{e}{The \ha\ and [\ion{N}{2}] $\lambda\lambda$6549,6585
    line fluxes are measured from the composite 
  $K$-band spectrum, which includes an overlapping but smaller sample of
  LAEs compared to the $H$-band composite measurements.}
\tablenotetext{f}{[\ion{N}{2}] $\lambda\lambda$6549,6585 2$\sigma$
    limit assuming a 1:3 doublet ratio.}
\label{table:lines}
\end{deluxetable}

The composite spectra were fit using a set of gaussian line profiles
and a linearly-varying continuum. For the $H$-band composite, five
emission lines are fit simultaneously: \hg, [\ion{O}{3}]
$\lambda$4364, \hb, [\ion{O}{3}] $\lambda$4960, and [\ion{O}{3}]
$\lambda$5008. The gaussian line profiles are constrained to be
centered on the vacuum wavelength $\lambda\sub{rest}$ of the
associated transition (see Table~\ref{table:lines}), and all lines are
constrained to have the same velocity width, but the amplitude of each
emission line is fit independently. With a linear continuum component,
there are a total of 8 free parameters in the fit. The results of the
fit are displayed in Fig.~\ref{fig:Hbandwide}, and the fit line fluxes
are given in Table~\ref{table:lines}.

The $K$-band composite is fit in a similar manner, but with only 3 fit
emission lines: [\ion{N}{2}] $\lambda$6549, \ha, and [\ion{N}{2}]
$\lambda$6585, for a total of 6 free parameters in the fit. The
results of this fit are also given in Fig.~\ref{fig:Kband} and
Table~\ref{table:lines}. 

Line flux uncertainties are estimated from the samples of bootstrap
spectra. For each bootstrap spectrum, the emission lines and continuum
level are fit as described above, where each free parameter in the
full composite fit is allowed to vary for each bootstrap spectrum as
well. For each bootstrap spectrum, the line fluxes and best-fit
velocity width are measured. The 1$\sigma$ uncertainty in the measurement of
these parameters in the composite spectrum is then estimated from the
distribution of values from the 300 bootstrap spectra; specifically, from the
central interval including 68\% of corresponding parameter values
among the bootstrap spectra (Table~\ref{table:lines}). Note that the
actual uncertainty in our measurement of a given emission line in the
composite spectrum is often significantly smaller than this value, but
we present these uncertainties to reflect the range of values
associated with a ``typical'' sample of similarly-selected LAEs.

In the same way, the
uncertainty in the line ratios within a band (e.g., N2, O3;
Table~\ref{table:subsamp}) are estimated by 
computing the corresponding line ratio for each bootstrap spectrum and
measuring the size of the interval encompassing 68\% of the bootstrap
line ratio measurements. Line fluxes and ratios are
calculated in the same manner for the low-\wla\ and high-\wla\
composite spectra. Note that this this strategy must differ for cross-band
line ratios (e.g., \ha/\hb), where a different number of individual spectra
contribute to the bootstrap composite for each emission line (see
Sec.~\ref{subsec:balmer} below). 

In Sec.~\ref{sec:ratios} below, we discuss the constraints inferred
from the emission line ratio measurements and limits in these
composite LAE spectra. 

\section{Measurements from optical line ratios}\label{sec:ratios}
\subsection{N2-BPT constraints} \label{subsec:BPT}

\begin{figure}
\center
\includegraphics[width=\linewidth]{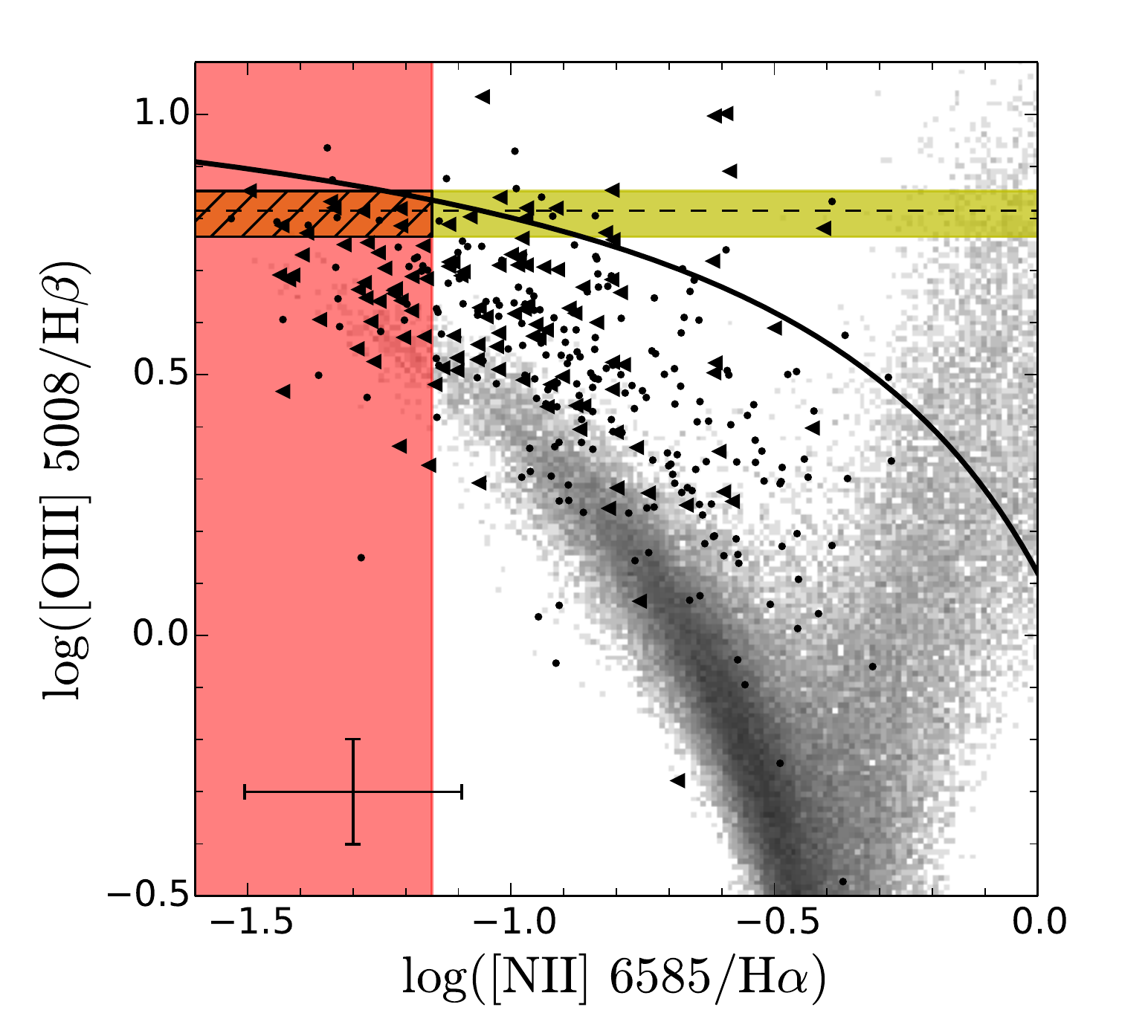}
\caption{N2-BPT diagram \citep{bal81} with SDSS
  $z\sim0$ galaxies (gray, log density scaling) and KBSS LBGs
  (\citealt{str16}; black points and 2$\sigma$ limits; cross in lower
  left shows typical errors). Yellow bar is the O3
  measurement for the LAE $H$-band composite spectrum
  (Fig.~\ref{fig:Hbandwide}), and pink region is current 2$\sigma$
  limit on N2 from the $K$-band composite (Fig.~\ref{fig:Kband}). 
  Black hatches denote the combined constraints from both
  measurements, which show that {\it the LAEs are consistent with the
  extreme high-ionization, low-metallicity tail of the LBG
  population}. For comparison, the black line shows the ``maximum 
  starburst'' curve from \citet{kewley01}.}
\label{fig:BPT}
\end{figure}

The \citet{bal81} ``BPT'' diagrams provide a simple means of
classifying the sources of ionizing radiation and physical gas properties
in galaxies. In particular, the N2-BPT diagram compares the ratio log([\ion{O}{3}]
$\lambda$5008/\hb) (hereafter, O3) to log([\ion{N}{2}] $\lambda$6585/\ha)
(hereafter, N2), which separate into two clear tracks for
low-redshift galaxies. These emission lines also have the advantage of
lying close to one another in wavelength, such that neither ratio is
strongly affected by dust attenuation or cross-band calibration
errors. Fig.~\ref{fig:BPT} displays the N2-BPT diagram for local
galaxies from the Sloan Digital Sky Survey (SDSS DR7;
\citealt{abazajian09}) along with high-redshift measurements described
below. 

In the N2-BPT diagram, star-forming galaxies occupy the locus of objects at low N2,
while AGN-dominated and composite objects form a ``fan'' that extends
toward high O3 and N2. The star-forming locus is extremely tight, with
90\% of star-forming galaxies falling within $\pm0.1$ dex of the
ridgeline \citep{kewley13a}. The small degree of scatter in this
sequence suggests that the intensity and shape of the stellar
ionizing fields and the properties of the ionized gas are
tightly coupled by one or more physical properties that vary along the
locus. In particular, these galaxies are known to follow a sequence in
metallicity \citep{dop00}, with low-metallicity galaxies or \ion{H}{2}
regions exhibiting high O3 and low N2 (the upper left of the N2-BPT),
while those with high metallicities occupy the lower right of the diagram.

As described in Sec.~\ref{sec:intro}, however, recent surveys at
higher redshifts indicate that typical galaxies at $z\approx2-3$
occupy a locus that is approximately as tight as that measured in the
local Universe \citep{ste14}, but offset toward higher O3 and/or N2
\citep{ste14,masters14,shapley15,san15,masters16,str16}. 
\citet{str16} use detailed studies of the nebular spectra of KBSS LBGs to determine
that this offset primarily corresponds to high nebular excitation
(i.e., a vertical shift in the N2-BPT plane) at a given stellar mass and
gas-phase metallicity. The black points in
Fig.~\ref{fig:BPT} represent LBGs from the KBSS, as described by
\citet{str16}. Triangles represent 2$\sigma$ upper limits on N2, and
the cross in the lower left shows the median uncertainty of the
points with detections. While measurement uncertainties
contribute significantly to the observed scatter, the KBSS LBGs appear
to roughly span the region between the SDSS locus and the black solid
line, which denotes the ``maximum starburst'' limit from \citet{kewley01}.

The measurements from the LAE composite spectra are displayed as
colored bands in the plot. The horizontal dashed line is the best-fit
O3 measurement from the composite $H$-band spectrum in
Fig.~\ref{fig:Hbandwide}, and the yellow region encompasses the 68\%
confidence interval on this value (accounting for the scatter among
the combined spectra through the bootstrap procedure described
in Sec.~\ref{subsubsec:makestack}). The red band corresponds to the
2$\sigma$ upper limit and confidence interval on N2
from the composite $K$-band spectrum in Fig.~\ref{fig:Kband}. The
average N2-BPT line ratios of our LAEs are thus localized to the far
upper left corner of the plot, where the two regions overlap (the
values of the line ratios and their uncertainties are given in
Table~\ref{table:subsamp}). 

Given the correspondence between the N2-BPT locus and nebular
metallicity, typical LAEs appear to be similar to the lowest-metallicity
LBGs in the KBSS sample. In fact, \citet{erb16} isolate the most
extreme 5\% of LBGs in the upper left of the N2-BPT
plane\footnote{Three of these ``extreme'' LBGs were previously described by \citet{ste14}.} (effectively
constructing a metallicity-selected sample of galaxies) and find
that they occupy almost exactly the same region as that defined by our
composite LAE constraints: O3 $\ge0.75$ and N2 $\le-1.1$. These
``extreme'' LBGs are found to have similar physical and spectroscopic
properties to the low-redshift population of rare, compact, high-excitation
galaxies known as ``Green Peas''
\citep{cardamone09,amorin12,jaskot13}, including high \lya\ equivalent
widths, \lya\ escape fractions, and ionization states (O32;
Table~\ref{table:ratios} \&
Sec.~\ref{subsubsec:lyavsexcitation}). In T15, we also showed that
these Green Peas (as presented by \citealt{hen15}) show similar
\lya\ and kinematic properties to the KBSS-\lya\ LAEs. Given that a simple selection
based on \lya\ emission and continuum faintness apparently isolates
the most extreme subset of objects with respect to $z\approx0$
galaxies and $z\approx2-3$ LBGs, it is worthwhile to consider the
mechanisms by which the \lya\ emission and nebular properties of these
galaxies are related; we discuss this topic in depth in Sec.~\ref{sec:lyaneb}.

\subsection{[\ion{O}{3}] auroral line and gas temperature} \label{subsec:4364}

\begin{figure}
\center
\includegraphics[width=\linewidth]{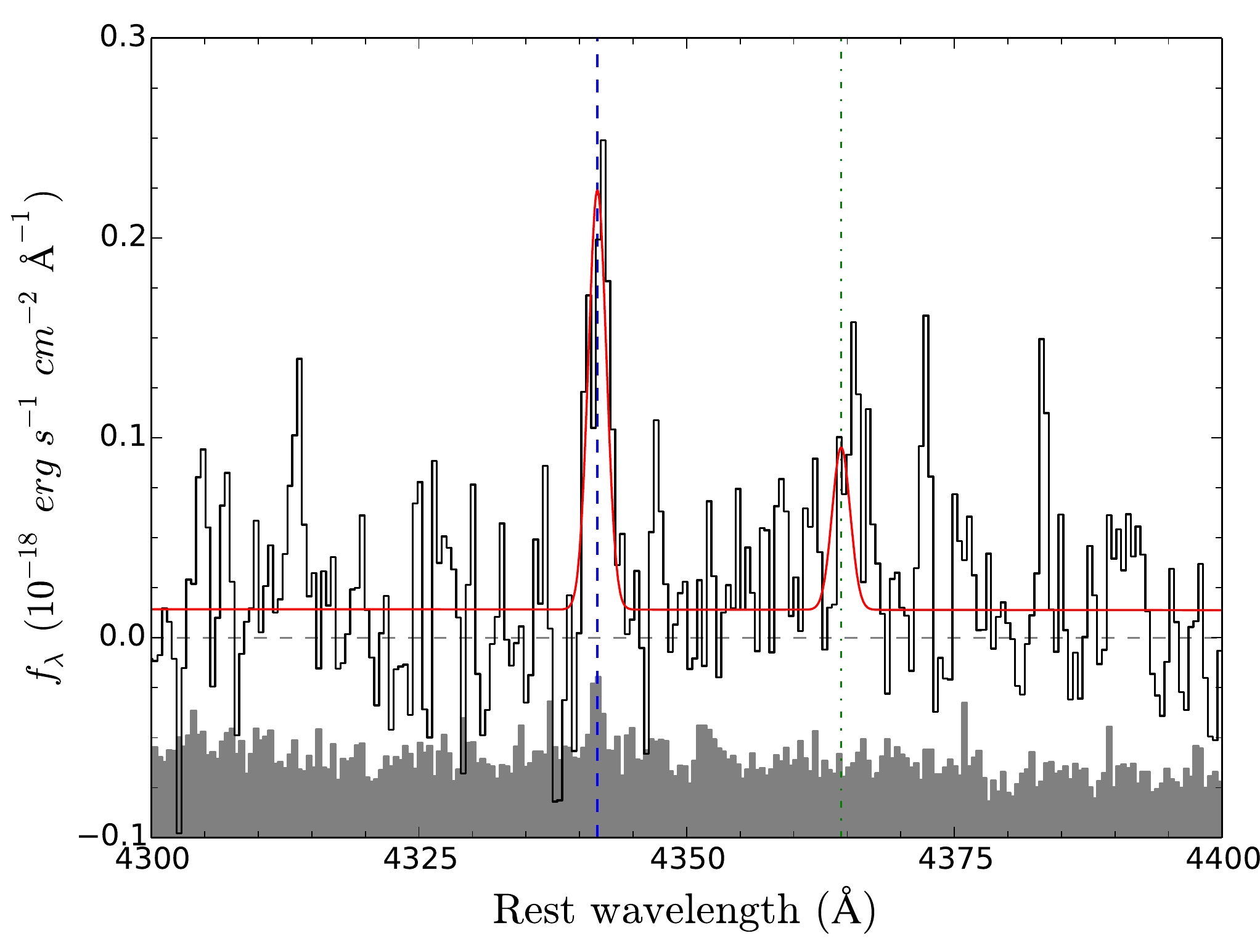}
\caption{Zoomed-in version of the composite $H$-band spectrum from
  Fig.~\ref{fig:Hbandwide} at the location of the \hg\ (blue dashed
  line) and [\ion{O}{3}] $\lambda$4364 (green dashed line) emission
  lines. Fit (red) and bootstrap uncertainty vector (grey) are as
  described in the above figures.}
\label{fig:hgo3}
\end{figure}

The [\ion{O}{3}] $\lambda$4364 transition is another emission line of
particular interest for studies of star-forming galaxies. Because this auroral emission line
corresponds to the transition from the second excited state to the
first excited state, its measurement in combination with the nebular
transition (from the first excited state to the ground state) provides a
direct measure of the electronic level populations and thus the temperature
of the \ion{O}{3} gas\footnote{More precisely, this measurement
provides the electron temperature $T_3$ in the region of the nebula
where \ion{O}{3} is the dominant ionization state of oxygen.}. As this
temperature is set in part by metal-line-dominated cooling in the
ionized nebular regions, the auroral and nebular line measurements can
also be converted into an estimate of the gas-phase
metallicity\footnote{Specifically, the abundance of oxygen, which
  dominates the cooling of gas at the temperatures, densities, and
  metallicities typical of star-forming regions.}, often described as
the ``direct method'' metallicity measurement.

The region of the $H$-band composite spectrum near the [\ion{O}{3}]
$\lambda$4364 auroral line is reproduced in more detail in
Fig.~\ref{fig:hgo3}. As described above, each of the five lines in the
$H$-band spectrum are fit simultaneously (along with the continuum)
and constrained to have the same velocity width and redshift. These
constraints significantly improve our ability to determine the  [\ion{O}{3}]
$\lambda$4364 flux. In particular, the nearby \hg\ emission line
provides a valuable cross-check of the wavelength and flux
calibrations at these wavelengths, which lie near the blue edge of the
MOSFIRE $H$ band.\footnote{As discussed in
  Sec.~\ref{subsubsec:makestack}, we mask out high-noise regions of
  the $H$ band spectra at $\lambda\sub{rest}<4400$\AA. This masking
  procedure improves the quality of our \hg\ and [\ion{O}{3}]
  $\lambda$4364 line fits, but does not change their inferred fluxes.}
Notably, the \hg\ line is detected with high 
significance, the emission is centered at the proper rest-wavelength,
and the measured flux is consistent with the expected \hg/\hb\ ratio
under case-B recombination and $\ebv\approx 0$, as discussed in
Sec.~\ref{subsec:balmer} below. For these reasons, we expect that
the derived [\ion{O}{3}] $\lambda$4364 flux is well-determined,
despite the relatively marginal (2.8$\sigma$) detection of the peak.

The [\ion{O}{3}] line fluxes are given in Table~\ref{table:lines}. As
described above, we use the $\lambda$4960 line to correct the
[\ion{O}{3}] $\lambda$5008 line flux because of the uncertain flux
calibration at the extreme red end of the MOSFIRE $H$ band. The ratio
of auroral to nebular \ion{O}{3} line flux $R\sub{O3}\equiv$ [\ion{O}{3}]
($\lambda$4364)/($\lambda\lambda$4960,5008) is therefore equivalent to
[\ion{O}{3}] ($\lambda$4364)/(4$\times\lambda$4960) in our measurement,
and we find $R\sub{O3}=2.0\pm0.7$\%. We use the iterative procedure
described by \citet{izo06} (reproduced from \citealt{all84})
to determine the \ion{O}{3} electron temperature $T_3$, finding a converged value
$T_3=1.78\pm0.33\times10^4$K, where the uncertainty reflects the
1$\sigma$ range of $R\sub{O3}$ ratios measured in our bootstrap spectra. 

\subsection{Gas-phase metallicity estimates}\label{subsec:zgas}

Given a measurement of the electron temperature, ionic abundances (and
the ``direct method'' oxygen abundance)
can be estimated via the ratios of additional metal-ion and hydrogen
emission lines, as further discussed by \citet{izo06}.\footnote{See
  also discussion in \citet{ste14}, wherein we present $T_e$
  measurements for three KBSS LBGs.} Using the
formulae fit in that paper, we 
calculate the ionic \ion{O}{3} abundance 12 + log(O$^{++}$/H$^{+}$) =
$7.68\pm0.16$. The estimation of the elemental O/H abundance requires
an ionization correction that can only be measured with the addition
of emission line measurements from other oxygen ions, which are not
currently measured for the LAE sample presented here. However, there
is a close relationship between O32 and O3 among the KBSS LBG galaxies
(see further discussion in Sec.~\ref{subsubsec:lyavsexcitation}), such
that the LBGs with O3 $\approx0.8$ have O32 $\approx0.7-1.0$ (that is,
nebular [\ion{O}{3}] emission 5$-$10$\times$ stronger than that
of [\ion{O}{2}]; \citealt{str16}). For this reason, the contribution of \ion{O}{2} to the
total oxygen abundance is likely to be small among the highly-excited
LAEs, and we calculate an ionization correction based on a likely
value of O32 $=0.7$. The temperature of the \ion{O}{2}
zone of the \ion{H}{2} region is typically lower than that of the
\ion{O}{3} zone, and the difference in temperature (the
``$T_2-T_3$'' relation) is typically found to be $T_2\approx
0.7\times T_3-3000$K by photoionization models
(e.g., \citealt{campbell86,garnett92,izo06,pilyugin09}), consistent with recent
direct observations of $T_2$ in \ion{H}{2} regions of local star-forming galaxies
\citep{brown14,berg15}.

Under these assumptions, the inferred ionic 
\ion{O}{2} abundance is 12 + log(O$^{+}$/H$^{+}$) = $7.17\pm0.19$,
where the uncertainty includes only the range of $T_3$ consistent with
our bootstrap spectra. If the true O32 ratio for our LAE spectra is
greater than the assumed value of 0.7, or if the \ion{O}{2}
temperature is higher than that predicted by our assumed $T_2-T_3$
relation, then the contribution of \ion{O}{2} to the total oxygen
abundance is even smaller. Conversely, \citet{and13} suggest that the
formula above overestimates $T_2$ by $\Delta T_2\approx1300$K;
applying such a shift would increase our estimate of 12 +
log(O$^{+}$/H$^{+}$) to 7.28.

Assuming 
that \ion{O}{3} and \ion{O}{2} are the dominant states of oxygen in
the nebular regions (and likewise that the neutral fraction of
hydrogen is negligible in these regions), the inferred ``direct'' oxygen
abundance is thus the sum of the above ionic abundances, and we
estimate a total oxygen abundance 12 + log(O/H)\sub{dir} =
7.80$\pm$0.17 ($Z\sub{neb}\approx0.13Z_\odot$). As
above, the uncertainty corresponds to the statistical uncertainty from
our bootstrap measurements; for comparison, assuming O32 = 1.0 or
using the \citet{and13} $T_2$ calibration would shift our inferred
oxygen abundance by $-$0.06 dex or +0.03 dex, respectively.

However, a larger source of systematic uncertainty may come from the
tendency of collisionally-excited lines (CELs) including the
\ion{O}{3} lines discussed above to overestimate the
volume-averaged electron temperature of a cloud. This effect may occur due to the temperature
sensitivity of the emissivity of CELs, which causes any
luminosity-weighted $T_e$ measurement to be biased toward the
highest-temperature regions of the nebula. Such a bias may cause metallicity estimates
from CELs to {\it underestimate} the metallicity relative to that
inferred from recombination lines (RELs) and stellar spectra. This
effect is seen in the detailed spectroscopic LBG study
described by \citet{ste16}, who find an offset between the nebular O/H
abundance inferred from the [\ion{O}{3}] ``direct'' method and
that obtained through comprehensive modeling of the nebular and stellar
spectra. \citeauthor{ste16} find an offset consistent with that
measured from CELs and RELs in local low-metallicity dwarf galaxies by
\citet{esteban14}:

\begin{equation}
\rm{log(O/H)\sub{REL}}-\rm{log(O/H)\sub{CEL}}=0.24\pm0.02\,\, \rm{dex}
\end{equation}

In constrast, \citet{bresolin16} find a low-metallicity REL-CEL
offset of similar magnitude, but they suggest that CELs are {\it more}
accurate than RELs by comparing both estimators to stellar
metallicities collected from the literature. Given that the results of 
\citet{ste16} appear to corraborate the \citet{esteban14} offset at
$z\approx2-3$, we apply a +0.24 dex correction to our ``direct''
abundance measurement described above in order to determine our final
estimate of the nebular gas-phase metallicity:

\begin{eqnarray}\label{eq:metallicity}
12+\rm{log(O/H)} & = & 8.04\pm0.19 \\
Z\sub{neb} & = & 0.22^{+0.12}_{-0.08} Z_\odot
\end{eqnarray}

\noindent where the uncertainty reflects both the statistical
uncertainties from our bootstrap analysis and the statistical uncertainty in the
\citet{esteban14} calibration, but it does not include the systematic
uncertainty in the application of the REL-CEL offset,
nor those associated with the \ion{O}{2} abundance discussed above.

As described in Sec.~\ref{subsec:BPT}, the O3 and N2 ratios are also
often used as gas-phase metallicity indicators (the ``strong-line''
metallicity indicators N2 and O3N2, Table~\ref{table:ratios}) through local
calibrations to $T_e$-based measurements. As discussed by
\citet{ste14,ste16}, these strong-line indicators are based on the adherence of
star-forming galaxies to their locus in the local N2-BPT plane, and
thus require recalibration at high redshift, where this locus is
offset toward higher values of nebular excitation. Lacking a direct
calibration of these relationships at $z\approx2$, we use the recent
calibration of O3N2 by \citet{str16}, which is based on a local set of
extragalactic \ion{H}{2} regions from \citet{pilyugin12}. Using this
relation, our best-fit measurement of O3, and our 2$\sigma$ upper
limit on N2, we obtain the following limit:

\begin{eqnarray}\label{eq:stromlines}
12+\rm{log(O/H)\sub{O3N2,Strom16}} < 8.17
\end{eqnarray}

\noindent which is consistent with the corrected direct
estimate in Eq.~\ref{eq:metallicity}. For comparison, the widely-used
N2 and O3N2 abundance calibrations by \citet{pettini04} also produce
estimates consistent with our direct-method determination:

\begin{eqnarray}\label{eq:PP04}
12+\rm{log(O/H)\sub{O3N2,PP04}} < 8.10 \\
12+\rm{log(O/H)\sub{N2,PP04}} < 8.24 \,\,.
\end{eqnarray}

\subsection{Balmer decrement and extinction measurements} \label{subsec:balmer}

The above inferences are based on line ratios (O3, N2, R\sub{O3}) that fall
within a single MOSFIRE band. However, not all the LAEs in our sample have
both $H$ and $K$ band detections, which means that cross-band line
ratios (such as the Balmer decrement, \ha/\hb) cannot be measured for the full 
sample of spectra. Eleven LAEs in our sample have $>$3$\sigma$ detections
of both \ha\ and \hb. Among these spectra, the average ratio is \ha/\hb\ =
2.92$\pm$0.45. The uncertainty is the 1$\sigma$ error on the
average estimated via a modified bootstrap technique similar to that
described in Sec.~\ref{subsubsec:makestack}
above, but modified so that the same randomized set LAEs contribute to
each bootstrap sample of both \ha\ (in the $K$ band) and \hb\ (in the $H$ band).

We estimate the average extinction from the Balmer decrement assuming
a \citet{car89} Milky-Way extinction curve and the tabulated intrinsic
\ha/\hb\ ratios from \citet{bro71}. Typically, extinction measurements
for high-redshift galaxies assume an electron temperature $T_e\approx
10^4$ K, corresponding to an intrinsic ratio \ha/\hb\ =
2.89\footnote{Some references prefer the value of 2.86 from \citet{ost06},
but this difference makes a negligible change in our inferred
extinction.}. However, our measurement of the [\ion{O}{3}]
$\lambda$4364 line implies a somewhat higher value of
$T_e\approx1.8\times10^4$ K, so we adopt the Balmer decrement value
for $T_e=2\times10^4$ K from \citet{bro71}: \ha/\hb\ = 2.74. Choosing
the higher intrinsic ratio would decrease our inferred extinction by a
small amount, as discussed below.

Under these assumptions, our Balmer decrement measurements correspond
to a reddening, $V$-band extinction, and \ha\ extinction as follows:

\begin{eqnarray}\label{eq:extinction}
\textrm{E}(B-V) & = & 0.06\pm0.12 \\
A_V & = & 0.20\pm0.39 \nonumber \\
A\sub{\ha} & = & 0.16\pm0.32 \nonumber \,\, .
\end{eqnarray}

Choosing an intrinsic ratio \ha/\hb\ = 2.89 would imply
$\ebv=0.01\pm0.12$, consistent with the above measurement. For
comparison, we also calculate the extinction inferred from the 
total $H$ and $K$ stacks using the uncorrected \ha\ and \hb\ values
from Table~\ref{table:lines} (despite corresponding to different
samples of LAEs). From these values, we calculate a Balmer decrement
\ha/\hb\ = 3.37, or \ebv\ $= 0.2$ under the assumptions above. This
value is slightly higher than that inferred from the matched samples
of detected \ha\ and \hb\ lines, likely reflecting the fact that our
current $K$-band spectra are shallower on average than those in the
$H$ band, such that bright \ha\ lines are over-represented in the full $K$
stack. We therefore take the extinction inferred from the matched
sample (Eq.~\ref{eq:extinction}) to best represent our full LAE
population, and the dust-corrections used to derive LAE star-formation
rates in Fig.~\ref{fig:sfr} are based on this value.

\section{The nebular origins of \lya\ emission} \label{sec:lyaneb}

In order to determine the physical properties of LAEs, it is important
to understand the physical drivers of their most salient
characteristic: strong \lya\ emission. Toward this end, we here
consider the relationship between the nebular properties described
above and the \lya\ emission of our LAE sample, as well as that of a
comparison sample of LBGs from the KBSS
\citep{ste14,ste16,str16}. {\it The KBSS and KBSS-\lya\ 
represent the richest current sample of combined \lya\ and
rest-optical spectroscopy for star-forming galaxies at any redshift},
so these surveys are a powerful tool for 
dissecting the physical differences between galaxies selected by \lya\
emission and those selected by continuum brightness, while also
establishing the variation in net \lya\ emissivity with galaxy
properties across the combined population of LAEs and LBGs.

\subsection{The BPT-\lya\ relation} \label{subsec:bptlya}

\begin{deluxetable*}{lcccccccc}
\tablecaption{LAE and LBG Subsamples}
\tablewidth{0pt}
\tablehead{
Sample & Subsample & $N\sub{obj}$ & $\langle W\sub{\lya}
\rangle$\tablenotemark{a} & SFR\tablenotemark{b} &
  N2 & O3 & \ha/\hb \tablenotemark{c} & E($B-V$)\tablenotemark{d}
}

\startdata
& all &  60 & 56.2\AA & 7.7$\pm$1.9 & \phn$<$ $-$1.15 & 0.82$\pm$0.05 & 2.92$\pm$0.45 & 0.06$\pm$0.12 \\
LAEs & $W\sub{\lya,phot} > 57$\AA &  30 & 79.0\AA & 4.5$\pm$1.0  & $<$ $-$0.94 & 0.90$\pm$0.09 & 2.63$\pm$1.03 & $<$ $-$0.29 \\
& 20\AA $< W\sub{\lya,phot} \leq 57$\AA &  30 & 24.7\AA & 14.4$\pm$4.4\phn & $<$ $-$1.07 & 0.76$\pm$0.07 & 3.19$\pm$0.38 & 0.15$\pm$0.11 \\[5pt]
\hline \\[-5pt]
& $W\sub{\lya,spec}>20$\AA &  48 & 47.3\AA & 19.3$\pm$2.2 & $-$1.12$\pm$0.12 & 0.73$\pm$0.02 & 3.77$\pm$0.19 & 0.27$\pm$0.05 \\
LBGs & $0$ $\leq W\sub{\lya,spec}\leq 20$\AA & 104\phn &  \phn7.8\AA & 21.0$\pm$1.4
& $-$1.00$\pm$0.06 & 0.64$\pm$0.02 & 3.95$\pm$0.12 & 0.32$\pm$0.03 \\
& $W\sub{\lya,spec}<0$ & 216\phn & \phn$-$7.6\AA\phs & 21.1$\pm$0.8 & $-$0.90$\pm$0.03 & 0.54$\pm$0.02 & 4.08$\pm$0.09 & 0.35$\pm$0.02
\enddata
\tablenotetext{a}{Composite spectroscopic rest-frame \lya\ equivalent
  width. $W\sub{\lya}>0$ indicates emission and $W\sub{\lya}<0$
  indicates absorption.}
\tablenotetext{b}{Dust-corrected \ha\ star-formation rate in \msun\
  yr$^{-1}$. For the LAEs, only objects with
  $K$-band spectra are included (Table \ref{table:laes}).}
\tablenotetext{c}{The Balmer decrement \ha/\hb\ is measured only for
  the subset of objects with $>$3$\sigma$ detections of both \ha\ and \hb\
  in their individual spectra (11 LAEs, of which 5 are in the
  low-$W\sub{\lya}$ group and 8 are in the low-$W\sub{\lya}$ group).}
\tablenotetext{d}{Color excess is estimated using a \citet{car89}
  extinction curve ($R_V=3.1$). Note that an intrinsic ratio \ha/\hb\ $=
  2.74$ ($T_e\approx 2\times 10^4$ K) is assumed for the LAEs, whereas
  \ha/\hb\ $= 2.89$ ($T_e\approx 10^4$ K) is assumed for the LBGs as
  described in Sec.~\ref{subsec:balmer}.}
\label{table:subsamp}
\end{deluxetable*}

\begin{figure*}
\center
\includegraphics[width=0.9\linewidth]{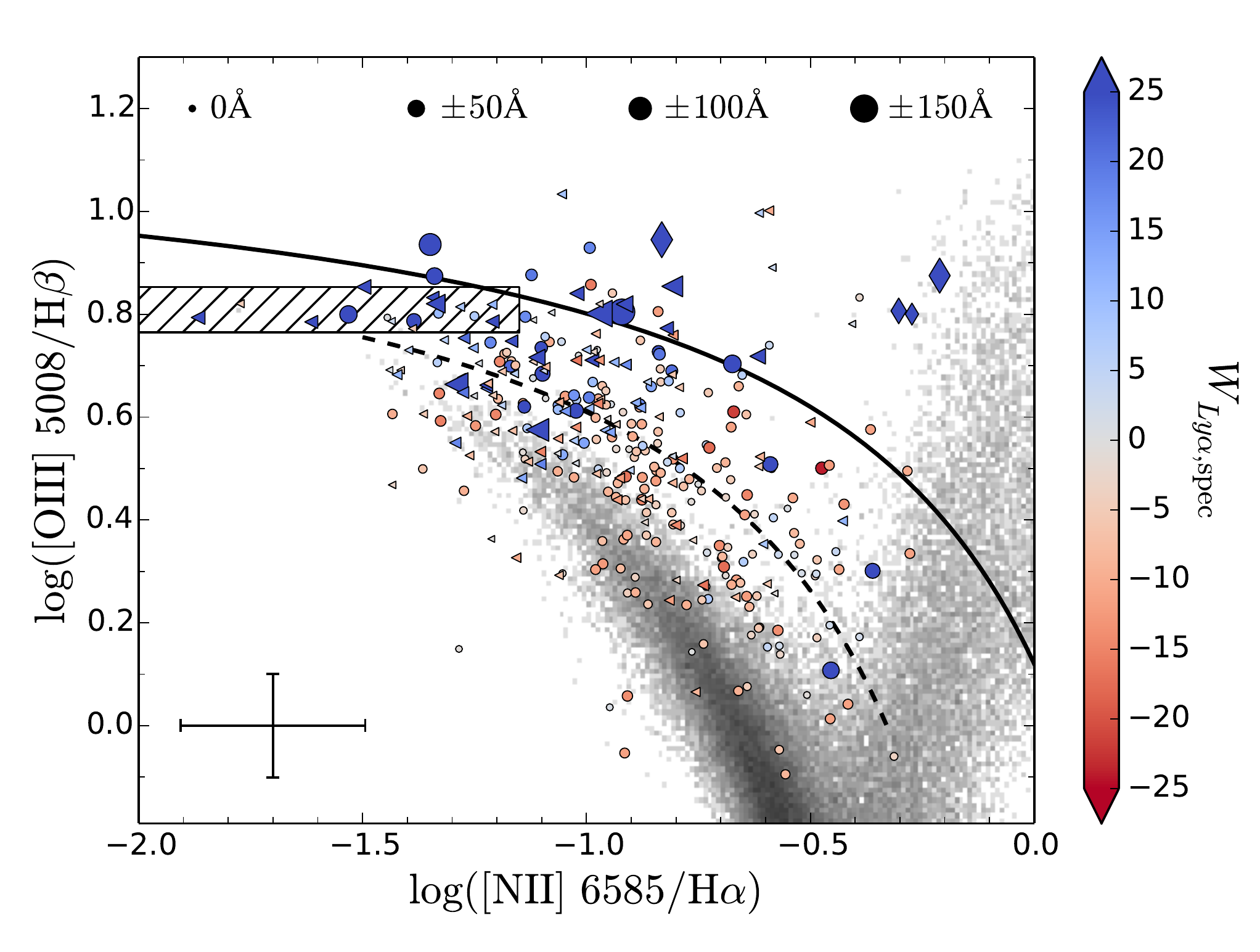}
\caption{N2-BPT diagram as in
  Fig.~\ref{fig:BPT}, but with KBSS LBGs color-coded by \lya\
  equivalent width ($W\sub{\lya}$). Blue points show \lya\ in emission, while red
  points show \lya\ in absorption. The sizes of the points correspond
  to the absolute value of $W\sub{\lya}$. Error bars in lower left
  correspond to the median 1$\sigma$ uncertainties on the detected
  KBSS points. Blue diamonds in the upper
  right corner correspond to KBSS objects with
  spectroscopically-identified AGN emission \citep{ste14}. The black
  hatched region shows the confidence interval for the LAE
  stacks. Solid black line denotes the ``maximum 
  starburst'' curve from \citet{kewley01}, while the dashed black line
  shows the \citet{str16} KBSS locus. There is a general trend for high-$W\sub{\lya}$ galaxies to
  occupy the upper left of the diagram (near the LAE sample), whereas
  low-$W\sub{\lya}$ objects occupy the lower right. The physical origins of this
  trend are analyzed throughout Sec.~\ref{sec:lyaneb}.}
\label{fig:BPTlya}
\end{figure*}

\begin{figure}
\center
\includegraphics[width=\linewidth]{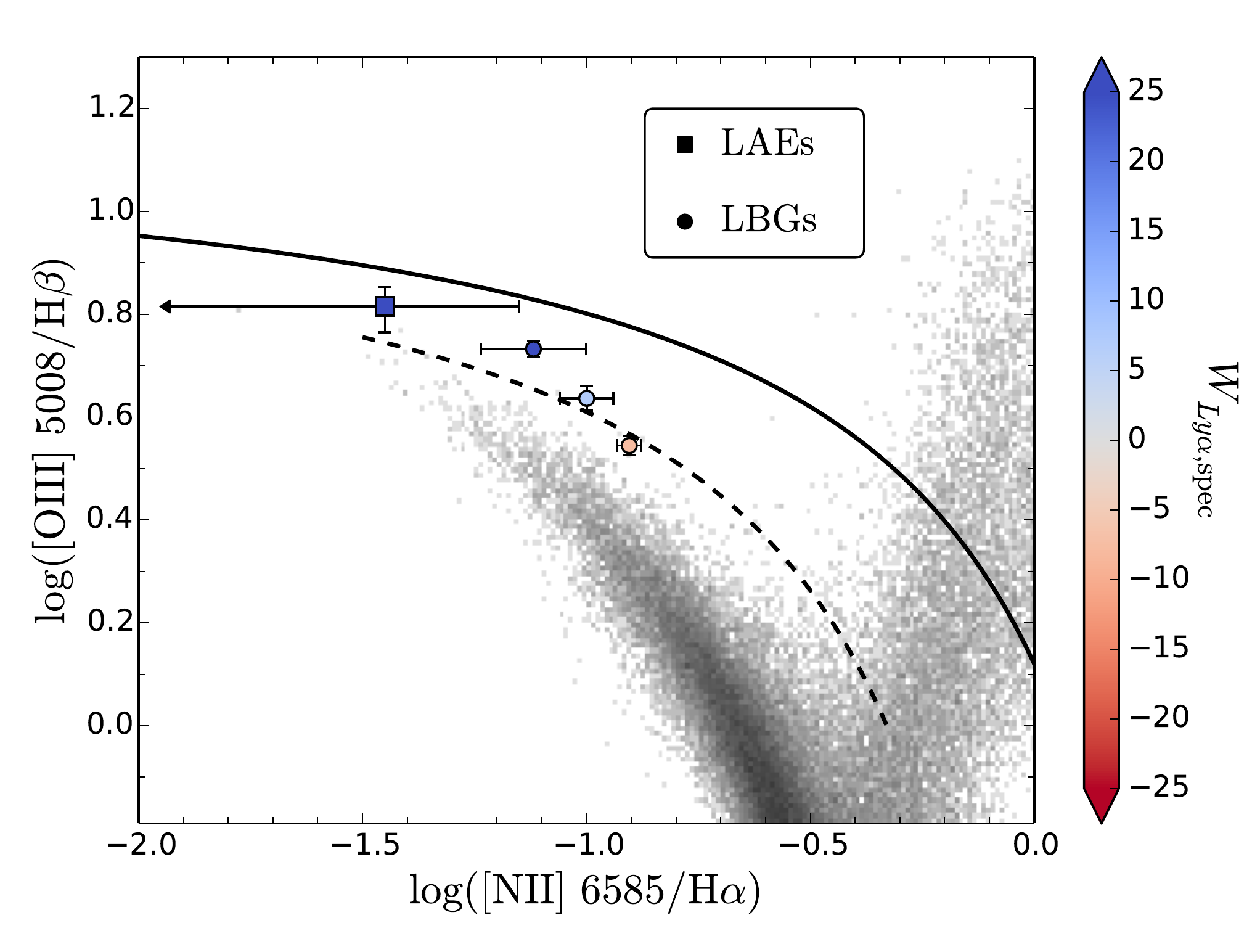}
\caption{N2-BPT diagram as in
  Figs.~\ref{fig:BPT}~\&~\ref{fig:BPTlya}, but with KBSS points
  stacked in three subsamples according to their value of
  $W\sub{\lya}$ (Table~\ref{table:subsamp}). Error bars correspond to
  the 1$\sigma$ uncertainties from bootstrap resampling. In analogy to the other
  points, the LAE stacks are presented with a point at the best fit
  value of O3 and the 1$\sigma$ limit on
  N2, with error bars indicating the 1$\sigma$
  uncertainty on O3 and the 2$\sigma$ upper limit on N2. The solid
  black line denotes the ``maximum starburst'' curve from
  \citet{kewley01}, while the dashed black line 
  shows the \citet{str16} KBSS locus. A strong
  trend is visible as \wla\ increases toward the low-metallicity end
  of the N2-BPT locus.}
\label{fig:BPTlyastack}
\end{figure}


As discussed above in Sec.~\ref{subsec:BPT}, the N2-BPT diagram
provides a useful discriminant of the physical properties of ionized
regions within a galaxy, which constrains the metallicity of both the gas
itself and the sources of ionizing radiation, including properties of
the stellar populations. 

Fig.~\ref{fig:BPTlya} displays the N2-BPT line ratios of 336 KBSS
galaxies with $>$5$\sigma$ ($>$3$\sigma$, $>$3$\sigma$) detections of \ha\
(\hb, [\ion{O}{3}] $\lambda$5008) and spectroscopic measurements of
their \lya\ equivalent widths, $W\sub{\lya,spec}$. While there is
considerable scatter in the nebular line ratios at a given value of
$W\sub{\lya,spec}$, there is also a clear trend such that 
\lya-emitting LBGs ($W\sub{\lya,spec}>0$, blue points) have high values of
O3 and low values of N2
(i.e., they lie in the upper-left region of the N2-BPT space), whereas
\lya\ absorbers ($W\sub{\lya,spec}<0$, red points) preferentially
occupy the opposite corner of parameter space. Objects with
spectra indicating AGN activity (e.g., broad nebular emission lines and/or
strong \ion{C}{4} or \ion{He}{2} UV emission) are denoted by diamonds
in the plot; these objects show high ratios of both O3 and N2 similar to
the spectra of low-redshift AGN. 

The region of the N2-BPT parameter space consistent with our composite
LAE spectra\footnote{The central 68\% confidence interval in
  O3 and the 2$\sigma$ upper limit on N2
  from our bootstrap analysis.} (as in Fig.~\ref{fig:BPT}) is displayed as
the hatched region in Fig.~\ref{fig:BPTlya}. The nebular line
properties of the composite LAE spectra are generally consistent with
those of the KBSS LBGs with the highest values of
$W\sub{\lya}$.

Fig.~\ref{fig:BPTlyastack} compares the composite LAE line ratios to
analogous stacked measurements of subsamples of the KBSS LBGs, which
are described in Table~\ref{table:subsamp}. The KBSS subsamples are
divided on the basis of $W\sub{\lya}$: \lya-absorbers
($W\sub{\lya,spec}<0$), weak \lya-emitters ($0<W\sub{\lya,spec}<20$\AA),
and strong \lya-emitters ($W\sub{\lya,spec}>20$\AA). For each
subsample, all spectra are combined for which the MOSFIRE $H$ and $K$
spectra cover the rest-wavelengths of all 4 of the N2-BPT diagnostic
lines: \hb, [\ion{O}{3}] $\lambda$5008, \ha, and [\ion{N}{2}]
$\lambda$6585. Objects spectroscopically identified as AGN (marked as
diamonds in Fig. ~\ref{fig:BPTlya}) are excluded from the stacks,
leaving a total of 368 LBGs. In order to ensure that each object
receives the same weighting in both the $H$ and $K$ composites while
maximizing their S/N, both spectra for each object are weighted by the
inverse-variance at the wavelength of the 
[\ion{N}{2}] $\lambda$6585 line (generally the weakest of the N2-BPT
diagnostic lines in this sample). The line ratios measured for each
composite spectrum are listed in Table~\ref{table:subsamp} and
displayed in Fig.~\ref{fig:BPTlyastack}. For comparison, the LAE
composite measurements are plotted in Fig.~\ref{fig:BPTlyastack} as a
point at the best-fit value of O3 and the
1$\sigma$ upper limit of N2, with error
bars reflecting the 68\% confidence interval on O3 and the 2$\sigma$
upper limit of N2. The value of $W\sub{\lya,spec}$ is measured 
for each LAE and LBG sample directly from the corresponding composite
UV spectrum by comparing the measured \lya\ line flux (without
correcting for \lya\ slit losses) to the UV continuum flux on the red
side of the \lya\ line, as is described for measurements of individual
LBG spectra in Sec.~\ref{subsec:lrisobs}.

As in Fig.~\ref{fig:BPTlya}, Fig.~\ref{fig:BPTlyastack} shows a clear
trend between the value of $W\sub{\lya}$ for a given subsample and its
position in the N2-BPT plane. The composite measurements parallel the
locus of SDSS N2-BPT measurements, albeit with an offset consistent
with previous studies of high-redshift star-forming galaxies, as
discussed above. Notably, our current limits on the typical
properties of faint LAEs appear consistent with the trend seen in the
LBG composites; the ``BPT offset'' of the full LAE composite
measurement may be slightly greater than that of the KBSS
composites, but the LAE measurement may actually be {\it more}
consistent with the SDSS locus depending on the (currently unmeasured)
typical LAE N2 ratio. Rather than
investigating the source of this offset, we therefore consider what
physical galaxy properties are changing {\it along the locus} of
$z\approx2-3$ LAEs and LBGs that accompany or drive the variation in
\lya\ emissivity.

The highest-$W\sub{\lya}$ objects and composite spectra in
our sample are those that lie nearest to the low-metallicity end of
the SDSS galaxy locus (see discussion in
Sec.~\ref{subsec:BPT}). Given the correlation between gas-phase 
metallicity and dust content, this may suggest that our observed trend is a
signature of the previously-studied tendency of LAEs to exhibit lower
dust attenuation with respect to continuum-selected
galaxies. In such a scenario, the variation in net \lya\ emissivity
along the N2-BPT locus (as parameterized by $W\sub{\lya}$) is primarily a variation in the
physics of \lya\ {\it escape}, which is expected to depend
sensitively on the distribution of gas and dust within the
interstellar medium of the host galaxies.

However, \citet{ste14} demonstrate that gas-phase
metallicity has more minor effects on the position of
individual galaxies within the locus of star-forming galaxies at
$z\approx2-3$ compared to $z\approx0$. Rather, the primary determinants of the N2-BPT line
ratios in LBGs galaxies appear to be the relative hardness of the incident radiation
field and the effective ionization parameter $n_\gamma/n_H$, the
dimensionless ratio of hydrogen-ionizing photons to 
hydrogen atoms within the ionized star-forming regions. This
ionization parameter may also be expressed as the factor $U\equiv 
\Phi_H/n_H c$, where $n_H$ is the number density of
hydrogen atoms (including ionized, neutral, and molecular) in the
star-forming regions and $\Phi_H$ is the surface flux of H-ionizing
photons incident on the illuminated face of the \ion{H}{2} region, as
defined by \citet{ost06}. A recent, thorough discussion of the ionization
parameter, its various definitions, and its observational constraints
in the star-forming regions of $z\approx2-3$ galaxies is given by
\citet{san16}. We refer 
to the ionization parameter as $U$ in the sections that follow in
order to compare our measurements to predictions from the Cloudy
photoionization code \citep{fer13}, which explicitly defines $U$ as
described above, but we note that the physical interpretation of this
ratio can become ambiguous when divorced from the specific 
plane-parallel or spherical ionization geometries assumed by
photoionization models (as discussed by \citealt{ste14}).

Crucially, the dependence of a galaxy's nebular line ratios on $U$ and
the radiation field hardness means that these ratios are strongly
determined by the overall {\it normalization} and {\it shape} of the
stellar radiation field at photon energies 1 Ryd $< E_\gamma \lesssim$
4 Ryd, which in turn means that the N2-BPT line ratios may be at least as
sensitive to properties of the young stellar populations within the
\ion{H}{2} regions as to the intrinsic properties of the ionized
gas. Specifically, \citet{ste14} argue that the nebular line spectra
of $z\approx2-3$ LBGs are best-explained by populations of hot
massive stars, and \citet{ste16} interpret combined deep rest-UV and
nebular line spectra of these LBGs in light of binary evolution models
in low-metallicity stellar populations, which 
naturally produce hotter, harder ionizing spectra than typical stellar
models over long ($\gg$Myr) timescales as described in Sec.~\ref{sec:intro}.

In the section below, we consider two possible modes by which the nebular spectra of these
galaxies may be tied to \lya\ emissivity: variation in the extinction
of \lya\ photons by interstellar dust (which may also appear as
reddening in the nebular spectra), or variation in the \lya\
production rate via recombination in ionized star-forming regions (which
may produce changes in the observed nebular excitation).

\subsection{Origin of the BPT-\lya\ relation}\label{subsec:lyacorr}


\subsubsection{\lya\ emission vs. dust attenuation}\label{subsubsec:lyavsdust}

\begin{figure}
\center
\includegraphics[width=\linewidth]{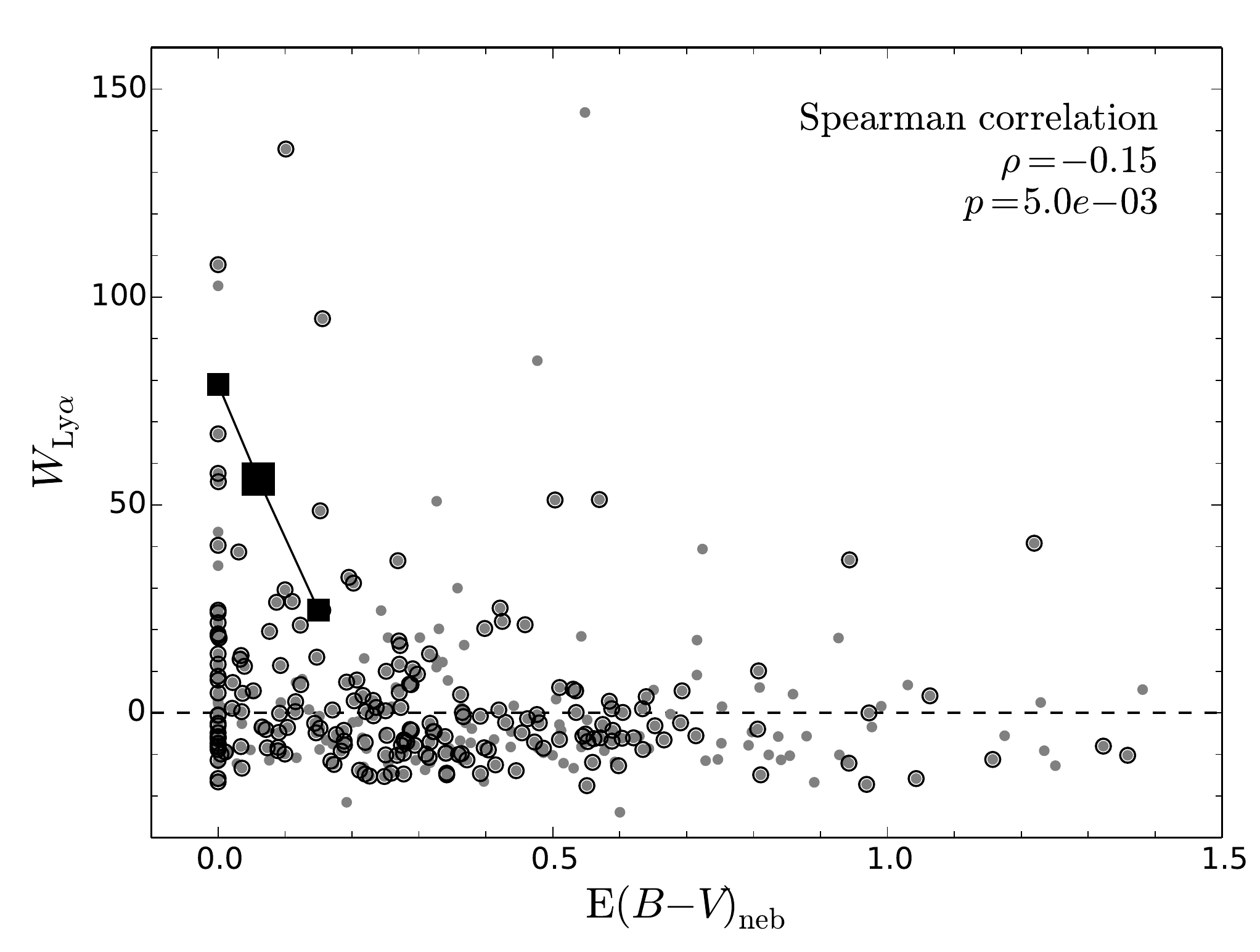}
\caption{The spectroscopic equivalent width of \lya\ 
  (\wla) vs. the nebular reddening \ebv\ inferred from the Balmer
  decrement measurement for individual KBSS LBGs (grey points) and
  KBSS-\lya\ LAE composite spectra (large black square indicates the full
  sample, while smaller squares denote the low-\wla\ and high-\wla\
  subsamples). Circled points are those with S/N $>5$ in \ha/\hb, the
  limit proposed by \citet{str16} for secure measurements of \ebv. The
  horizontal dashed line demarcates the boundary 
  between \lya-absorbers ($\wla<0$) and \lya-emitting LBGs
  ($\wla>0$). The results of a non-parametric Spearman correlation
  test on the LBG values are displayed in the upper right, which
  indicates that \wla\ has only modest dependence on the inferred
  reddening and attenuation by dust. The LAEs have much higher values
  of \wla\ than reddening-matched LBGs.}
\label{fig:ebv_vs_wlya}
\end{figure}

We investigate the modulation of \lya\ emissivity by dust by comparing
the relationship between 
\wla\ and the nebular reddening \ebv\ estimated from the
Balmer decrement (as described in Sec.~\ref{subsec:balmer}) for each
of the KBSS LBG spectra and the LAE composites (including the
low-$W\sub{\lya}$, high-$W\sub{\lya}$, and combined subsamples). The
inferred nebular reddening and \lya\ equivalent width for each LBG
(from the set of 336 objects with full N2-BPT and \lya\ line coverage) and
LAE composite is shown in Fig.~\ref{fig:ebv_vs_wlya}. An association
of high-\wla\ LBGs with low values of \ebv\ is visible, although
there are several LBGs with high values of both \wla\ and \ebv. A
non-parametric Spearman rank-correlation test finds a weak
negative correlation ($\rho=-0.15$) between \wla\ and \ebv\ for the
LBG sample with moderate
significance ($p=5\times 10^{-3}$). \citet{str16} find that some KBSS LBGs
exhibit unphysical dust-corrected line ratios (e.g., in the R23-O32
plane) when corrections are applied based on low-S/N measurements of
\ebv\sub{neb}, suggesting a limit of \ha/\hb\ $>5$$\sigma$ for
reliable estimates of the dust attenuation.\footnote{Specifically,
  \citet{str16} find that some low-S/N objects are scattered toward
  unphysically low O32 and high R23 values. In addition, there is a subset of
objects with robust line detections for which the \ha/\hb\ appears to
overestimate \ebv, likely indicating cases where the \citet{car89}
extinction curve is inappropriate.} The 200 LBGs in the 
sample that meet this cut (including the uncertainty in the
cross-band calibration) are circled in Fig.~\ref{fig:ebv_vs_wlya}; 
they occupy a very similar distribution to the lower-S/N observations,
with a comparable correlation ($\rho=-0.21$) and significance
($p=3\times10^{-3}$). The LAE composite spectra (black squares)
appear to show a much stronger relationship with \ebv,  but we
have insufficient data to quantify this trend among the LAEs
alone. Notably, however, the LAE \wla\ measurements are clearly 
inconsistent with the distribution of LBG points at similar values of
\ebv: at fixed reddening, the LAE composites have significantly higher
values of \wla\ than the LBG points. It appears, therefore, that a
difference in dust attenuation is insufficient to explain either the variation of
\wla\ among the KBSS LBG sample or the differences between the LAE and
LBG populations.

Although it is expected that absorption
by dust is the primary mechanism for the destruction of \lya\ photons
in galaxies, there have been previous observational indications
that dust and \lya\ emission can co-exist. While no previous sample of
$z\approx2-3$ galaxies has had sufficient measurements of both rest-UV
and rest-optical emission line spectra to quantify the relationship
between \wla\ and nebular \ebv\ in detail, studies of the broadband
spectral energy distributions of LAEs have found galaxies exhibiting
both strong \lya\ emission and large inferred {\it stellar} \ebv\
\citep{kor10,hag14,matthee16}, including objects with
$\ebv\sub{stars}\gtrsim0.4$. The 14 ``extreme'' LBGs in
the \citet{erb16} sample have higher \wla\ and lower \ebv\sub{neb} than average KBSS
LBGs, but include individual objects with inferred reddening as
high as $\ebv\sub{neb} \approx 0.34$ in the sample of objects with
\ha/\hb\ S/N $>10$ (or as high as $\ebv\sub{neb} \approx 0.95$ with no
S/N cut). In the full sample of KBSS LBGs presented here, there are
galaxies with $\wla>20$\AA\ which exhibit reddening as high as
$\ebv\sub{neb} = 1.2$, even with the \citet{str16} cut on
S/N. Conversely, there is a substantial population of LBGs with low
values of \ebv\ and low or no  net \lya\ emission: the average LBG
with \ebv\sub{neb} consistent with our full LAE composite ($\ebv\sub{neb}\lesssim0.06$,
approximately the lowest quartile in the LBG \ebv\sub{neb} distribution) is
actually a net {\it absorber} of \lya\ photons in slit spectroscopy (median
$W\sub{\lya,spec}=-2.0$\AA)\footnote{Note, however, that the
 escape of scattered \lya\ photons at large galacto-centric radii can cause
 galaxies with net (spatially-integrated) \lya\ emission to show net
 absorption in slit spectroscopy \citep{ste11}.}.

While neither \ebv\sub{stars} nor \ebv\sub{neb} is a perfect proxy for the
attenuation of \lya\ photons by dust, \ebv\sub{neb} has the advantage
of tracing the attenuation of photons from the same star-forming
regions where \lya\ photons are expected to originate (rather than the
diffuse interstellar dust distribution traversed by photons from
spatially-extended populations of stars).\footnote{While \ebv\sub{neb}
  is seen to be greater than \ebv\sub{stars} in typical galaxies
  samples, \citet{price14} demonstrate that this discrepancy is
  minimized in high sSFR galaxies similar to those discussed here.} We
suggest, therefore, that {\it our 
observations are the strongest evidence yet that low dust content,
while associated with \lya\ escape, is neither necessary nor
sufficient for producing strong \lya\ emission in galaxy spectra.} The
trend in \wla\ with position on the N2-BPT diagram is therefore
unlikely to be a product of the association of gas-phase metallicity
with dust-to-gas ratio.

\subsubsection{\lya\ emission vs. nebular excitation}\label{subsubsec:lyavsexcitation}
 
\begin{figure}
\center
\includegraphics[width=\linewidth]{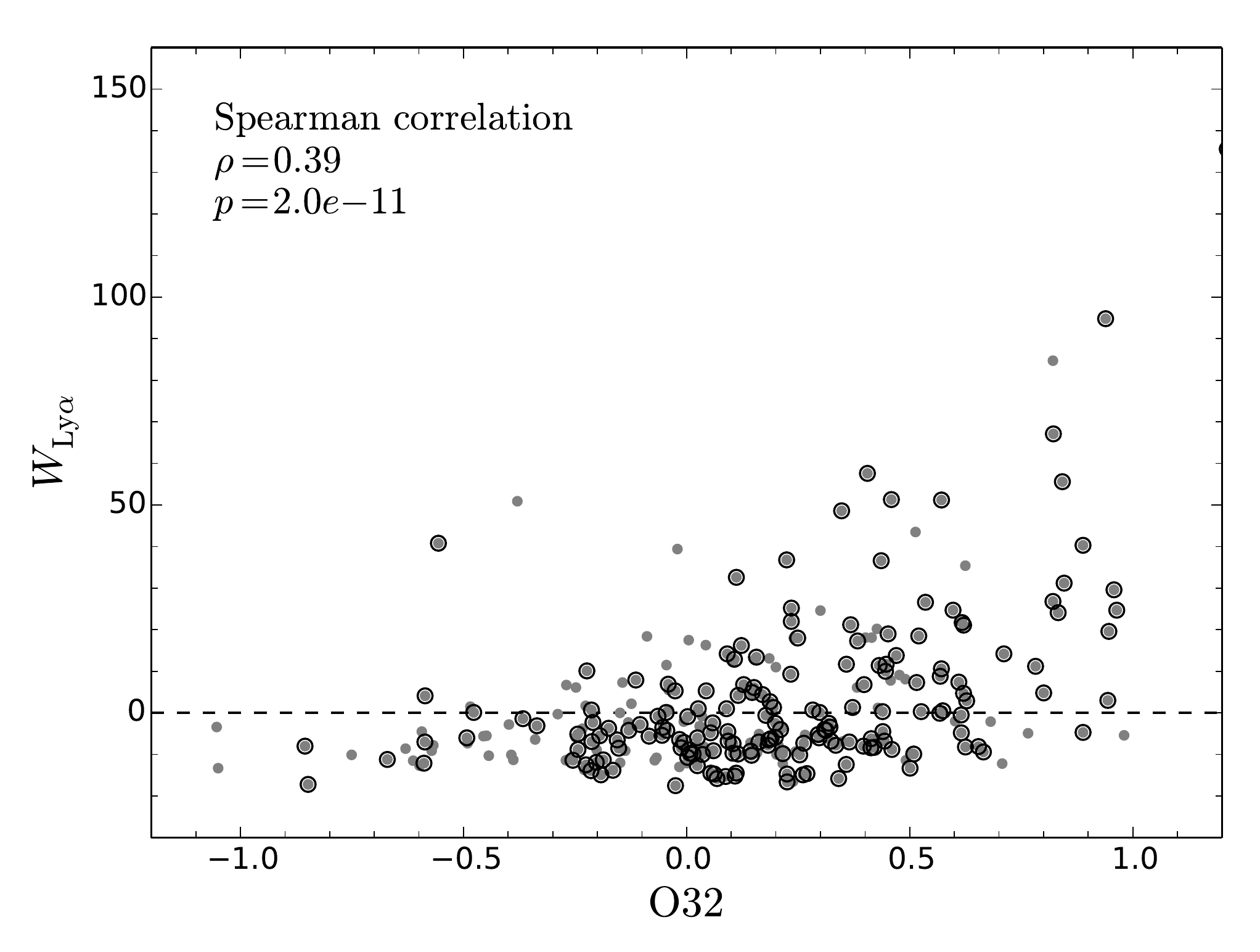}
\caption{The spectroscopic equivalent width of \lya\ (\wla) vs. the
  dust-corrected O32 ratio for individual KBSS LBGs (grey
  points). Circled points are those with more secure dust corrections,
  as in Fig.~\ref{fig:ebv_vs_wlya}. Despite substantial scatter, there
  is a stronger trend than is seen for \wla\ vs. \ebv.}
\label{fig:o32_vs_wlya}
\end{figure}

\begin{figure}
\center
\includegraphics[width=\linewidth]{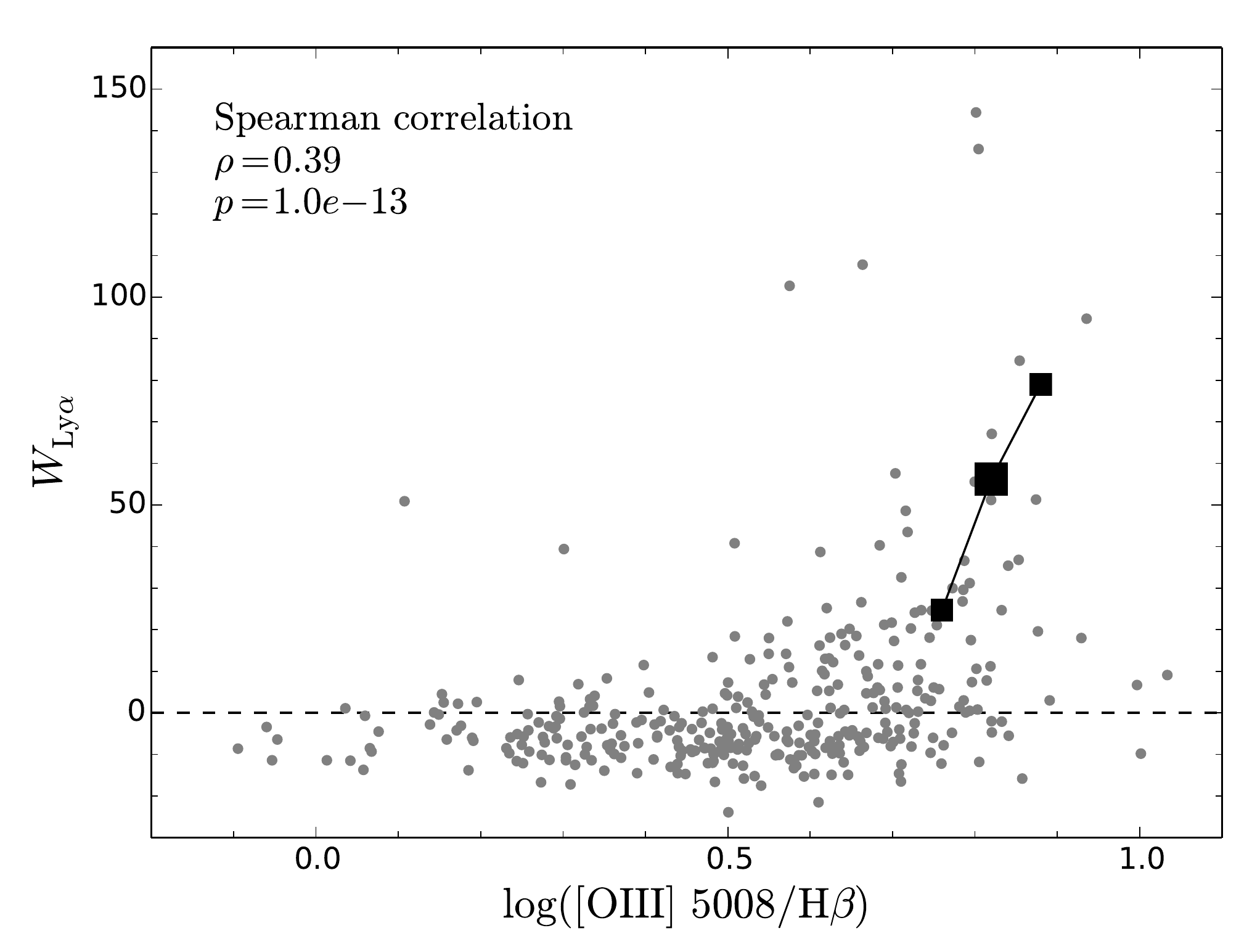}
\caption{The spectroscopic equivalent width of \lya\ ($W\sub{\lya}$)
  vs. the [\ion{O}{3}]/\hb\ ratio--a measure of nebular excitation--for individual KBSS LBGs (grey
  points) and LAE composites (large black square indicates the full
  sample, while smaller squares denote the low-\wla\ and high-\wla\
  subsamples), as in Fig.~\ref{fig:ebv_vs_wlya}. A strong trend is
  present, with the LAE stacks displaying similar excitation-matched
  \wla\ values to the LBGs, indicating that the observed \wla\ is
  strongly modulated by the physics governing \lya\ {\it production}
  within \ion{H}{2} regions.} 
\label{fig:o3hb_vs_wlya}
\end{figure}

 We now consider the second mode by which the nebular spectra of
galaxies may be linked to their \lya\ emission: the ionization and
recombination processes within their star-forming regions. There are
multiple reasons why the ionization and excitation states of gas in
\ion{H}{2} regions may be associated with \lya\ emission. Stronger
sources of ionizing photons (e.g., hotter populations of massive
stars) will both increase the typical ionization state of their
surrounding gas and result in a larger production rate of \lya\
photons (as well as other products of recombination
emission). Secondly, density-bounded \ion{H}{2} regions (those in
which the star-forming cloud becomes completely ionized) will 
be more transparent to escaping \lya\ photons than those that are
surrounded by thick shells of neutral gas. Similarly, such
density-bounded \ion{H}{2} regions may exhibit high
average ionization ratios (e.g., O32; Table ~\ref{table:ratios}), as
discussed in Sec. ~\ref{sec:discussion}

 We have not obtained measurements of the [\ion{O}{2}]
$\lambda\lambda$3727,3729 emission-line doublet (which lies in the
MOSFIRE $J$ band at $z\approx2.1-2.6$) for our LAE sample, but we can
investigate the trend between O32 and \wla\ in our comparison sample
of KBSS spectra. Fig.~\ref{fig:o32_vs_wlya} shows the O32 ratio for
the subset of the N2-BPT LBG sample that also have a $>$3$\sigma$
detection of [\ion{O}{2}] $\lambda\lambda$3727,3729 (275 objects; 82\% of the objects
included in the full N2-BPT sample). The O32 ratios have been
dust-corrected assuming the \ebv\sub{neb} measurements described above
and a \citet{car89} extinction curve. As in Fig.~\ref{fig:ebv_vs_wlya},
circled points are those meeting a 5$\sigma$ cut on the \ha/\hb\ dust
correction and MOSFIRE $J$-$H$ cross-calibration.

Substantial scatter is present among the values of \wla\ at fixed O32,
although a Spearman rank-correlation test shows a much stronger
trend ($\rho=0.39$; $p=2\times 10^{-11}$) than is seen in the
\wla-\ebv\ relationship. The objects with the most secure dust
corrections and cross-band flux calibrations (175 objects) show a still stronger correlation
($\rho=0.45$; $p=3\times 10^{-10}$). The median \wla\ measurement
among LBGs in the upper quartile of O32 (O32 $>0.46$) is $\wla=8.5$\AA, indicating
that these objects are net \lya\ emitters (unlike the
lowest quartile of LBGs in \ebv). Among the
KBSS-LBG sample, it therefore appears that O32 is a better predictor
of strong \lya\ emission than \ebv.

While we cannot directly measure O32 for the LAE samples presented
here, the O3 ratio is a related
measure of the nebular excitation properties of galaxies. Although O3
is sensitive to the gas-phase oxygen abundance at very low metallicities
($Z\lesssim0.2Z_\odot$, Sec.~\ref{sec:models}), it is much more sensitive to
the ionization parameter and hardness of the incident spectrum at more
intermediate sub-solar metallicities typical of the KBSS LBGs
($0.3Z_\odot\lesssim Z \lesssim 0.9Z_\odot$;
\citealt{ste14,ste16,str16}). Furthermore, the O32 and O3 ratios are closely
correlated; the Spearman rank correlation between both values is $\rho=0.74$
for the 175 KBSS LBGs with the highest-confidence dust-corrected O32
values. The O3 ratio also has the advantage of requiring no dust
correction or cross-band calibration, as both lines lie near each
other in the MOSFIRE $H$ band at $z\approx 2-2.6$.

The O3 ratio is therefore a useful discriminant for comparing the
excitation properties of the LBG and LAE samples and their variation
with \wla, as is shown in Fig.~\ref{fig:o3hb_vs_wlya}. A Spearman
test among the 336 LBGs in the N2-BPT sample yields $\rho=0.39$
and $p=10^{-13}$ for the O3-\wla\ correlation, approximately as strong as the
O32-\wla\ correlation. Furthermore, the KBSS galaxies with O3 $> 
0.82$, consistent with the full LAE composite measurement, are
typically strong \lya\ emitters (median $\wla=14.6$\AA). {\it In
general, the LAEs have quite similar values of \wla\ to
excitation-matched samples of KBSS LBGs, in contrast to the
distribution of attenuation-matched LBGs in
Fig.~\ref{fig:ebv_vs_wlya}}. Similarly, there are very few
\lya-emitting KBSS LBGs with low O3 values, including only two objects
with O3 $< 0.5$ and $\wla > 14$\AA.

The substantial scatter in \wla\ at fixed excitation is not surprising,
given the multiplicity of factors that govern \lya\
escape. Nevertheless, it appears that the net \lya\ emission that escapes
star-forming galaxies at small galactic radii (that is, the \lya\
emission to which slit spectroscopy is most sensitive) remains closely
coupled to the properties of their ionized birthplaces despite the
subsequent interactions of these photons with the surrounding interstellar and
circumgalactic media.

\section{Photoionization Model Comparison} \label{sec:models}


\subsection{Model parameters}\label{subsec:params}

The nebular spectra of
star-forming galaxies are sensitive to a broad range of physical
parameters, including the electron density $n_e$, ionization parameter
$U$, gas-phase metallicity and elemental abundance patterns; the
stellar metallicity and abundance patterns, ages, initial-mass function (IMF),
and evolutionary properties of the embedded stars; as well as the
foreground extinction. Many of these properties can produce degenerate
effects on galaxy spectra, particularly when only a few nebular
lines are observed. Given that our current measurements are limited to
the brightest lines in the $H$ and $K$ atmospheric windows, we use the
trends established among the brighter LBG samples in
Sec.~\ref{subsec:lyacorr} and the more detailed modeling presented by
\citet[hereafter S16]{ste16} and \citet{str16} to constrain the range of
physically-motivated model parameters.

With this in mind, we run a grid of Cloudy\footnote{We use Cloudy
  v13.02 for consistency with the modeling of KBSS LBGs by
  \citet{ste16} and \citet{str16}.} \citep{fer13} photoionization
models over a range of physical gas parameters and sources of incident
radiation consistent with these previous studies. We vary the
ionization parameter $U$ over the range $-3\le$ 
log$U$ $\le -1.5$, where log$U$ is varied in steps of $\Delta$log$U$ =
$0.1$. The gas-phase metallicity is varied over the range
$0.1\le Z\sub{neb}/Z_\odot\le1.0$ in steps of $\Delta$$Z=0.1$, as well
as $0.01\le Z\sub{neb}/Z_\odot\le 0.1$ in
steps of $\Delta$$Z=0.01$. The Cloudy models
assume a solar abundance pattern from \citet{asp09}, but we scale the
output nitrogen-based line ratios according to the relation between
N/O and O/H suggested by \citet{str16}: 

\begin{equation}\label{eq:no}
\textrm{log(N/O)} = -13.48+1.45 \times [12+\textrm{log(O/H)}] \,.
\end{equation}

\noindent However, this relationship is not intended to be
applied at the lowest metallicities,
where many surveys of local low-metallicity galaxies and \ion{H}{2}
regions have found that the N/O vs. O/H relation becomes flat, with
log(N/O) $\approx -1.5$ at low O/H. We therefore impose a minimum
value of log(N/O) $= -1.5$, which applies at all 
inferred oxygen abundances  less than 12 + log(O/H) = 8.26
($Z\sub{neb}=0.4Z$). 

We fix the electron density $n_e=300$ cm$^{-3}$ as in S16,
consistent with the measured electron densities from the KBSS and
MOSDEF surveys ($n_e\approx 200-360$ cm$^{-3}$;
\citealt{ste14,ste16,san16,str16}). While individual galaxy spectra
in these surveys exhibit a wide range of $n_e$, \citet{str16}
demonstrate that there is no trend between inferred $n_e$ and offset
from the low-redshift N2-BPT locus, and we likewise find no
correlation between inferred $n_e$ and stellar mass or
\lya\ equivalent width.

We perform our photoionization modeling using incident stellar
radiation fields from the latest version of two spectral-synthesis
codes: Starburst99 \citep{lei14} and BPASSv2 \citep{eld16,sta16}, the
latter of which includes explicit modelling of the effects of binary
interactions on stellar evolution.  Several models from each model
suite are described in detail in S16. In that paper, deep rest-UV 
and rest-optical spectra are modeled simultaneously to constrain the
properties of the stellar populations in KBSS LBGs, finding that the
composite spectra require $Z_*\ll0.008$ and are best fit by
$Z_*=0.001-0.002$ ($Z_*/Z_\odot=0.07-0.14$). As in S16, we explicitly
decouple the metallicity of our stellar models ($Z_*$) from the
gas-phase metallicity input to the photoionization code
($Z\sub{neb}$); the physical rationale for this choice is dicussed below.

We adopt the best-fit $Z_*=0.07Z_\odot$ BPASSv2 model from S16 as our fiducial stellar
population, which is denoted as BPASSv2-z001-300bin in that paper. The
model assumes a \citet{sal55} IMF (power law index of $-2.35$) and a
stellar mass range of $0.5\le M_*\le 300$. For 
comparison, we also consider the Starburst99 model denoted as
S99-v00-z001 in S16, which has the
default \citet{kro01} IMF\footnote{Note that the low-mass behavior of the IMF
causes significant differences in the total stellar masses and
star-formation rates inferred for \citet{sal55} and \citet{kro01}
IMFs, but has a negligible effect on the predicted UV spectrum of a
stellar population.} (power law index of $-2.30$) and
a stellar mass range of $0.5\le M_*\le 100$. We also consider models
with $Z_*=0.002$ ($Z_*=0.14 Z_\odot$)\footnote{These models are
  denoted BPASSv2-z002-300bin and S99-v00-z002 in S16.} that are
otherwise identical to the $Z_*=0.001$ models. All our stellar models
assume a continuous star-formation history of 100 Myr. As
noted in S16, these models do not change appreciably in the far-UV
after a few $\times$10$^7$ years, which is approximately the central
dynamical time of the KBSS-\lya\ LAEs and thus the shortest likely
timescale for galaxy-scale bursts of star formation in these systems.

As shown in
S16, small differences in the IMF power-law index do not significantly
affect the output nebular line ratios: increasing (flattening) the
slope of the Starburst99 IMF from $-2.3$ to $-2.0$ or even $-1.7$ does
not reproduce the high O3 and R23 ratios observed in the LBG
composite spectrum by S16. Changing the upper-mass cutoff of the IMF
(i.e., from 100 \msun\ to 300 \msun) does harden the EUV spectrum,
although to a lesser extent than the inclusion of massive
stellar binaries. Even the Starburst99 models from
\citet{lei14} that include rapid stellar rotation do not produce
the significant changes to the EUV spectral shape that are required to
explain the nebular line ratios of the KBSS LBGs.  In addition, models
which exclude binary interactions (e.g., both the Starburst99 models
and the BPASS non-binary models) do not generate \ion{He}{2}
$\lambda$1640 emission (which is seen in the S16
rest-UV LBG composite) in continuous star-formation over long
($>$10Myr) timescales. In fact, S16 find
that the BPASSv2-z001-300bin model is the only spectrum considered
that produces the full suite of observed rest-UV and rest-optical line
measurements for the LBG stack. Nonetheless, in order to investigate
the dependence of the inferred nebular LAE properties on the assumed
stellar population, we consider both the Starburst99 and BPASSv2
models in our analysis. 

\subsection{Modeling the composite spectra}\label{subsec:bptmodel}

\begin{figure*}[p!]
\centerline{
\includegraphics[width=0.5\linewidth]{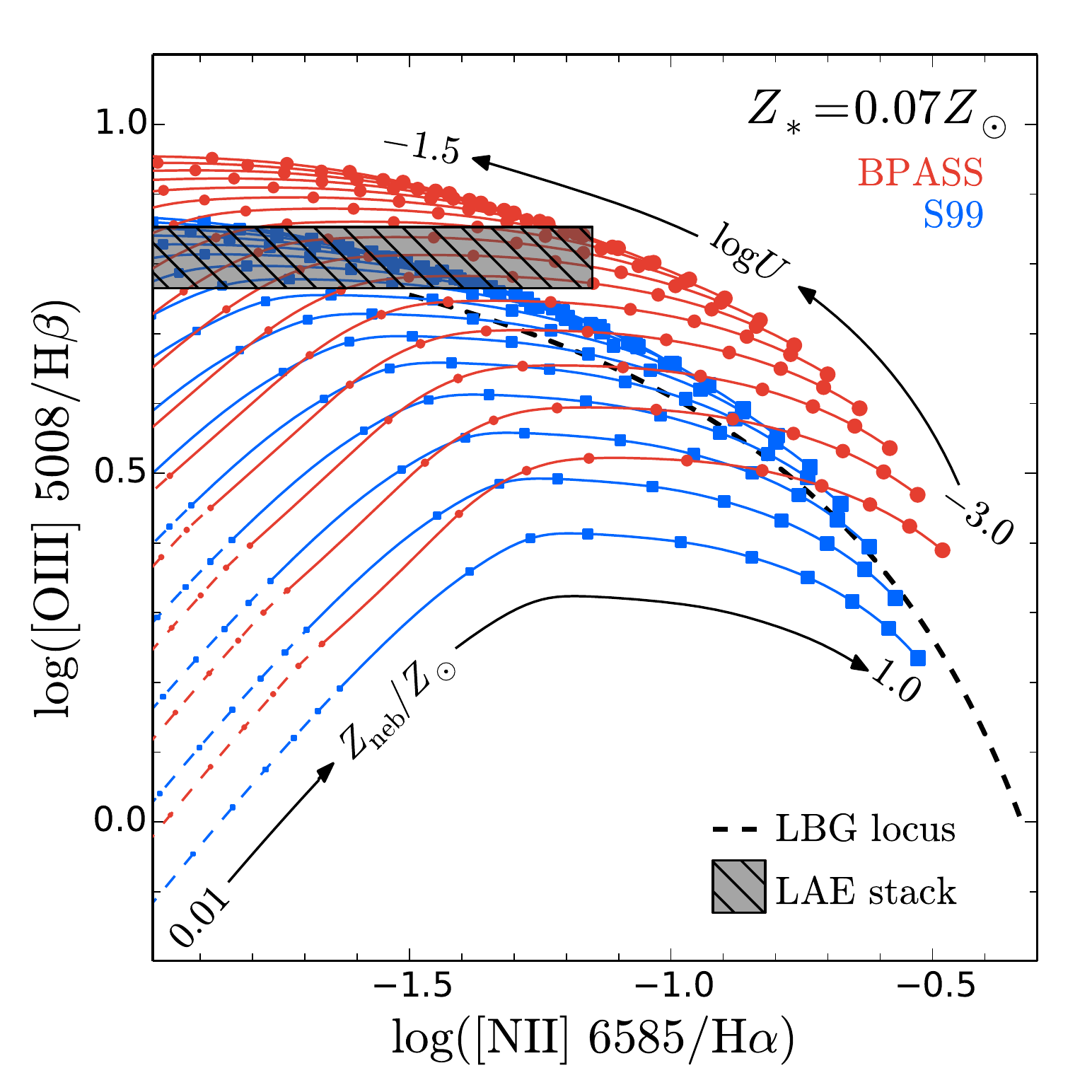}\includegraphics[width=0.5\linewidth]{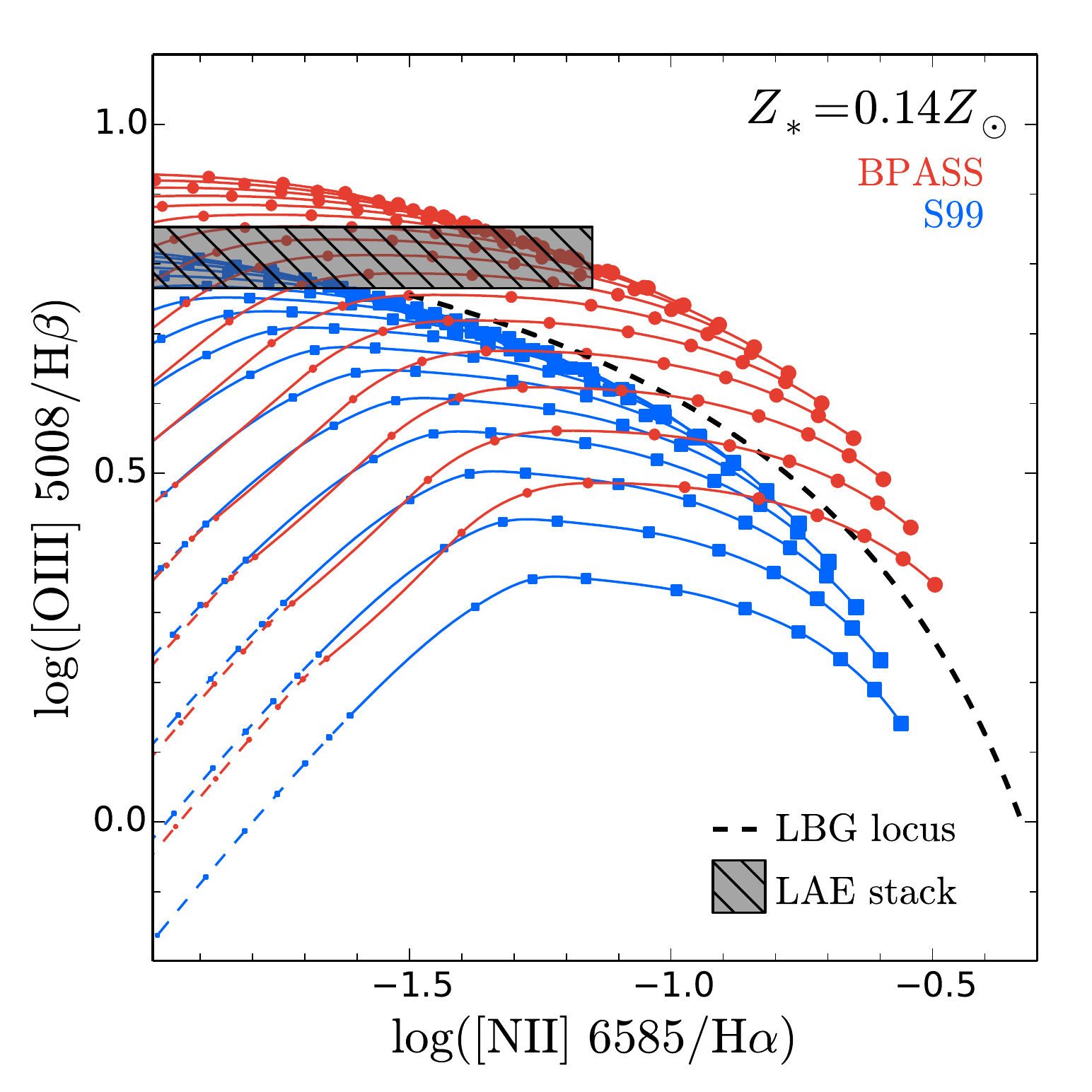}}
\centerline{
\includegraphics[width=0.5\linewidth]{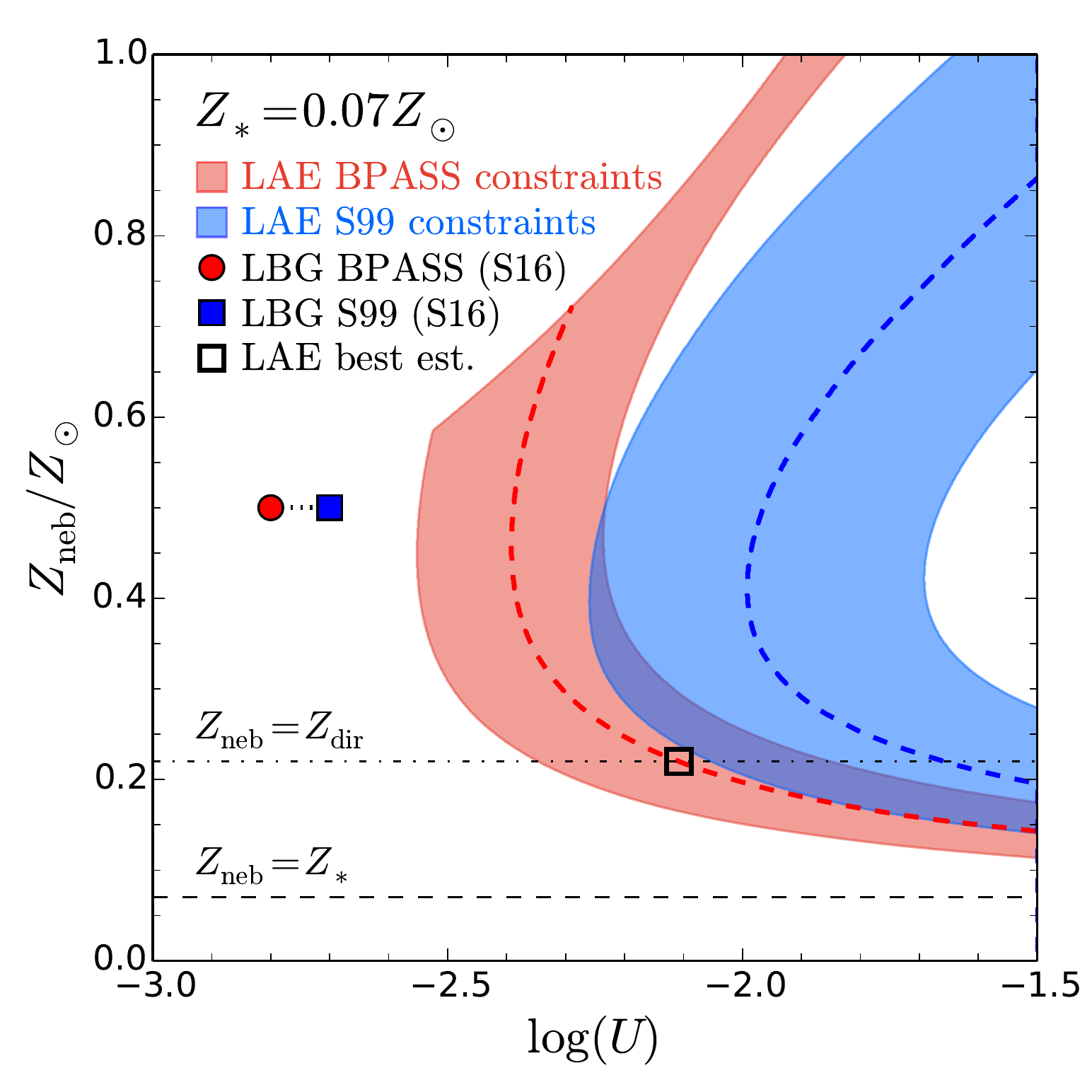}\includegraphics[width=0.5\linewidth]{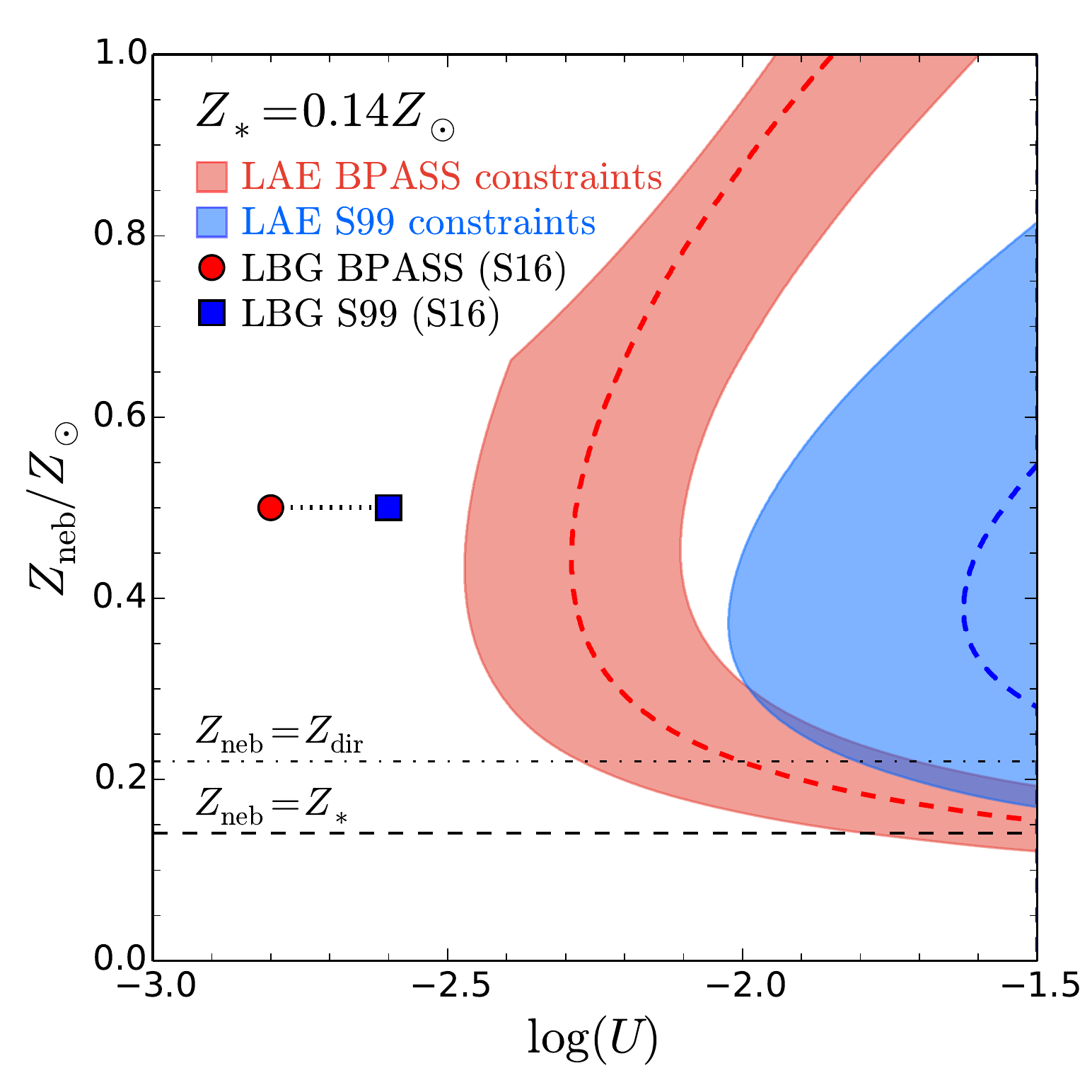}}
\caption{({\it Top}) N2-BPT predictions for the Cloudy photoionization models
  given an input spectrum from BPASS (red) or S99 (blue) stellar populations. The left panel
  uses a $Z_*=0.07 Z_\odot$ ($Z_*=0.001$) input model, while the right panel
  has $Z_*=0.14 Z_\odot$ ($Z_*=0.002$). Grey hatched region
  denotes the measurement of O3 and 2$\sigma$ limit on N2 from the
  full LAE composite spectra
  (Figs.~\ref{fig:Hbandwide}~\&~\ref{fig:Kband}). Photoionization
  models are plotted as a 
  function of $Z\sub{neb}$ for each value of $U$, with $-3\le$
  log$U$ $\le -1.5$ (with $U$ increasing from the bottom to the top of
  the panel) and $0.01\le Z\sub{neb}/Z_\odot\le1.0$. Marker size
  increases with increasing metallicity, and $Z\sub{neb}<0.1$ is shown
  with a dashed line. ({\it Bottom}) Constraints on the
  gas-phase metallicity $Z\sub{neb}$ and ionization parameter $U$ from the LAE
  composite spectra and the Cloudy models presented in the top
  panels. Red regions correspond to the parameters consistent with the
  LAE N2-BPT limits assuming a BPASS stellar population (with stellar
  metallicity as above), while the blue regions correspond to
  parameter constraints assuming a S99 stellar population. Dashed
  curves correspond to parameters that reproduce the best-fit
  measurement of O3 from the LAE composite spectrum. The horizontal
  dot-dash line corresponds to the gas-phase metallicity estimated via the
  auroral [\ion{O}{3}] transition (Sec.~\ref{subsec:4364}), and the
  black box represents our best estimate of the LAE nebular and stellar
  parameters: ($Z_*$, $Z\sub{neb}$, log$U$) = (0.07, 0.022, -2.1). The dashed
  line represents the gas-phase metallicity equal to the input stellar
  metallicity in each panel: it is evident that the two metallicities are difficult to
  reconcile assuming solar abundance patterns. Red circles (blue squares)
  represent the best-fit values of $Z\sub{neb}$ inferred for the LBG
  composite spectra by \citet{ste16} using the same stellar BPASS
  (S99) model spectra, although the S99 models fail to reproduce other
  nebular ratios discussed in that paper. The LAE spectra require
  significantly enhanced ionization parameters with respect to the LBG
  samples.}
\label{fig:bpt-cloudy}
\end{figure*}

The Cloudy predictions for both the S99-v00-z*** models (hereafter
``the S99 models'') and the BPASSv2-z***-300bin models (hereafter
``the BPASS models'') in the N2-BPT plane are given in the top panels
of Fig.~\ref{fig:bpt-cloudy} for a range of gas-phase metallicities and
ionization parameters. Values of $Z\sub{neb}$ between grid points are
interpolated using a cubic spline. For both stellar metallicities considered, the
BPASS model produces higher average excitation (O3) values at fixed
$Z\sub{neb}$ and $U$. Note that O3 increases with $Z\sub{neb}$ at low
$Z\sub{neb}$ for all models up to a maximum at $Z\sub{neb}\approx0.4$
and then decreases as $Z\sub{neb}$ is increased further. N2, however,
increases monotonically with increasing $Z\sub{neb}$.

The constraints on the typical LAE excitation are plotted as the shaded
box in each panel (as above, we use the 68\% confidence interval in O3
and the 2$\sigma$ upper limit on N2). For the S99 models, only the
highest ionization parameters are able to reproduce the best-fit LAE
line ratios (particularly for $Z_*=0.14Z_\odot$ models), while the
BPASS stellar models are relatively insensitive to $Z_*$ and are able
to reproduce the LAE measurements over a range of values of
$Z_*$, $Z\sub{neb}$, and $U$.

The bottom panels of Fig.~\ref{fig:bpt-cloudy} show these constraints
explicitly. In each panel, the red shaded region corresponds to the
values of ($U$, $Z\sub{neb}$) that produce N2 and O3 ratios
consistent with the LAE composite measurements using the BPASS stellar
models. Similarly, the blue region corresponds to the range of
allowed parameters assuming the S99 input stellar spectra. The blue
and red dashed curves correspond to the values ($U$, $Z\sub{neb}$)
that produce the best-fit value of O3 = $0.82$
(Table~\ref{table:subsamp}) and are consistent with the upper limit
on N2. As in the top panels, values are interpolated in between grid
points using a cubic spline. Formally, values of $Z\sub{neb}\gtrsim
Z_\odot$ are allowed for both of the $Z_*=0.07Z_\odot$ stellar models,
although the BPASS model cannot reproduce the best-fit observed O3
ratio and N2 limit with $Z\sub{neb}>0.7$. Stronger constraints on the
N2 ratio are needed to cross-check the typical gas-phase metallicity
($Z\sub{neb}\approx0.2Z_\odot$) inferred from the [\ion{O}{3}]
$\lambda$4364 measurement in Sec.~\ref{subsec:4364}. However, we note
that $Z\sub{neb}$ less than $\sim$0.2 would require extremely high
values of the ionization parameter (log$U\gtrsim-2.0$).

Given that the galaxies in the LBG sample have luminosities and masses
$\sim$10$\times$ larger than the typical LAEs, we find it likely
that the LAEs have average stellar metallicities at least as low as the
metallicity favored for the LBGs in S16 ($Z_*=0.07Z_\odot$). Under this assumption, our
LAE observations strongly favor values of $Z\sub{neb}$ significantly
higher than the modeled stellar metallicities, similar to the effect
seen in the S16 and \citet{str16} LBG samples. As described in those works, this
apparent discrepancy is likely a manifestation of the same non-solar abundance
ratios in both the stars and gas. The shape of the model stellar spectrum
is most strongly dependent on those elements that dominate the opacity
of the stellar photosphere (i.e., iron), which are produced by Type Ia
SNe at late times. Conversely, the nebular metallicity constraints are
most sensitive to the species which dominate the cooling of the
$\sim$$10^{4}$ K gas, dominated by oxygen that is released in the
comparatively prompt Type II SNe. S16
and \citet{str16} therefore argue that the apparent $Z\sub{neb}/Z_*$
ratio should be interpreted as a non-solar O/Fe abundance ratio, and
these discrepancies signify the $\alpha$-enhancement of young galaxies
and star-forming regions at high redshift. Our LAE measurements are
consistent with this interpretation.

The ionization parameter of the gas has even stronger constraints from
our LAE compsite spectra. Using the 
BPASS models, the best-fit O3 ratio can only be 
reproduced with log$U > -2.4$ (log$U > -2.3$) for $Z_*=0.07Z_\odot$
($Z_*=0.14Z_\odot$). The dependence of the minimum value of $U$ on stellar
metallicity reflects the fact that the same nebular excitation can be
achieved with a lower value of $U$ (that is, fewer photons per
hydrogen atom) for a harder incident spectrum. This effect is seen
more strongly for the S99 models, which have significantly softer EUV
spectra: the $Z_*=0.07Z_\odot$ S99 model requires log$U > -2.0$ to reproduce
the best-fit O3 value regardless of $Z\sub{neb}$, and the
$Z_*=0.14Z_\odot$ S99 model requires log$U > -1.65$. If
$Z\sub{neb}=0.22Z_\odot$ is assumed, log$U \gg-1.7$ is required.

\begin{figure*}
\centerline{
\includegraphics[width=0.33\linewidth]{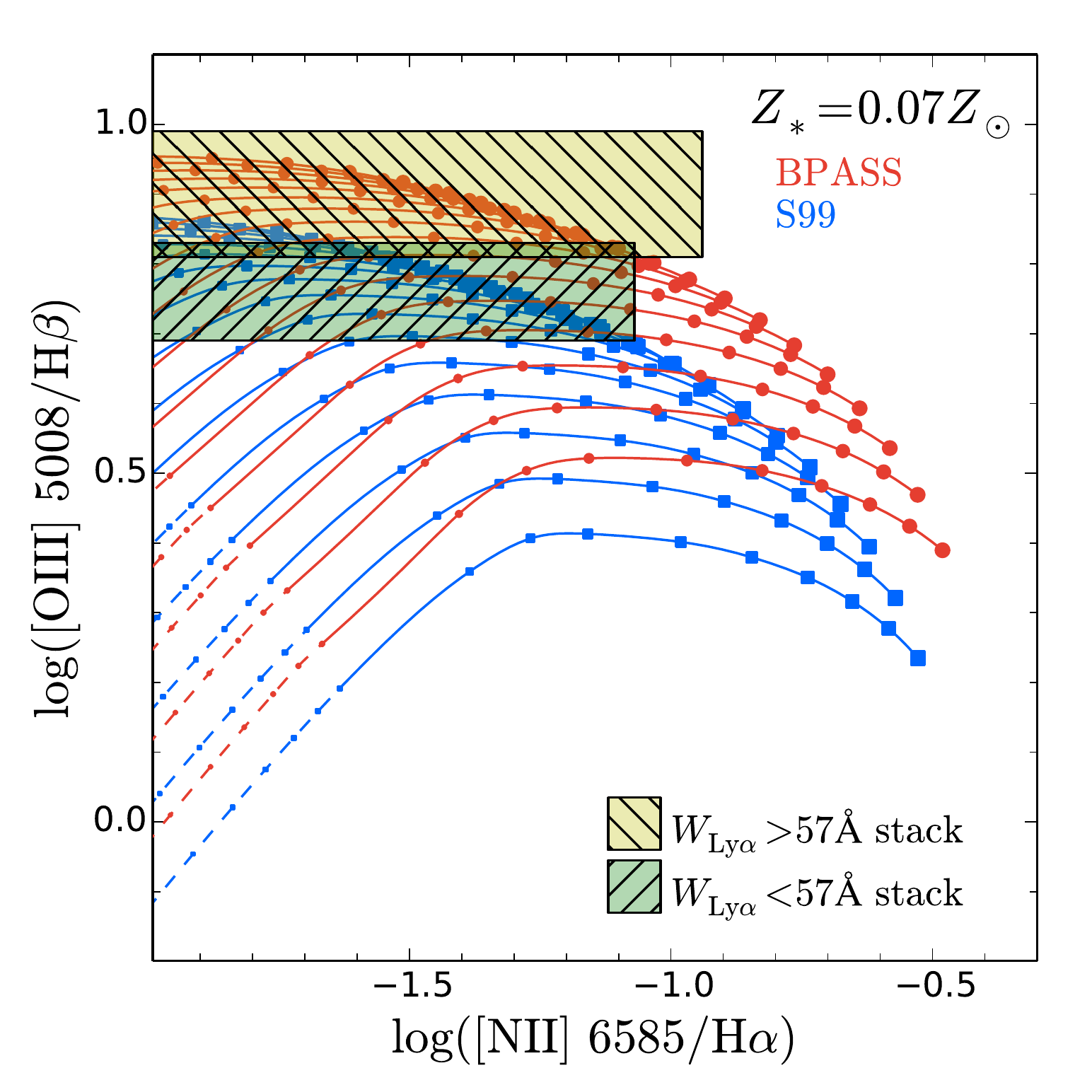}\includegraphics[width=0.33\linewidth]{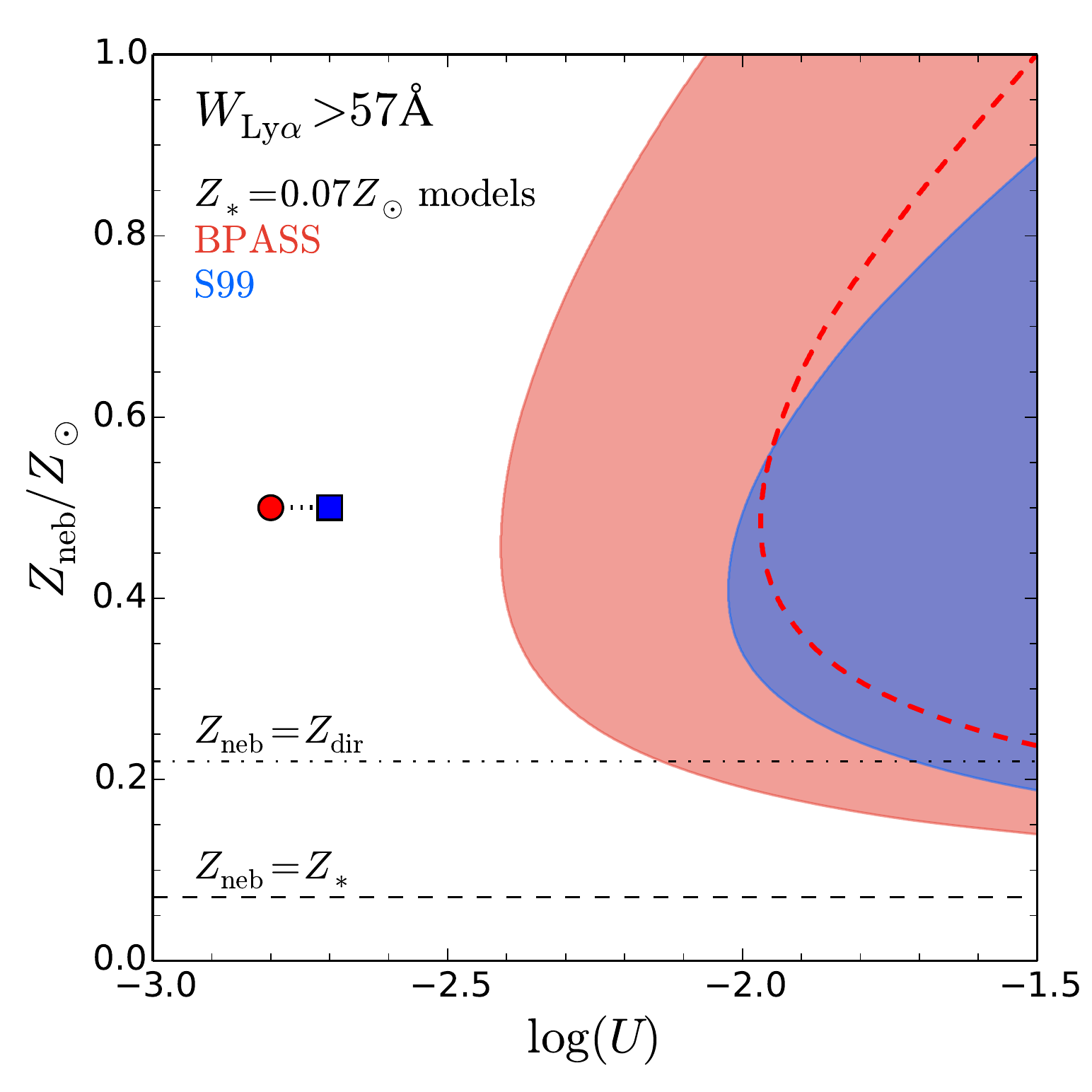}\includegraphics[width=0.33\linewidth]{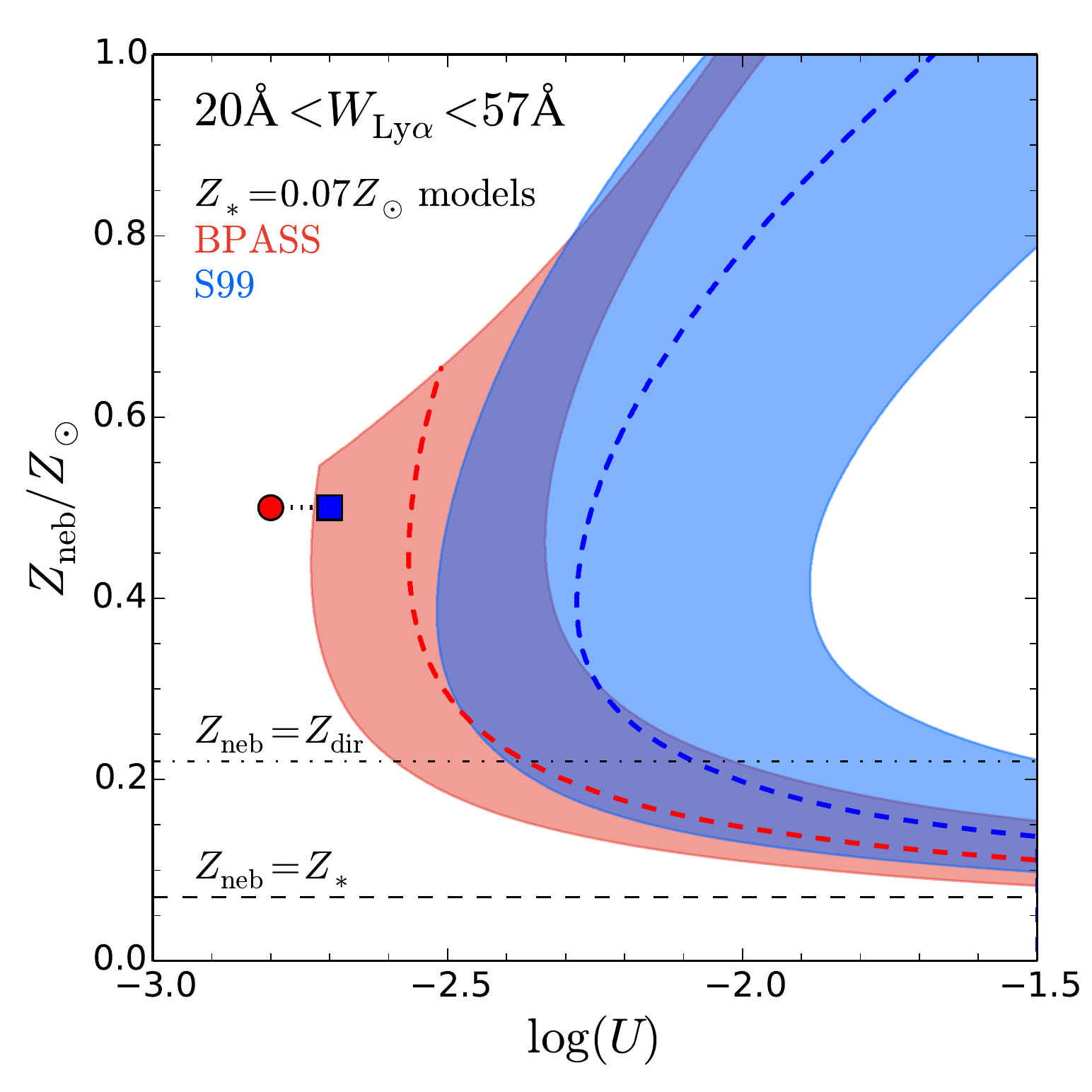}}
\caption{({\it Left}) N2-BPT predictions for the Cloudy
  photoionization models as in the top row of
  Fig.~\ref{fig:bpt-cloudy}. Green (yellow) shaded region denotes the
  constraints from the low-\wla\ (high-\wla) composite LAE
  spectra. The input spectra have $Z_*/Z_\odot=0.07$
  as in the left panels of Fig.~\ref{fig:bpt-cloudy} ({\it Center}) Constraints on 
  $Z\sub{neb}$ and $U$ for the high-\wla\ sample corresponding
  to the green region in the left panel. Annotation matches the bottom panels of
  Fig.~\ref{fig:bpt-cloudy}. ({\it Right}) Constraints on 
  $Z\sub{neb}$ and $U$ for the low-\wla\ sample (yellow region
  in the left panel).}
\label{fig:bpt-cloudy-wlya}
\end{figure*}

S16 infer tight constraints on both $U$ and $Z\sub{neb}$ from their
LBG composite spectra, and these values are also plotted on the bottom
panels of Fig.~\ref{fig:bpt-cloudy}. The best-fit values ($U$,
$Z\sub{neb}$) are shown as calculated using the same BPASS and S99
model spectra used for the LAE constraints in each plot. As described
above, the S99 models provide a poor match for full set of line ratios
in the S16 spectrum, but we note that the required ionization
parameter for the LAE composite measured here is significantly
higher than the best-fit value for the S16 LBG composite even if we
remain agnostic as to the choice of stellar population model.

In Fig.~\ref{fig:bpt-cloudy-wlya}, we display similar constraints for
the high-\wla\ and low-\wla\ LAE samples. Only the
$Z_*=0.07Z_\odot$ models are displayed, since the BPASS predictions change
only slightly with $Z_*$ and the S99 models with $Z_*=0.14Z_\odot$ struggle to
reproduce the measurements from the full composite spectrum. The
models in the left panel of Fig.~\ref{fig:bpt-cloudy-wlya} are the
same as those presented in the first panel of
Fig.~\ref{fig:bpt-cloudy}, but the hatched regions correspond to the
constraints on N2 and O3 from the high-\wla\ and low-\wla\ LAE
composite spectra (Table~\ref{table:subsamp}). 
The low-\wla\ LAE subsample (green region in the left panel of
Fig.~\ref{fig:bpt-cloudy-wlya}) has lower average excitation,  such
that it can be reproduced by both BPASS and S99 stellar models
with intermediate values of $U$. The constraints on $Z\sub{neb}$ and $U$
for this sample are displayed in the center panel of
Fig.~\ref{fig:bpt-cloudy-wlya}. The allowed range of
ionization parameters shifts closer to the LBG values from S16 (the
red circle and blue square in the same panel) compared to the values
estimated from the full LAE stack.

\begin{figure}[hbt]
\center
\includegraphics[width=\linewidth]{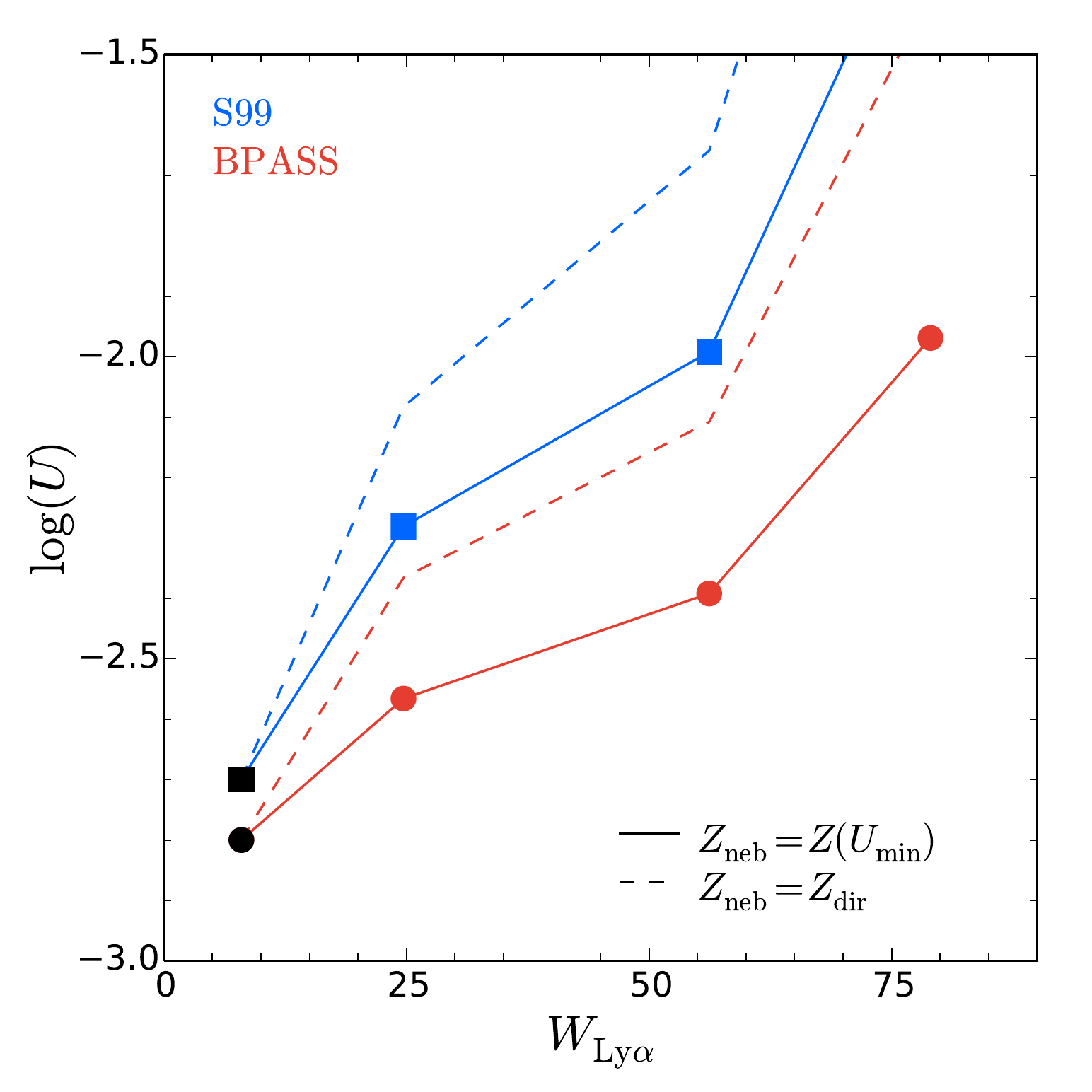}
\caption{The ionization parameter $U$ inferred from the
  composite spectrum of a given galaxy sample vs. the stacked
  spectroscopic equivalent width \wla\sub{,spec} of that sample. Black
  points refer to the LBG measurements of \citet{ste16}, while colored
  points correspond to the full, low-\wla, and high-\wla\ LAE samples
  presented in this paper. Circles and solid red lines indicate the minimum
  value of $U$ (for any metallicity $0.01\le
  Z\sub{neb}/Z_\odot\le1.0$) consistent with N2 limits and the
  best-fit measurement of O3 as predicted by the BPASS stellar
  model. Squares and solid blue lines are the analogous curve for the S99
  models, and a linear extrapolation is assumed for log$U>-1.5$. Dashed lines reflect the assumption of
  $Z\sub{neb}=0.22Z_\odot$ for each LAE sample. In general, the differences between the S99 and
  BPASS models become more apparent as \wla\ increases.}
\label{fig:minlogU_vs_wlya}
\end{figure}

Similarly, the allowed range of $U$ shifts even {\it farther} from the S16
measurements for the high-\wla\ sample. Only the most
extreme S99 models are able to reproduce O3 line ratios consistent
with the high-\wla\ constraints (yellow region in the left panel), and
none of the S99 models reproduce the best-fit value for that
sample. Even the BPASS models require extremely high ionization
parameters (log$U\approx-2.0$) to reproduce the best-fit value of O3
at $\wla>57$\AA, although values as low as log$U\approx -2.5$ are
allowed by the constraining region (which accounts for sample variance
in the composite spectrum). Together with the low-\wla\ and full 
sample of LAEs, the constraints indicate that the characteristic
nebular ionization parameter for a population of galaxies grows with
the typical \lya\ equivalent width among that population.

This trend can be  seen more clearly in
Fig.~\ref{fig:minlogU_vs_wlya}. Here, we show the minimum value of the
ionization parameter $U$ consistent with the best-fit O3 ratio for
each of the LAE samples (marginalizing over the gas-phase
metallicity) as a function of the median \lya\ equivalent width \wla\
of that sample. For comparison, we also include the best-fit value of
$U$ from S16 with the measured value of \wla\ from the composite UV
LBG spectrum. The
minimum $U$ inferred from both the BPASS and S99 $Z_*=0.07Z_\odot$ models
are shown as solid curves. We also include the variation in $U$
assuming the nebular metallicity inferred from the REL-corrected
direct method (Sec.~\ref{subsec:4364}) as dashed curves (although the true
metallicity is likely to vary among the \wla\ subsamples). As in
Figs.~\ref{fig:bpt-cloudy}~\&~\ref{fig:bpt-cloudy-wlya}, the increase in
$U$ with \wla\ is clearly visible, as is the fact that the softer S99
models require higher values of $U$ to reproduce the observational
constraints. In fact, the deviation between the two stellar population
models grows with \wla, indicating that \lya-emitting populations may
be particularly advantageous galaxy samples for discriminating among
these stellar models. If the high-\wla\ LAEs have lower oxygen abundances than
the intermediate and low-\wla\ samples, as might be expected from the
BPT-\wla\ relation seen in the LBG sample (Fig.~\ref{fig:BPTlya}), the
slope of the $U$-\wla\ relation may be even steeper than 
displayed in Fig.~\ref{fig:minlogU_vs_wlya}, which has necessarily
marginalized over this variation. Conversely, harder stellar spectra
resulting from lower iron abundances among high-\wla\ LAEs could
reduce the necessary value of $U$. Future observations that probe the ionization state of 
the gas with less dependence on metallicity (i.e., via O32 or
[\ion{Ne}{3}]/[\ion{O}{2}]) and further cross-checks of the gas-phase
oxygen abundance (e.g., from N2) will therefore provide more powerful
tests of the physical properties of stars and their surrounding
gaseous regions. 

\subsection{Low metallicity objects}\label{subsec:lowz}

\begin{figure}
\center
\includegraphics[width=\linewidth]{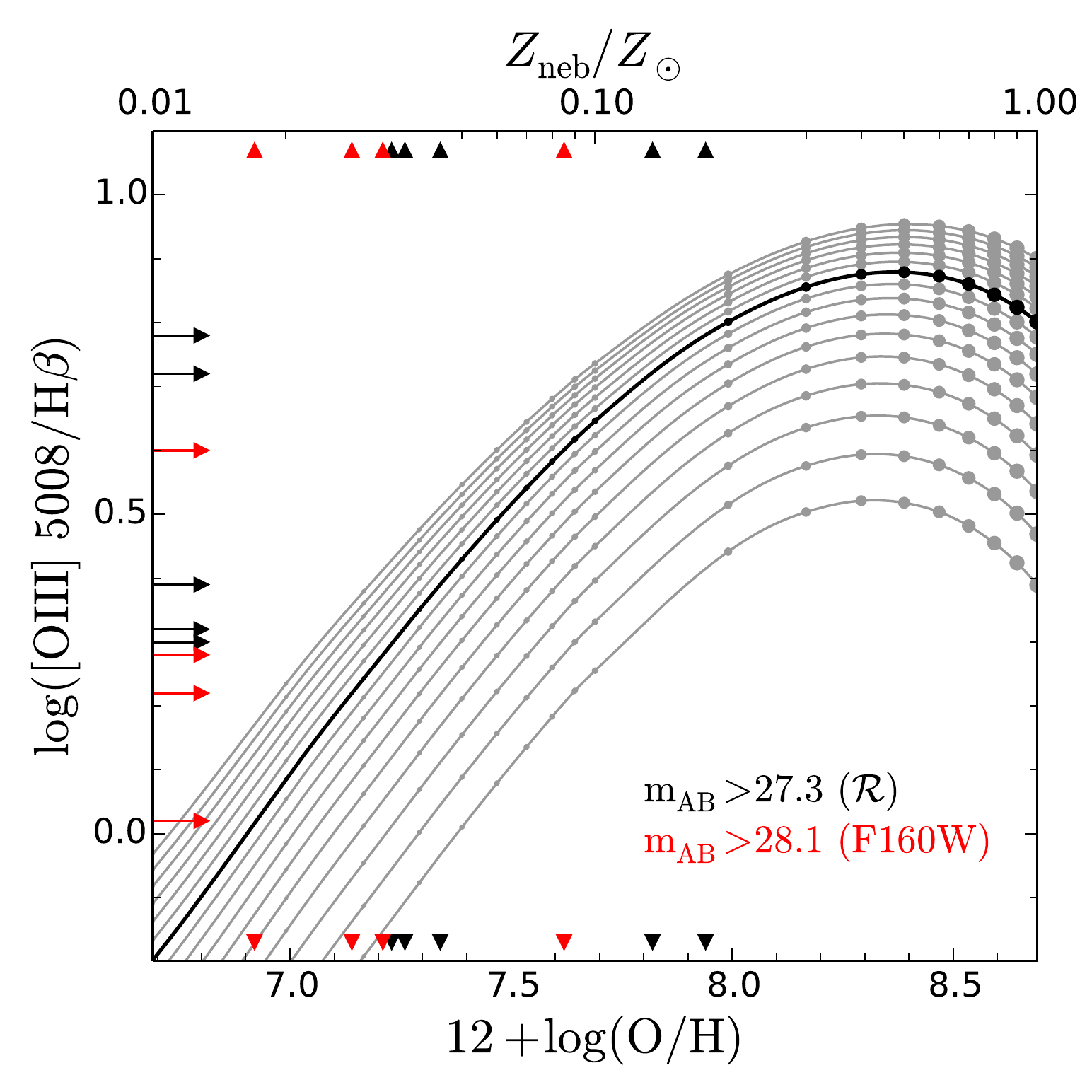}
\caption{Inferred oxygen abundances for the faintest LAEs in our
  sample. Red and black arrows on the left side are the
  measured O3 ratios for the continuum-undetected LAEs in
  Fig.~\ref{fig:o3_rmag}. These values are converted to
  abundances (shown as triangles on the top and bottom axes)
  assuming the log$U=-2.1$ BPASS photoionization model (black
  curve). Grey curves show the full range of modeled ionization
  parameters; choosing a higher (lower) log$U$ curve will yield a
  lower (higher) inferred oxygen abundance at fixed O3.}
\label{fig:lowZ}
\end{figure}

As shown in Figs.~\ref{fig:bpt-cloudy}~\&~\ref{fig:bpt-cloudy-wlya}, the relationship between $Z\sub{neb}$
and O3 is double-valued, and its shape at $0.3\lesssim Z\sub{neb}\lesssim0.9$
makes it very difficult to constrain the gas-phase oxygen abundance
from O3 measurements alone. However, at low values of $Z\sub{neb}$,
the relationship becomes much steeper, and the abundances of objects
that can be assumed to occupy the low-$Z\sub{neb}$ or
high-$Z\sub{neb}$ branches may be constrained much more easily.

Fig.~\ref{fig:o3_rmag} shows that the O3 ratios of our individual
LAEs exhibit a marked downturn at the lowest continuum
luminosities. Based on the photoionization models discussed above,
these lower values of O3 could be explained by the faintest LAEs having 1)
significantly higher stellar metallicities, 2) lower ionization parameters, and/or
3) lower gas-phase metallicities (i.e., oxygen abundance) compared to
the typical LAEs in our sample. 

Given the fact that the LAEs exhibit {\it higher} ionization
parameters than the continuum-bright LBGs and seem to require similar
(or lower)
stellar metallicities, we suggest that the observed downturn is caused
by low oxygen abundances among the faintest LAEs. This interpretation
is supported by other arguments as well: \lya-emitting galaxies
that are faint at $\lambda\sub{rest}\approx 1900-4500$\AA\ are likely
to have low stellar masses and high specific star-formation rates,
both qualities that correlate with low oxygen abundances in other
galaxy samples. Finally, the shape of the $Z\sub{neb}$$-$O3 relation would
predict a sharp downturn, as O3 is fairly flat at
$Z\sub{neb}\gtrsim0.3$ but decreases quickly for
$Z\sub{neb}\lesssim0.2$. As the typical inferred gas-phase metallicity of our
LAE sample is $Z\sub{neb}\approx0.22Z_\odot$, we would expect LAEs with
below-average oxygen abundances to exhibit significantly lower values
of O3.

We therefore interpret the O3 values of the faintest LAEs according to
the following assumptions: they have ionization parameters at least as
high as that inferred for the typical LAEs in our sample
(log$U\approx-2.1$) and have stellar populations similar to the BPASS
$Z_*=0.07Z_\odot$ models that match the LAE and LBG composite spectra
discussed above. We also assume that each of these LAEs occupies the
lower-metallicity track ($Z\sub{neb}<0.4Z_\odot$) where the
$Z\sub{neb}$$-$O3 relation is double-valued (this assumption is
discussed further below).

Fig.~\ref{fig:lowZ} shows the oxygen abundances that are inferred for
our 9 continuum-undetected LAEs with measured O3 ratios based on these assumptions. The measured
O3 ratios are shown as arrows on the left axis, while the inferred
oxygen abundances (nebular metallicities) are shown on the bottom
(top) axis. As in Fig.~\ref{fig:o3_rmag}, black points are LAEs
undetected in our ground-based continuum images ($\mathcal{R}>27.3$),
while the red points are also undetected in deep {\it HST} images
($m\sub{AB, F160W}>28.1$). The $Z\sub{neb}$$-$O3 curve with
log$U=-2.1$ is shown in black (with points as in
Figs.~\ref{fig:bpt-cloudy}~\&~\ref{fig:bpt-cloudy-wlya}), while the
analogous curves for other ionization parameters are shown in
grey.

The inferred oxygen abundances for the continuum-undetected LAEs are
in the range 12 + log(O/H) = $6.9-8.0$, with six LAEs below 12 +
log(O/H) = 7.4 ($Z\sub{neb}<0.05Z_\odot$).  Increasing the assumed
ionization parameter would require even lower oxygen abundances to
produce the observed values of O3. While some individual objects may
have lower ionization parameters than that assumed, we suggest that it
is unlikely that the {\it average} ionization parameter is
significantly lower among these LAEs than the typical LAEs in our
sample for the reasons discussed above. We have assumed each
of our LAEs occupies the low-metallicity track of the
$Z\sub{neb}$$-$O3 curve; for all but two of the continuum-undetected
LAEs, the high-metallicity track would predict significantly
super-solar average metallicities, which would be extremely suprising
for galaxies that otherwise appear to be young, low-mass galaxies.

\section{Discussion}\label{sec:discussion}

The above results indicate that our population of faint LAEs exhibit 
distinct physical properties with respect to the population of
brighter, more massive, and more rapidly star-forming LBGs. In this
section we compare these properties to those of other extreme,
low-mass galaxy samples, arguing that galaxies selected via emission
lines (including \lya) are representative of the low-mass,
low-luminosity population. We also discuss the mechanisms by which
the nebular properties may be linked to galaxy mass and \lya\
production and escape. 

\subsection{Comparison to other extreme high-$z$ galaxy samples}\label{subsec:compare}

As discussed in Sec.~\ref{sec:intro}, some recent samples of
galaxies selected by their extreme nebular emission lines show
similar stellar masses and other properties to the KBSS-\lya\ LAEs described
here.  The three ``extreme'' LBGs with $T_e$ measurements presented by
\citet{ste14} have $M_*=10^{9.4-9.7}$, SFR = 30$-$60
\msun\ yr$^{-1}$, and O3 = 0.79$-$0.9. \citet{mas13,mas14} present
samples of 14 and 22 EELGs with median dynamical masses $\langle 
M\sub{dyn}\rangle\approx10^{9.1}$\msun. This mass is ostensibly
similar to that estimated for the KBSS-\lya\ LAEs in T15 ($\langle
M\sub{dyn}\rangle\approx10^{8.9}$\msun), but \citeauthor{mas13} assume
a geometric factor $C\equiv r\sub{eff}\sigma^2/G M\sub{dyn}=3$, while
T15 assumes $C=5$. A consistent choice of estimator would suggest that
the \citet{mas13,mas14} EELGs have dynamical masses $\sim$0.5 dex
higher than the LAEs described here. Similarly, the median \citet{mas14} SFR
(9 \msun\ yr$^{-1}$ inferred from SED fitting) is roughly double our
median LAE SFR (5 \msun\ yr$^{-1}$ from dust-corrected
\ha). \citet{masters14} present an even higher-SFR sample of EELGs,
with average SFR = 29 \msun. 

Despite these differences, the EELG samples share many physical properties with
the KBSS-\lya\ LAEs, including low inferred
metallicities. \citet{mas14} calculate oxygen abundances for 7 EELGs
from the direct $T_e$ method or strong-line indicators, finding 12 +
log(O/H) = 7.45$-$8.25 and a median value of 7.90. \citet{masters14}
find higher typical abundances via the N2 estimator for their EELG
sample, with median 12 + log(O/H) = 8.34 and individual values ranging
from 8.67 (roughly solar) to $<$7.82. Our derived $T_e$ metallicity is
consistent with these median values within 1$\sigma$, although neither
EELG sample utilizes the +0.24 dex correction we assume for our $T_e$
measurement (Sec.~\ref{subsec:4364}). If we neglect this factor for
consistency, then our estimated oxygen abundance 12 +
log(O/H) = 7.80$\pm$0.17 is lower than either EELG sample (but still
consistent with the \citealt{mas14} median within 1$\sigma$). Similarly, our
lowest-metallicity objects appear to be at least as low as the
individual EELGs, with at least 5 LAEs in our sample having lower
metallicities than the lowest EELGs according to
our analysis in Sec.~\ref{subsec:lowz}. \citet{masters14}
observe a turnover in the R23-N2 plane at 12 + log(O/H) $\lesssim$
8.4, mirroring the turnover seen in the O3 distribution of
our faintest LAEs
(Figs.~\ref{fig:o3_rmag}~\&~\ref{fig:lowZ}). Similarly,
\citet{henry13b} find a downturn in both R23 and O3 at
$M_*\lesssim10^{8.8}\msun$ in a sample of $z\sim 2$ EELGs. This
evidence suggests that both the typical and extreme metallicities
observed in our LAE sample are at least as low as those measured by
other emission-line selections, which is likely consistent with the
lower masses and SFRs inferred for our sample, as well as their higher redshifts.

While galaxy selections based on line emission are effective at
identifying low-mass, low-metallicity galaxies, it is important to
consider whether all such galaxies are selected by this method; only
$\sim$50\% of $L\sim L_*$ LBGs exhibit \lya\ in emission
\citep{sha03,ste10}. A relatively unbiased selection of intrinsically-faint galaxies may be achieved
through gravitational lensing, which has proven fruitful at selecting
galaxies with $L\ll L_*$ at $z\approx1.5-3$ (e.g.,
\citealt{alavi14,stark14}). \citet{stark14} present 
17 UV continuum selected gravitationally lensed galaxies with masses
$M_*\approx2.0\times10^6-1.4\times10^9$\msun. While not selected based
on line emission, these galaxies exhibit emission in many UV metal
lines including \ion{N}{4}], \ion{O}{3}], \ion{C}{4}, \ion{Si}{3}],
and \ion{C}{3}]. \citet{stark14} find that \lya\ emission closely
tracks the emission of other high-excitation lines such as
\ion{C}{3}], as was previously seen by \citet{sha03}. Furthermore, all 11 galaxies for which \lya\ fell
within the observed passband exhibited \lya\ emission, with 10/11
galaxies having $\wla\gtrsim20$\AA. This result is consistent with
recent work by \citet{oyarzun16}, who find that the fraction of \lya\
emitters increases as galaxy mass decreases, with $\sim$86\% of
galaxies with $10^{7.6}<M_*/\msun\leq10^{8.5}$ at $3<z<4.6$ showing
\lya\ in emission with a typical equivalent width $\wla\approx
36$\AA. The high incidence of \lya\ emission at low galaxy masses
therefore suggests that our sample of faint LAEs  is likely to be
fairly representative of the total population of faint, low-mass,
star-forming galaxies at $z\approx2-3$.

\citet{stark14} also conducted photoionization modeling of the rest-UV
emission lines in 4 galaxies from their sample. This subset has
inferred stellar masses $10^{7.46} \leq M_*/\msun \leq 10^{8.06}$
and oxygen abundances 7.29 $\leq$ 12 + log(O/H) $\leq$ 7.82 (0.04
$\leq Z/Z_\odot \leq$ 0.13). These abundances are similar to those
inferred for our faintest LAEs. \citet{stark14} find ionization
parameters $-2.16\leq $log$U\leq-1.84$ using
(single-star) stellar models from \citet{bruzual03}; these values are similar
to those inferred for our LAE sample using the S99 models. Based on
our analysis in Sec.~\ref{sec:models}, we expect that \citet{stark14}
would have inferred lower ionization parameters using stellar models
including intrinsically-harder ionizing spectra similar to the BPASS
models we employ here (discussed further in Sec.~\ref{subsec:origins} below).

Finally, recent rest-optical galaxy photometry at $z=3-7$ has
indicated that extreme line emission is ubiquitous in the earliest
galaxy populations (e.g.,
\citealt{stark13,schenker13a,smit14}). In particular, \citet{smit14} find that 
bright, lensed galaxies ($M\sub{UV}\approx-20$) at $z\sim7$ have
typical rest-frame equivalent widths of EW([\ion{O}{3}]+\hb) $>$ 637\AA,
including 4 extreme objects with EW $\approx$ 1500\AA. The KBSS LAE
composite has EW([\ion{O}{3}]+\hb) $=$ 829\AA, comparable to these
high-$z$ samples and significantly higher
than the LM1 LBG composite from \citet{ste16}, which has EW([\ion{O}{3}]+\hb) $=$
238\AA. The 4 LAEs in our sample that are undetected in HST/WFC3
(Fig.~\ref{fig:o3_rmag}) require EW([\ion{O}{3}]+\hb) $\gtrsim$
1000$-$1700\AA, which mirrors the extreme subsample of \citet{smit14}
and further indicates the resemblance between the faintest galaxies at
$z\approx 2-3$ and their progenitors at the highest redshifts. 

%

In summary, the KBSS-\lya\ LAEs presented in this paper appear
to have similar nebular properties to the faintest galaxies and most
extreme line-emitters selected by other surveys at comparable
redshifts (as well as the reionization epoch), although our sample is the
first to include a large sample of rest-optical spectra of galaxies in this
luminosity and redshift regime. 

\subsection{Origins of strong line emission in faint galaxies}\label{subsec:origins}

We now turn our discussion to the physical interpretation of our
observations: why is \lya\ emission so strongly linked to the nebular
line properties and continuum luminosities of galaxies?

The strong nebular line emission of faint galaxies is partially a
result of the relationship between mass and metallicity; as discussed
above, the low metallicities of low-mass galaxies naturally produce
strong emission in many high-excitation nebular lines. However, the
connection between \lya\ emission and the excitation state of nebular
gas is less clear.

One natural explanation of this link may be the shape of the incident
stellar spectra that drive the line emission from the nebular
gas. Both high excitation (parameterized by the O3 ratio) and high \wla\
are indicators of an intrinsically hard incident spectrum. The intrinsic \lya\
luminosity of an ionization-bounded \ion{H}{2} region is approximately
proportional to the total production of ionizing photons, so \wla\ is
a measure of the ratio of ionizing to non-ionizing incident UV flux (modulo the
subsequent scattering and absorption of the \lya\ photons). Similarly, the
O3 ratio is strongly dependent on the shape of the ionizing spectrum,
which may be conceptualized as the ratio of $\sim$1$-$2 Ryd photons to
$\sim$2$-$4 Ryd photons. Stellar populations that produce high numbers
of ionizing photons to non-ionizing photons will typically also
produce harder spectra within the hydrogen-ionizing band, which may generate
the observed relation.

\begin{figure}
\center
\includegraphics[width=\linewidth]{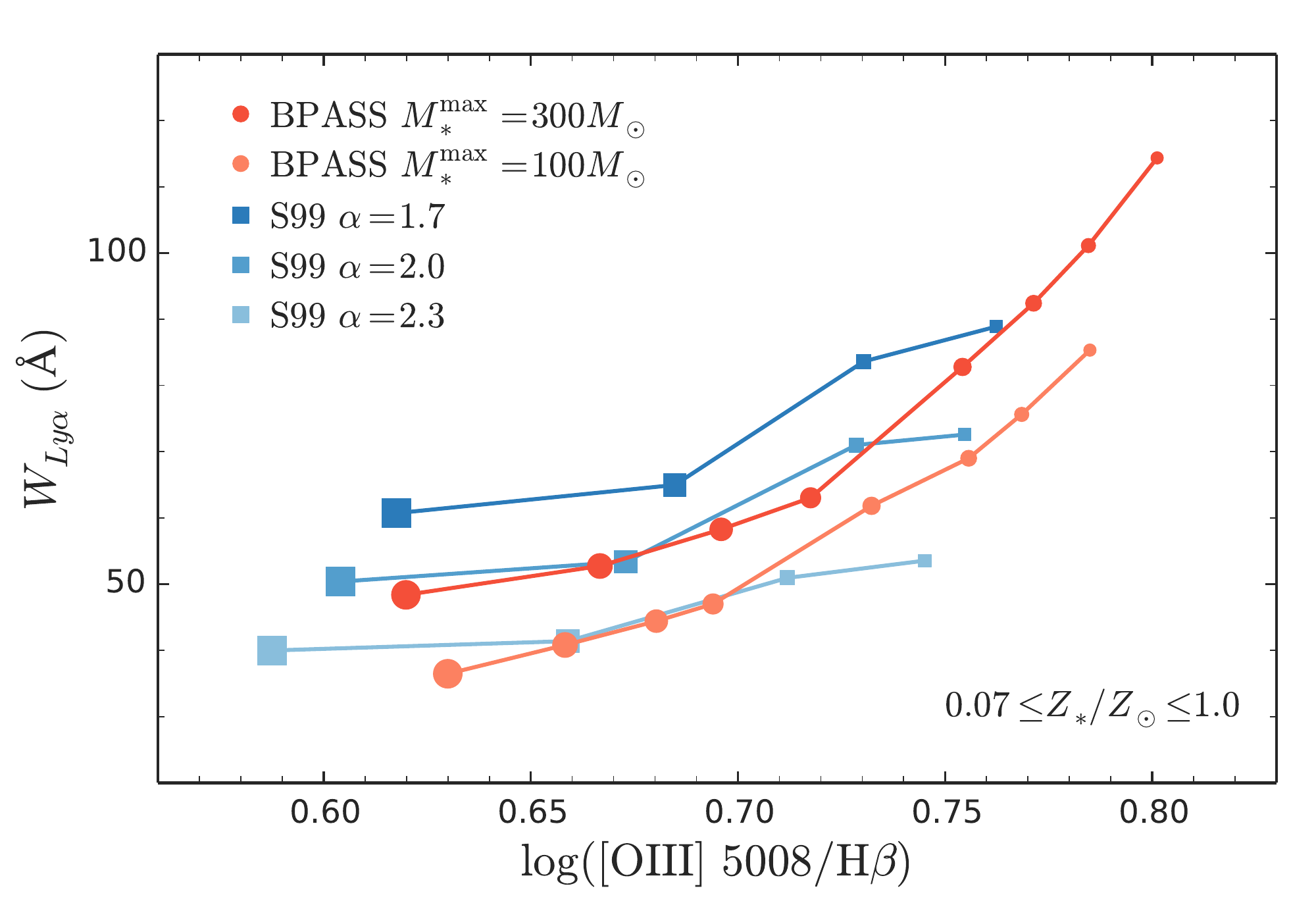}
\caption{Predicted \wla\ and O3 ratios from Cloudy photoionization
  models for BPASS (circles) and S99 (squares) model spectra of varying stellar
  metallicities ($0.07Z_\odot\le Z_* \le Z_\odot$ with symbol size increasing with
  $Z_*$). The predictions are computed for a fiducial nebular 
  metallicity $Z\sub{neb}=0.2Z_\odot$ and ionization parameter
  log$U=-2.1$. We plot multiple IMFs for each model, varying the upper
  mass cutoff for the BPASS models ($M\super{max}_*=300\msun$ or
  $100\msun$) and the IMF slope for the S99 models ($\alpha=-2.3$,
  $-2.0$, or $-1.7$). In all cases, the harder stellar spectra produce
  higher \wla\ and O3 ratios, providing a qualitative explanation for
  the BPT-\lya\ relation and other nebular trends explored in this
  paper. Notably, however, the standard-IMF S99 models do not produce
  values of \wla\ consistent with typical LAE measurements.}
\label{fig:wlya_vs_o3hb}
\end{figure}

\begin{figure}
\center
\includegraphics[width=\linewidth]{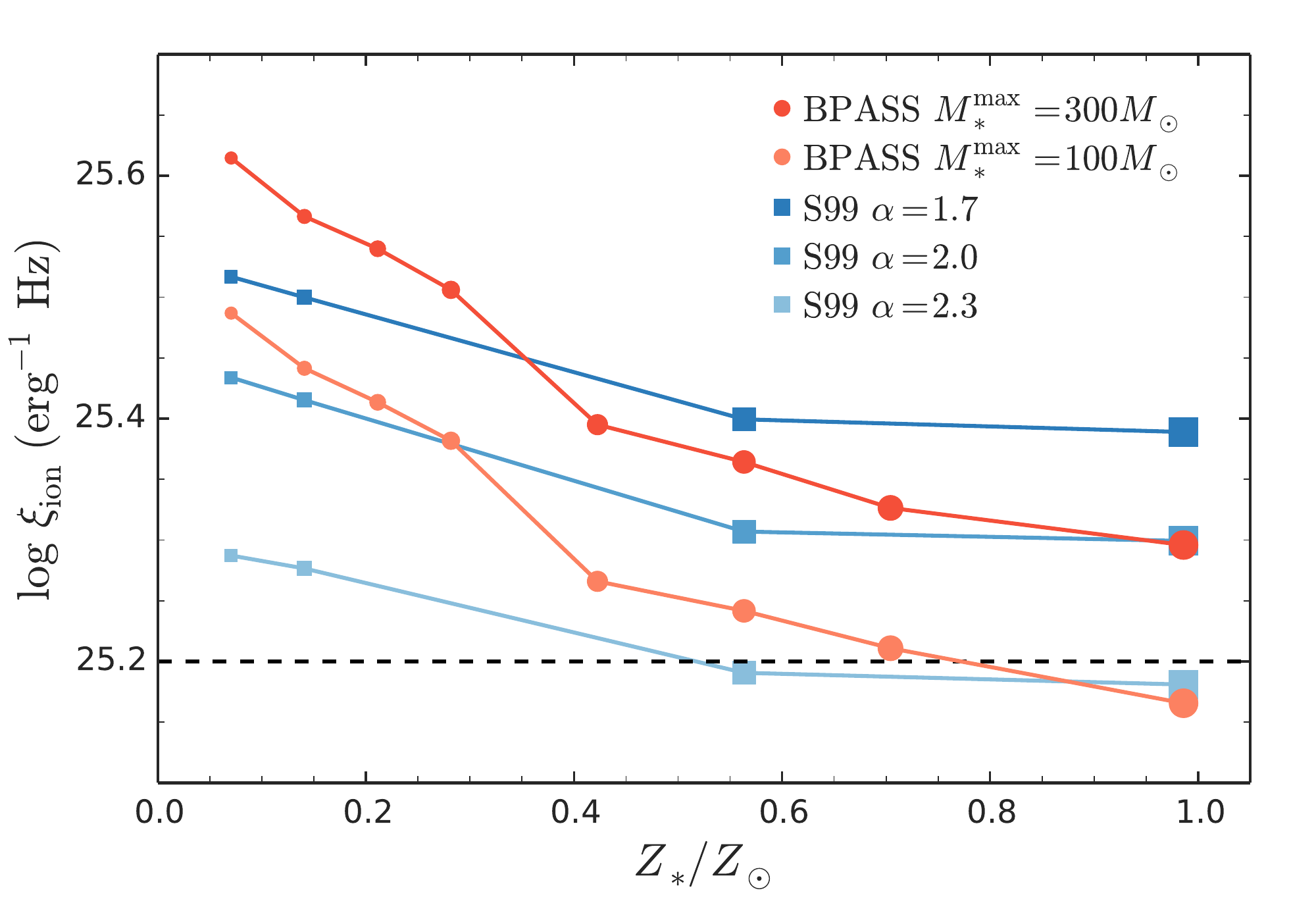}
\caption{Predicted $\xi\sub{ion}$ ratios for the same BPASS and S99
  model spectra presented in Fig.~\ref{fig:wlya_vs_o3hb}.
 The same
  low-metallicity stellar models that produce acceptable fits to the
  N2-BPT and \wla\ constraints (i.e., the BPASS models) likewise
  predict $\xi\sub{ion}$
 ratios significantly higher than the
  ``canonical'' values assumed by typical models of reionization
  (e.g., \citealt{rob15}; dashed line at
  log$(\xi\sub{ion}/[\mathrm{erg}^{-1} \mathrm{Hz}])=25.2$). These
  stellar models therefore naturally 
  produce high ionization parameters and reduce the
  requirements for ubiquitously high LyC escape fractions
  among primeval galaxies.}
\label{fig:xi_vs_zstar}
\end{figure}

 We provide an example of the above argument using photoionization
models in Fig.~\ref{fig:wlya_vs_o3hb}. For each of the stellar models,
we consider a variety of stellar metallicities and IMF parameters, but
the general relation remains the same: models that produce high O3
ratios likewise predict high intrinsic values of \wla. The trend is
especially clear for the BPASS models, for which the stellar
metallicity strongly modulates the total ionizing photon production
(Fig.~\ref{fig:xi_vs_zstar}) and thereby the secondary production of
\lya\ photons. 

The parameter $\xi\sub{ion}$ is defined as the number of
hydrogen-ionizing photons produced by a stellar population per unit 
UV luminosity (typically normalized at 1500\AA). A value
log($\xi\sub{ion}/$erg$^{-1}$ Hz) = 25.2 has been assumed by recent work modeling
reionization (e.g., \citealt{rob13,rob15}), which is
appropriate for a single-star population with a typical IMF in
steady-state (e.g., the $\alpha=2.3$ S99 model in
Fig. ~\ref{fig:xi_vs_zstar}). The BPASS models produce much higher values of
$\xi\sub{ion}$ at low metallicity, even with the the same IMF
($M_*\super{max}=100\msun$). Because reionization constraints impose a
requirement on the total galactic emission of ionizing photons, which is
proportional to $\xi\sub{ion}\times f\sub{esc,LyC}$, these models therefore
reduce the global LyC escape fraction required to reionize the
Universe (e.g., \citealt{ma16_fesc}).

In addition, these high values of $\xi\sub{ion}$ also provide a natural
explanation for the high ionization parameters inferred for our LAE
sample. At fixed gas density, the ionization parameter $U$ is
proportional to the number density of ionizing
photons\footnote{Technically, $U$ in the Cloudy photoionization code
  is proportional to the number density of photons at the Lyman
  limit.}. While the expected value of $U$ in high-$z$ \ion{H}{2}
regions may not be obvious {\it a priori}, a variation of $\sim$0.7
dex (as appears to be present between the LBG and LAE samples;
Fig.~\ref{fig:bpt-cloudy}) would be extremely difficult to explain without
a significant difference in $\xi\sub{ion}$ between the two populations. While
a very young stellar population age could elevate
$\xi\sub{ion}$ in individual galaxies, our measurements of high $U$
for a large sample of LAEs--and their correspondence to typical
galaxies in their mass and luminosity range, as argued above--suggests
that these faint galaxies efficiently produce large numbers of
ionizing photons {\it in steady state}.  

The strong dependence of $\xi\sub{ion}$ on $Z_*$ among the BPASS
models provides an alternative mechanism to modulate the ionization
parameter. \citet{san16} identify an  
anti-correlation between gas-phase metallicity and $U$ among
continuum-selected galaxies from
the MOSDEF survey. The similar trend seen here
among the KBSS LAEs and LBGs which may indicate
that the effect of metallicity on the hardness of the stellar
spectra drives the large observed variation in $U$ with galaxy luminosity.

Lastly, as discussed in Sec.~\ref{subsec:bptmodel}, the hardness of
the ionizing field and the ionization parameter produce somewhat
degenerate effects on the nebular excitation (O3): the softer S99
models require significantly higher values of $U$ to produce the same
values of the O3 ratio. By this token, both the {\it normalization}
and the {\it shape} of the BPASS ionizing spectra are favorable for high nebular excitation. Firstly,
their intrinsic hardness can produce high O3 with values of $U$ that
are only moderately elevated with respect to more luminous
populations. Secondly, their high $\xi\sub{ion}$ values at low
metallicities naturally produce the large total number of ionizing
photons required explain these $U$ values at apparently
fixed\footnote{While current galaxies samples (KBSS, MOSDEF) show a
  large dispersion in inferred gas density, no trend is seen with
  metallicity. In fact, 2 of the 3 extreme-excitation LBGs in
  \citet{ste14} have above-average inferred densities ($n_e\approx
  600-1600$ cm$^{-3}$), which would require {\it even higher} values
  of $\xi\sub{ion}$ to produce the inferred ionization parameters.} gas density.

To summarize, invoking a stellar population model in which the spectra
become significantly harder at low metallicities (whether via the
specific BPASS model suite or a set of similar output model spectra)
will self-consistently reproduce the strong \lya\ and nebular line ratios
observed for faint, high-redshift galaxies.

\subsection{\lya\ emission and LyC leakage}\label{subsec:LyC}

\begin{figure}
\center
\includegraphics[width=\linewidth]{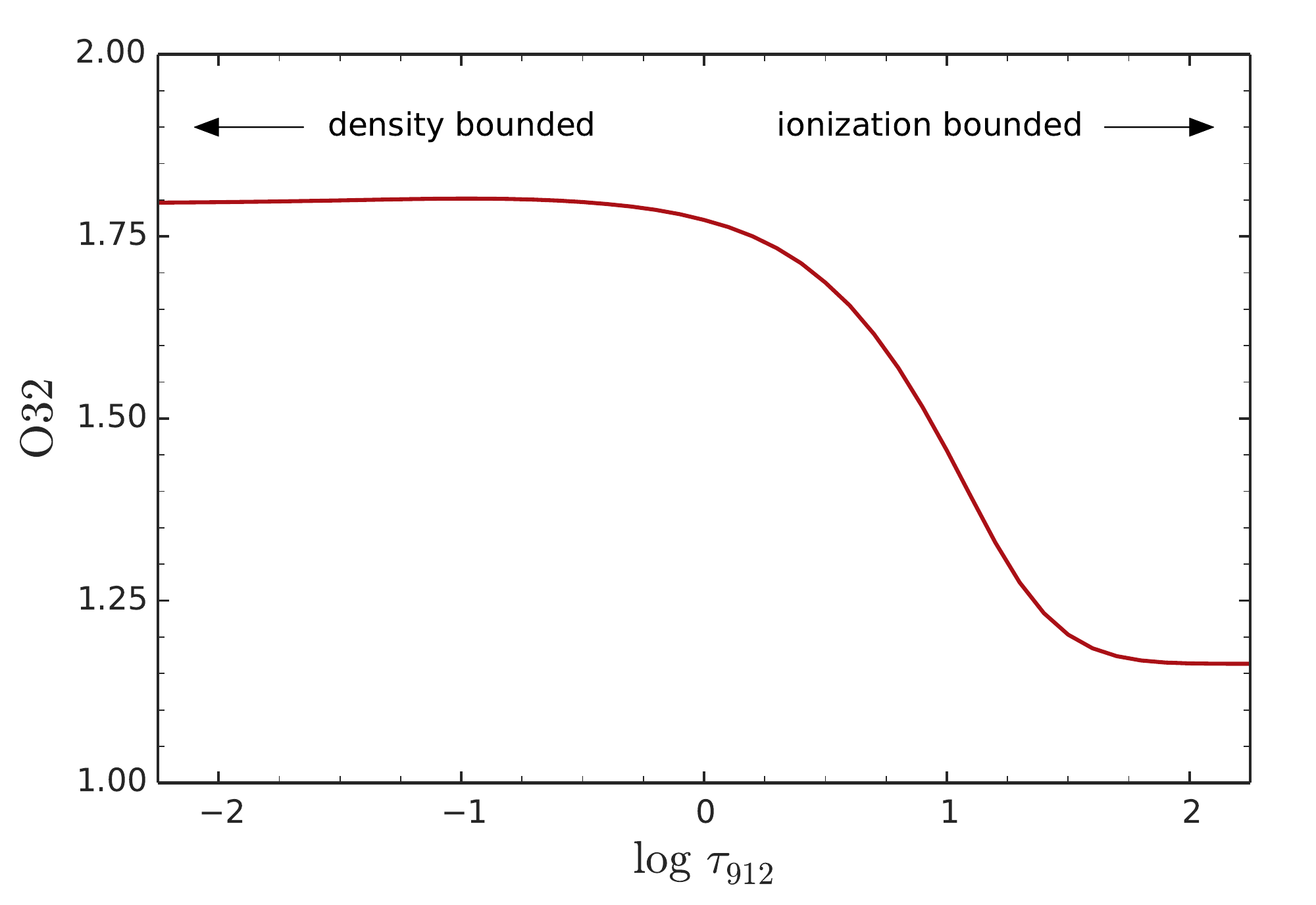}
\caption{Predicted O32 ratios from Cloudy photoionization
  models for a plane-parallel slab of gas with varying total optical
  depth at the Lyman limit ($\tau_{912}$). Incident radiation field is that of the
  BPASS $Z_*=0.07Z_\odot$ model, with log$U=-2.1$ and $Z\sub{neb}=0.2$ as in
  the preceding figures. Optically thin (that is, density-bounded;
  $\tau_{912}\lesssim 1$) ionized regions are characterized by elevated O32 ratios.}
\label{fig:O32_vs_lytau}
\end{figure}

As discussed in Sec.~\ref{subsubsec:lyavsexcitation}, there are
indications that \lya-emitting galaxies have elevated O32 ratios with
respect to typical galaxies. Some studies have suggested that these
ratios are indicative of density-bounded \ion{H}{2} regions, in which
the entire star-forming cloud becomes ionized. Such regions would be
expected to lack a partially-ionized boundary wherein low-ionization
species may dominate the local line emission and thereby lower the
luminosity-averaged measurement of the ionization ratio (see
discussion by e.g., \citealt{jaskot13,nak13,nak14}). Some evidence for this
behavior is seen in local observations of \ion{H}{2} regions using the
ionization parameter mapping technique (IPM;
\citealt{pellegrini12,zastrow13}). In particular, the IPM measurements
of the Large Magellanic Cloud by \citet{pellegrini12} indicate that
optically-thick \ion{H}{2} regions are surrounded by ionization fronts
that are dominated by low-ionization species (e.g., high
[\ion{S}{2}]/[\ion{S}{3}] ratios), whereas optically-thin (i.e.,
density-bounded) regions are dominated by high-ionization species
throughout. As an illustrative example of this phenomenon,
Fig.~\ref{fig:O32_vs_lytau} reproduces the results of a Cloudy
photoionization simulation in which a slab of gas with a variable,
uniform column density is illuminated by a BPASS model
spectrum.\footnote{Specifically, we run a set of plane-parallel Cloudy
  simulations with the $Z_*=0.07Z_\odot$ BPASS input spectrum,
  $Z\sub{neb}=0.2Z_\odot$, log$U = -2.1$, and a variable stopping
  criterion based on the total optical depth at $\lambda=912$\AA.} In the
density-bounded case (optically thin at the Lyman limit;
$\tau_{912}\ll1$), the O32 ratio is elevated by $\sim$0.6 dex with
respect to the ionization-bounded case
($\tau_{912}\gg1$).\footnote{Note that the normalization of the curve
  in Fig.~\ref{fig:O32_vs_lytau} is dependent on the precise choice of
incident radiation field and log$U$, but the predicted {\it difference} in
O32 between the density-bounded and ionization-bounded cases is
insensitive to these parameters.}

\begin{figure}
\center
\includegraphics[width=\linewidth]{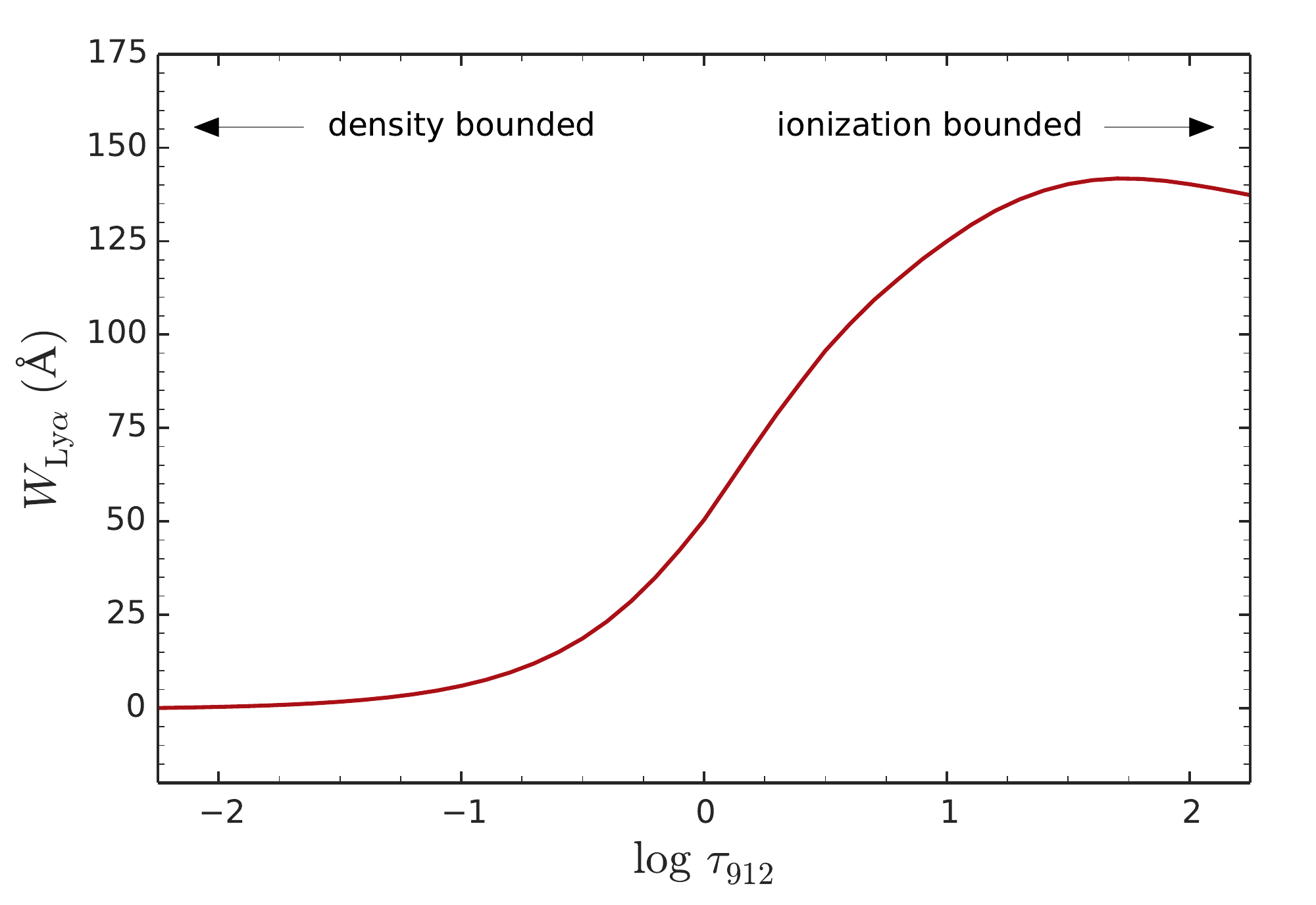}
\caption{Predicted \wla\ ratio for the same photoionization models
  presented in Fig.~\ref{fig:O32_vs_lytau}. Optically-thin ionized
  regions produce little \lya\ emission, which peaks at column
  densities of \ion{H}{1} that are moderately optically thick to
  ionizing photons. Note that the Cloudy photonization code is not
  optimized to reproduce the behavior of \wla\ at the highest values of
  $\tau_{912}$, where \lya\ transmission is dominated by resonant
  scattering in a dusty medium. }
\label{fig:wlya_vs_lytau}
\end{figure}

Such density-bounded regions are highly interesting for studies of
EoR galaxies because they produce large local escape fractions of
ionizing photons (and potentially of \lya\ photons, although \lya\
{\it production} may be suppressed; Fig.~\ref{fig:wlya_vs_lytau}). There are
observational indications that galaxies with high O32 ratios are more
likely to leak ionizing (LyC) photons: several recently-discovered LyC
leakers at low-redshift \citep{izotov16a,izotov16b} show high ionization states (O32
$>0.6$). However, these galaxies were selected as likely LyC emitters
based on these same nebular line ratios, and other low-$z$ LyC
leakers show smaller values of O32
\citep{leitet13,borthakur14,leitherer16}. At $z\sim3$, two confirmed
LyC leakers are known: Ion2 \citep{debarros16} and
Q1549-C25 \citep{shapley16}. Ion2 was selected based on
broadband continuum colors according to the method of
\citet{vanzella15} and is associated with extreme line ratios (O32
$>$ 1 \& EW\sub{rest}([\ion{O}{3}]) = 1500\AA;
\citealt{vanzella16}). Conversely, Q1549-C25 appears to have more
moderate line emission (EW\sub{rest}([\ion{O}{3}]+\hb) = 256\AA)
typical of KBSS LBGs.

Given the diversity of known LyC-leakers, is unclear that real high-$z$ density-bounded \ion{H}{2}
regions must necessarily have high O32 ratios. Harder ionizing
photons have smaller ionization cross-sections and longer mean free
paths than lower-energy photons, which could cause \ion{H}{2} regions
to become hotter and more highly ionized at large radii. This effect may be stronger at high
redshift, where we have argued that low-metallicity stellar
populations are likely to produce hard ionizing spectra. In addition,
any dominance of low-ionization species within ionization
fronts of individual \ion{H}{2} regions need not translate to a low O32
ratio in the luminosity-weighted, galaxy-averaged observations
conducted for distant galaxies. 

Furthermore, \lya\ production is extremely weak in \ion{H}{2}
regions with uniformly low gas column densities
(Fig.~\ref{fig:wlya_vs_lytau}). In more realistic simulations of \lya\
production and escape in a clumpy, multi-phase ISM, \cite{dijkstra16}
find that both ionizing and \lya\ photons escape through
low-column-density sightlines, in agreement with observational
indicators that a patchy distribution of gas is required for efficient photon
escape. In T15, we found that the escape fraction of \lya\ photons was significantly
related to a galaxy's covering fraction of neutral (or low-ionization)
gas. Similarly, the low-$z$ LyC leaker discovered by
\citet{borthakur14} has a very low value of O32 $=
-0.5$, but also exhibits a low covering fraction of
low-ionization gas. Photons may also be scattered or extinguished far
from their original \ion{H}{2} regions; \citet{ste11} found that star-forming galaxies emit \lya\
radiation at large radii even when they appear to be \lya\ absorbers in slit
spectroscopy. However, the galaxies with high spatially-integrated
values of \wla\ are those with centrally-concentrated \lya\
emission, suggesting that the bulk of \lya\ (and potentially LyC)
absorption occurs at small galactocentric radii.

Together, these
observations indicate that the distribution of \ion{H}{1} gas on many
spatial scales plays a strong role in scattering and extinguishing
Lyman radiation, a fact which likely produces much of the scatter in the
relationships between \wla\ and nebular properties explored in
Secs.~\ref{sec:lyaneb}$-$\ref{sec:models}. Nonetheless, both the
nebular properties of the KBSS-\lya\ LAEs and their neutral gas
distributions presented in T15 indicate that they are strong
candidates for Lyman-continuum leakage and ideal analogs of EoR
galaxies.

\section{Conclusions} \label{sec:conclusions}

In this paper, we have shown that the \lya-emitting properties of
galaxies (including 60 faint LAEs and 368 brighter LBGs) are closely
linked to the properties of their embedded star-forming regions, as
probed by rest-optical and rest-UV spectroscopy. This sample of galaxies from the
KBSS and KBSS-\lya\ represents the largest set of combined \lya\ and
nebular-line spectroscopy of galaxies at any redshift and spans a
large range of galaxy properties, including luminosities, SFRs,
masses, extinctions, ionization states, and net \lya\ emission.

By constructing composite spectra from our LAE sample, we also set new
constraints on the physical properties of $L\approx0.1L_*$ galaxies at
$z\approx2-3$.  The primary conclusions of this work are summarized below:

\begin{enumerate}[leftmargin=*]

\item Faint LAEs have extremely high nebular excitation
  (parameterized by the [\ion{O}{3}] $\lambda$5008/\hb\ ratio)
  consistent with most extreme LBGs in current surveys, which are
  likely the low-metallicity tail of the galaxy distribution. {\it
    Fig.~\ref{fig:BPT}; Sec.~\ref{sec:ratios}}
\item A 2.8$\sigma$ detection of the [\ion{O}{3}] $\lambda$4364 auroral
  emission line suggests that the \ion{H}{2} regions in these LAEs
  have high electron temperatures with respect to more massive LBGs
  ($T\sub{$e$,LAE}=1.78\pm0.33\times10^4$K). Calculating a metallicity
  via the ``direct'' ($T_e$) method gives a typical LAE oxygen abundance
  of 12 + log(O/H) = 8.04$\pm$0.19
  ($Z\sub{neb}=0.22^{+0.12}_{-0.08} Z_\odot$) after correcting for
  the $-$0.24 dex offset of metallicities derived from this collisionally-ionized
  transition. This metallicity is consistent with our limits 
  from the O3N2 and N2 strong-line indicators.  {\it
    Fig.~\ref{fig:hgo3}; Secs.~\ref{subsec:4364}$-$\ref{subsec:zgas}}
\item The continuum-faintest LAEs in our sample show
  evidence for the turnover in the [\ion{O}{3}] $\lambda$5008/\hb\
  ratio that occurs at very low metallicities, and six LAEs appear to
  have oxygen abundances 12 + log(O/H) $\approx$ 6.9$-$7.4
  ($Z\sub{neb}\approx0.02-0.05$). {\it
    Figs.~\ref{fig:o3_rmag}~\&~\ref{fig:lowZ}; Sec.~\ref{subsec:lowz}}
\item Across a broad range of LBG and LAE \lya\ emissivities, the
  locations of galaxies in the N2-BPT plane vary systematically with \lya\
  equivalent width. We find that the variation of dust attenuation
  with metallicity plays a minor role in this relation, but the
  ionization and excitation state of the star-forming regions are much
  more strongly correlated with the net \lya\ emission. In particular,
  the photoionization models of the LAEs require ionization parameters
  log$U\approx-2.1$ for typical LAEs, 0.7 dex higher than typical
  LBGs. {\it
    Figs.~\ref{fig:BPTlya}$-$\ref{fig:minlogU_vs_wlya}; Sec.~\ref{sec:lyaneb}}
\item The rest-frame optical spectra of the KBSS-\lya\ LAEs
  indicate stellar populations that produce harder
  spectra than typical stellar population models. Successful stellar
  models include those with very low stellar 
  metallicities (which are consistent with the low Fe/O abundances reported
  elsewhere), and those that include the effects of binary interaction
  on the evolution of massive stars. In particular, we are unable to
  reproduce the properties of the highest-\wla\ LAEs except with
  stellar models that include binary evolution. {\it
    Figs.~\ref{fig:bpt-cloudy}$-$\ref{fig:minlogU_vs_wlya};
    Sec.~\ref{sec:models}}
\item In general, the nebular properties of faint LAEs at
  $z\approx2-3$ indicate the production and escape of large numbers of ionizing
  photons, which we argue is typical of low-luminosity galaxies at
  high redshift. Combined with our previous work indicating the low
  covering fraction of neutral gas around these faint galaxies, we
  suggest that analogous galaxies at $z\gtrsim 6$ may be more effective at
  reionizing the Universe than is typically assumed in current
  models. {\it Figs.~\ref{fig:wlya_vs_o3hb}$-$\ref{fig:wlya_vs_lytau};
  Sec.~\ref{sec:discussion}}
\end{enumerate}

Future work will include measuring additional global properties of
these galaxies, including the variation of nebular excitation and
\lya\ production with the physical sizes, masses, and star-formation
rates of individual LAEs. Additional indicators of the nebular
properties (including ionization state and strong-line metallicity
indicators) will provide valuable cross-checks of the measurements
inferred here. In a few years, JWST will be able to obtain similar spectra of
$L\sim0.1L_*$ galaxies without being limited by ground-based
atmospheric windows, facilitating further analysis of the variation of
these galaxy properties across populations and redshifts.

\acknowledgments

\noindent We thank Eliot Quataert, Mariska Kriek, and Dawn Erb for extremely
useful discussions. In addition, we thank the organizers of the {\it 
  Escape of Lyman radiation from galactic labyrinths} conference at the
Orthodox Academy of Crete in April 2016; this paper was much improved
by the talks and discussion that took place at that meeting. We are
also grateful for the insightful comments of our anonymous referee.
This paper uses data collected through Keck program 2015B\_U42M, and we
are indebted to the staff of the W.M. Keck Observatory who keep the 
instruments and telescopes running effectively. We also wish to extend
thanks to those of Hawaiian ancestry on whose sacred mountain we are
privileged to be guests. This work has been supported in part by the US
National Science Foundation through grants AST-0908805 and
AST-1313472. RFT receives for support from the
Miller Institute for Basic Research in Science at the University of
California, Berkeley.


\bibliographystyle{apj_new}
\bibliography{thesis}

\end{document}